\documentclass[12pt]{article}
\usepackage[dvips]{graphicx}
\usepackage{epsfig,cite}
\usepackage {amsmath}
\usepackage{amssymb}
\usepackage{slashed}
\newcount\timecount
\newcount\hours \newcount\minutes  \newcount\temp \newcount\pmhours
\hours = \time
\divide\hours by 60
\temp = \hours
\multiply\temp by 60
\minutes = \time
\advance\minutes by -\temp
\def\hour{\the\hours}
\def\minute{\ifnum\minutes<10 0\the\minutes
            \else\the\minutes\fi}
\def\clock{
\ifnum\hours=0 12:\minute\ AM
\else\ifnum\hours<12 \hour:\minute\ AM
      \else\ifnum\hours=12 12:\minute\ PM
            \else\ifnum\hours>12
                 \pmhours=\hours
                 \advance\pmhours by -12
                 \the\pmhours:\minute\ PM
                 \fi
            \fi
      \fi
\fi
}

\def\monthname{\relax\ifcase\month 0/\or January\or February\or
   March\or April\or May\or June\or July\or August\or September\or
   October\or November\or December\else\number\month/\fi}

\def\bold#1{\setbox0=\hbox{$#1$}%
     \kern-.025em\copy0\kern-\wd0
     \kern.05em\copy0\kern-\wd0
     \kern-.025em\raise.0433em\box0 }

\def\beq{\begin{equation}}
\def\eeq{\end{equation}}

\def\ga{\mathrel{\raise.3ex\hbox{$>$\kern-.75em\lower1ex\hbox{$\sim$}}}}
\def\la{\mathrel{\raise.3ex\hbox{$<$\kern-.75em\lower1ex\hbox{$\sim$}}}}
\def\gev{{\rm \, Ge\kern-0.125em V}}
\def\tev{{\rm \, Te\kern-0.125em V}}
\def\gyr{{\rm \, G\kern-0.125em yr}}

\def\tbt{\tan \beta}
\def\tanb{\tbt}

\def\gappeq{\mathrel{\rlap {\raise.5ex\hbox{$>$}}
{\lower.5ex\hbox{$\sim$}}}}
\def\lappeq{\mathrel{\rlap{\raise.5ex\hbox{$<$}}
{\lower.5ex\hbox{$\sim$}}}}
\def\Toprel#1\over#2{\mathrel{\mathop{#2}\limits^{#1}}}

\def\stau{\widetilde \tau}
\def\snutau{\widetilde {\nu}_\tau}

\def\mpl{M_{\rm Pl}}
\def\mchi{m_{\tilde \chi}}

\def\m12{m_{1\!/2}}

\newcommand\iso[2]{\mbox{${}^{#2}${\rm #1}}}
\def\he#1{\iso{He}{#1}}
\def\be#1{\iso{Be}{#1}}
\def\li#1{\iso{Li}{#1}}
\def\bor#1{\iso{B}{#1}}

\def\stau{\tilde{\tau}}

\def\mgrav{m_{3/2}}
\def\grav{\widetilde{G}}
\def\qbar{\overline{q}}
\def\mpl{M_{P}}
\def\mchi{m_{\chi}}
\def\cosw{\cos \theta_W}
\def\sinw{\sin \theta_W}

\def\bea{\begin{eqnarray}}
\def\eea{\end{eqnarray}}

\def\pfrac#1#2{\left(\frac{#1}{#2}\right)}
\def\avg#1{\langle #1 \rangle}
\def\beqar{\begin{eqnarray}}
\def\eeqar{\end{eqnarray}}

\def\lya{\mbox{Ly$\alpha$}}
\def\xlya{\mbox{$X$Ly$\alpha$}}
\def\galph{\Gamma_{2p \rightarrow 1s}}

\def\anion{{\cal A}}

\usepackage{lscape}
\usepackage[dvips]{graphicx}

\def\avg#1{\left\langle #1 \right\rangle}

\def\beq{\begin{equation}}
\def\eeq{\end{equation}}

\begin{document}

\begin{titlepage}
\pagestyle{empty}
\rightline{KCL-PH-TH/2012-36, LCTS/2012-18, CERN-PH-TH/2012-223}
\rightline{UMN--TH--3117/12, FTPI--MINN--12/29}
\vspace{0.5cm}
\begin{center}
{\large {\bf Metastable Charged Sparticles and the Cosmological \li7 Problem}}

\end{center}
\vspace{0.5cm}
\begin{center}
{\bf Richard~H.~Cyburt}$^{1}$, {\bf John~Ellis}$^{2,3}$,
{\bf Brian~D.~Fields}$^{4}$, \\
\vskip 0.1in
{\bf Feng Luo}$^{2,5}$, {\bf Keith~A.~Olive}$^{5,6}$,
and {\bf Vassilis~C.~Spanos}$^{7}$\\
\vskip 0.2in
{\small {\it
$^1${Joint Institute for Nuclear Astrophysics (JINA), National Superconducting
Cyclotron Laboratory (NSCL), Michigan State University, East Lansing,
MI 48824, USA}\\ 
$^2${Theoretical Physics and Cosmology Group, Department of Physics, King's College London, London~WC2R 2LS, UK}\\
$^3${TH Division, Physics Department, CERN, CH-1211 Geneva 23, Switzerland}\\
$^4${Departments of Astronomy and of Physics, \\ University of Illinois, Urbana, IL 61801, USA}\\
$^5${School of Physics and Astronomy, \\
University of Minnesota, Minneapolis, MN 55455, USA}\\
$^6${William I. Fine Theoretical Physics Institute, School of Physics and Astronomy,\\
University of Minnesota, Minneapolis, MN 55455,\,USA}\\
$^7${Institute of Nuclear Physics, NCSR ``Demokritos", GR-15310 Athens, Greece}} \\
}
\vspace{1cm}
{\bf Abstract}
\end{center}
{\small
We consider the effects of metastable charged sparticles on
Big-Bang Nucleosynthesis (BBN), 
including bound-state reaction rates and chemical effects.
We make a new analysis of the bound states of negatively-charged
massive particles with the light nuclei most prominent in BBN,
and present a new code to track their abundances, paying particular attention to that of \li7.
Assuming, as an example, that the gravitino is the lightest supersymmetric
particle (LSP), and that the lighter stau slepton, ${\tilde \tau_1}$,
is the metastable next-to-lightest sparticle
within the constrained
minimal supersymmetric extension of the Standard Model (CMSSM), 
we analyze the possible effects on the standard BBN abundances of ${\tilde \tau_1}$
bound states and decays for representative values of the gravitino mass.
Taking into account the constraint on
the CMSSM parameter space imposed by the discovery of the Higgs boson at the LHC,
we delineate regions in which
the fit to the measured light-element abundances is as good
as in standard BBN. We also identify regions of the CMSSM parameter
space in which the bound state properties, chemistry 
and decays of metastable charged sparticles can solve the cosmological \li7 problem.}


\vfill
\end{titlepage}

\section{Introduction}

The agreement of standard Big-Bang Nucleosynthesis (BBN) calculations
with the measured abundances of the light elements imposes important 
constraints on scenarios for new physics that involve massive
metastable particles~\cite{Lindley:1984bg} -\cite{Vasquez:2012dz}.
 If these particles are neutral, only the effects of
particles produced in the showers following their decays need to be taken
into account, but in the case of negatively-charged metastable particles $X^-$,
the formation of $(AX^-)$ bound states should also be considered, and are
very important~\cite{Maxim,HHKKY,CEFOS,Bird08,kkm3,jed08,grant2,Kawasaki08,Pospelov08,grant4,pp2}. 

The emergence of the primordial `lithium problem'
adds cosmological motivation to
the studying the effects of metastable particles on BBN.
As reviewed in refs.~\cite{Fields11,pp2,jp,Steigman07},
WMAP and other observations \cite{wmap7} have determined precisely
the cosmic baryon density and thus pinned down 
the one free parameter of standard BBN \cite{cfo2}.
Using this as an input, BBN makes precise predictions
for light-element abundances, and those of deuterium and \he4 are
in good agreement with obervations.  But the BBN expectations for \li7/H
based on the WMAP baryon density
are {\em higher} than the observed abundances by
factors of $2-4$, amounting formally to a $4-5\sigma$ discrepancy;
this is the cosmological lithium problem \cite{CFO,Coc03}.
Nuclear uncertainties \cite{Coc03,cfo4,coc05,Boyd10} and/or resonances
are all but excluded as solutions to 
the problem \cite{Cyburt09,Boyd10,Chakraborty10,Broggini12,O'Malley11,Kirsebom11}.  
There is the possibility that depletion plays a role in altering the \li7 abundance \cite{dep}.
However, these solutions typically have difficulty in explaining the thinness of the 
\li7 plateau \cite{li7obs} as well as the observation of \li6 in some halo stars \cite{li6obs}. The temperature scale used in the \li7 abundance determination has also been considered \cite{mr,hos} and it seems unlikely that a significant change in the \li7 abundance
is possible within reasonable uncertainties in the effective temperature.
We note, however, recent observations of lithium
in the interstellar medium of the metal-poor Small Magellanic 
Cloud test these systematics and are consistent with the halo-star
results~\cite{Howk12}.
Thus, the cosmological lithium problem seems increasingly likely to be 
real, and to point to new physics during or after BBN.

In a previous paper~\cite{CEFLOS1.5}, we extended analyses of the effects of particle showers
in the decays of metastable particles to include the most relevant uncertainties in nuclear reaction
rates. We applied our analysis to scenarios within the constrained minimal
supersymmetric extension of the Standard Model (CMSSM, see Appendix~B for its 
specification) in which the lightest
neutralino $\chi$ is the lightest supersymmetric particle (LSP), and the heavier,
neutral gravitino is metastable. Not only did we find regions of this CMSSM
parameter space where the cosmological light-element abundances agreed
with the measured values at least as well as in standard BBN, but we also
identified regions of this CMSSM parameter space where the cosmological \li7 
problem is alleviated and even potentially solved.
In this paper, we extend the analysis of~\cite{CEFLOS1.5} to include the $(AX^-)$
bound-state effects expected in the case of a negatively-charged metastable 
particle $X^-$. 

Bound-state effects were also discussed
in~\cite{CEFOS}, and here we update and supersede that analysis
incorporating qualitatively and quantitatively new rates and processes that were
not available at the time. To this end, we first review
our calculations of bound-state properties, and then turn to their effects on BBN. 
These include calculations of (1) bound-state recombination, which fixes the
abundances of various exotic `ions' such as $(p X^-), (\he4 X^-), (\be7 X^-)$, etc.,
and (2) bound-state catalysis, which causes additional changes in light-element 
production and destruction rates beyond the non-thermal reactions considered 
in~\cite{CEFLOS1.5}. Our new calculations of bound-state properties such as 
binding energies and charge radii are in reasonable agreement with other work, and
we use them to discuss the effects of uncertainties in the nuclear
inputs. For this purpose, we have compiled a complete and up-to-date list
of the relevant reactions, tabulated below. We have verified, 
using a simple driver code, that our recombination
rates give $(p X^-)$ and $(\he4 X^-)$ abundances in good agreement with
previous results~\cite{Maxim}.
We have then updated the BBN code used in~\cite{CEFOS, CEFLOS, CEFLOS1.5}
to include the recombination and catalysis rates, in a more complete,
accurate and systematic way than previously.

As an example of the application of our code, we consider the case of a
supersymmetric model in which the gravitino is the LSP, and the lighter stau
slepton, ${\tilde \tau_1}$ is the metastable NLSP. We work within the
framework of the CMSSM, and seek regions of its parameter space
where the consistency with the measured light-element abundances
of standard BBN calculations is at least maintained, and also look for
regions where the cosmological \li7 problem may be alleviated or even solved. 
We find that this is possible for generic values of the CMSSM parameters where the lifetime of
the NLSP $\sim 10^3$~s (as in \cite{jed,kkm2}), 
and that there are more extended regions of parameter
space where the cosmological \li7 is at least no worse than in standard BBN calculations.

The paper is organized as follows.
In Section~2 we discuss the relevant properties of $X^-$ bound states, including the Coulomb
radii of several nuclides and our three-body model for the $(\alpha + \alpha + X)$ system,
and various choices for binding energies.
In Section~3, we discuss relevant nuclear interaction rates involving bound states,
and our implementation of them in the BBN network.
In Section~4, we briefly describe the chemical reactions involving bound states.
In Section~5, we introduce the supersymmetric framework we use to produce our 
numerical results. In particular, we consider models where
the lighter stau slepton is our candidate for the metastable charged particle $X^-$, and 
results for stau lifetimes are summarized in Section~6.
Our main results are given in Section~7, which includes a discussion of the light element
abundance observations and the abundances we find from BBN with
stau bound states. 
Our results are summarized in Section~8.  


\section{Bound-State Properties}

Before considering the impact of a new electromagnetically-charged particle
on BBN predictions, we first study the properties of its bound states with
light nuclei. To this end, we solve the time-independent Schr\"{o}dinger 
equation with an interaction potential given by the Coulomb potential 
between a finite-sized nuclide and a point-like $X^-$ particle. The Coulomb 
potential is determined by the charge and the rms charge radius 
of the nuclide of interest. Here we adopt the latest rms charge radii measurements
of He, Li and Be isotopes, and assume the charge distribution of
the nuclide to be Gaussian~\footnote{As long as one reproduces the rms charge radius, 
the detailed form of the charge distribution is not important for our purposes.}.  

When solving the Schr\"{o}dinger equation, we first define
dimensionless variables for the energy and a typical length scale.  This
makes the equations a function of a single parameter, the ratio of the
rms charge radius $R_{\rm c}$ to the Bohr radius $R_{\rm B}$. We have verified that our solutions
interpolate smoothly between Coulombic bound-state energies at small
$R_{\rm c}/R_{\rm B}$ and harmonic oscillator energy levels at large $R_{\rm c}/R_{\rm B}$, and
that our numerical solutions match analytic solutions. 

Numerical results for bound states for the nuclides of interest for
BBN are shown in Table~\ref{tab:bsp}. We note that the rms charge
radii of many of the nuclides considered have been determined
experimentally. However, in other cases we must rely on
phenomenological estimates, and Table~\ref{tab:bsp} gives some ranges
in these cases. In particular, the charge radii for the two nuclei
relevant for crossing the $A=8$ divide, namely \be8 and the first
excited state of \be9, are not known experimentally. Given that \be8
is a barely bound state of two $\alpha$ particles (\he4 nuclei) that
are hardly touching, it is expected that the rms charge radius of \be8
should be close to twice the rms charge radius of \he4, i.e., $R_{\rm
  c,8}=3.362$~fm.  This estimate is in good agreement with the value
given in~\cite{Kamimura09}, namely $R_{\rm c,8}=3.39$~fm, after
correcting for the poor binding energy determination in~\cite{Arai}.
For comparison, we also consider the value $R_{\rm
c,8}=2.50$~fm~\cite{Pospelov07}; an estimate chosen close to the \be9
charge radius, though likely an underestimate.

\begin{table}[htb]
\small
\begin{center}
\caption{\it Properties of $X^-$ bound states with relevant nuclides.
The Table lists relevant nuclides together with 
their (unbound) masses and rms charge radii $R_{\rm c}$
in Coulombic parameterizations of the potentials.  
Bound-state binding energies $B_A$ come from our
2-body calculations based on the given
charge radii (except where otherwise noted).
Experimental values and ranges of $R_{\rm c}$ are listed, where available for some nuclides, and ranges of theoretical
estimates of $R_{\rm c}$ for other nuclides. \label{tab:bsp}}
~~\\
\begin{tabular}{|c|c|c|c|}
\hline\hline
Nuclide & Mass (amu) & $R_{\rm c}$ (fm) & $B_A$ (MeV) \\
\hline
$^1$H & 1.00782503 & $0.8750\pm0.0068$~\cite{prms} & 0.02493 \\
$^2$H & 2.01410178 & $2.1303\pm0.0010$~\cite{drms} & 0.04879 \\
$^3$H & 3.01604927 & $1.63\pm0.03$~\cite{trms} & 0.07264 \\
$^3$He & 3.01602931 & $1.9506\pm0.0014$~\cite{herms} & 0.2677 \\
$^4$He & 4.00260325 & $1.681\pm0.004$~\cite{arms} & 0.3474 \\
$^6$Li & 6.01512228 & $2.517\pm0.030$~\cite{lirms} & 0.8000 \\
$^7$Li & 7.01600405 & $2.39\pm0.030$~\cite{lirms} & 0.8893 \\
$^7$Be & 7.01692925 & $2.647\pm0.015$~\cite{berms} & 1.2879 \\
$^8$Be & 8.00530509 & 3.390~\cite{Kamimura09} & 1.1679 \\
 &  & 2.50~\cite{Pospelov07} & 1.408~\cite{Pospelov07} \\
 &  & N/A$^*$ & 0.492 \\
$^9$Be & 9.01218213 & $2.519\pm0.012$~\cite{berms} & 1.4699 \\
$^9$Be$^*$ & 9.01398998 & 2.519 & 1.4700 \\
& &  2.880~\cite{Arai} & 1.3527 \\
& & 3.390 & 1.2173 \\
$^8$B & 8.037675026 & 2.65 & 1.8547 \\
\hline\hline
\end{tabular}\\
~~\\
{
  ${}^{*}$Our result for $B_A$ in this case is based on our three-body calculation.  
}
\end{center}
\label{tab:BE}
\end{table}

The $(\be8 X^-)$ state can play an important role in primordial \be9
synthesis, if the state exists and has appreciable abundance.  As we
will see in the next section, the significance of $(\be8 X^-)$ depends
not only to the qualitative issue of whether this system is bound, but
also on the value of the binding energy.  For this species, therefore,
we have gone beyond the 2-body model that has been used to calculate
the other binding energies appearing in Table \ref{tab:BE}.  To
analyze the bound state properties of the 3-body system
($\alpha+\alpha+X$), we have utilized the Variational Monte Carlo
(VMC) method~\cite{vmc}.  This method assumes a suitable form of
wavefunction with parameters to be determined by Monte Carlo
variation.  Parameters are randomly chosen after some initial guess;
the expectation value of the Hamiltonian is then computed and, if it is lower
than the initial expectation value, the current parameters are set as
the adopted values.  This variation is repeated until some convergence
criteria are met, such as small (or no) changes in the expectation
value of the Hamiltonian with decreasing step size.  The complexity of
the adopted wave function can be increased, with added parameters,
allowing for another test of convergence.
Note that, as with the 2-body case, the
3-body binding energy is expressed with respect to
free \be8$+X$; this is larger than the binding
relative to free $\alpha + \alpha + X$ by
$m(\be{8,\rm free}) -2m_\alpha = -Q_{\rm 8,free} = 92 \ \rm keV$.
 
We have adopted wavefunctions with the forms of exponentials with
Gaussian cutoffs and extra Gaussian terms with decreasing dispersion
with respect to the cutoff scale.  The method was validated with the
three-body systems of the neutral He-atom and the negatively charged
H-atom, using finite-sized Coulomb potentials with the $p$ and
$\alpha$ rms charge radii.  The VMC method reproduced the binding
energies for these Coulomb-only systems quite quickly, agreeing with
the observed values.  For the $\alpha+\alpha+X$ system, we similarly
adopt finite-sized Coulomb interactions, plus an added nuclear
$\alpha+\alpha$ interaction from~\cite{Al66,Al67}.  The adopted
nuclear $\alpha+\alpha$ potential reproduces elastic scattering data
and the \be8 ground state resonance energy.  We find that
$\be8+X$ is bound, but with $B_8 = 492 \ \pm \ 50$ keV.  This
is more fragile than found when assuming a two-body $(\be8 X^-)$
structure with an appropriate rms charge radius.  The Coulomb
repulsion of the two $\alpha$ particles loosens the binding of the
three-body system.

For $(\be8 X^-)$ to allow substantial \be9 production, two conditions
must be satisfied:
\begin{enumerate}

\item
Production of $(\be8 X^-)$ must be possible and effective, and

\item
Production of \be9 must proceed resonantly through
the first excited state in $(\be9^\ast X^-)$, that is,
$(\be8 X^-) + n \rightarrow (\be9^\ast X^-) \rightarrow \be9 + X^-$
\cite{Pospelov07}.

\end{enumerate}
The requirement that $(\be8 X^-)$ production is possible
merely demands that this state
is stable,
as we and others have found.   But, as we will see,
in the early Universe $(\he4 X^-)$ is the dominant bound state.
Consequently, $(\be8 X^-)$ production occurs via
$(\he4 X^-)  + \he4 \rightarrow (\be8 X^-) + \gamma$.
This channel is only effective if it has $Q > 0$,
which demands that
\beq
\label{eq:8be_lim}
B_8 > B_8^{\rm min} = B_4 - Q_{\rm 8,free} = 0.439 \ \rm MeV \, .
\eeq
We find that all estimates of $B_8$ satisfy this constraint,
though our 3-body result does so by a much smaller
margin than the others.

The second requirement for \be9 production depends on the position of
the first excited state of $(\be9^* X^-)$, to which we now turn.  In
ordinary \be9, this is a cluster state that is poorly described by
shell-model calculations.  Its structure is that of two $\alpha$
particles and a neutron and, to first order, this state is the \be8 $+
\, n$ ground state.  Therefore, a first guess for the rms charge radius
would be $R_{\rm c,9^*}=R_{\rm c,8}$.  However, the presence of the
neutron impacts the structure, and the excited state of \be9 should
have a radius larger than the ground state. This constrains our
estimate of the impact of the neutron on the rms \be9 charge radius to
the range $R_{\rm c,9^*} \in (2.519, 3.39)$~fm, where the low value
is just the ground-state value $R_{\rm c,9}$, and the upper value is
the estimate of $R_{\rm c,8}$ given in~\cite{Kamimura09}.
This range
includes the result for the
rms charge radius given in~\cite{Arai},
namely $R_{\rm c,9^*} = 2.88$~fm.
If one assumes the same relative shift in the rms charge radius as for
\be8, one finds $R_{\rm c,9^*}=3.11$~fm. However, it may be larger,
given that the level energy calculated is less accurate in the \be9
case than in the \be8 case.

\begin{figure}[h]
\begin{center}
\includegraphics[width=0.8\textwidth,angle=0]{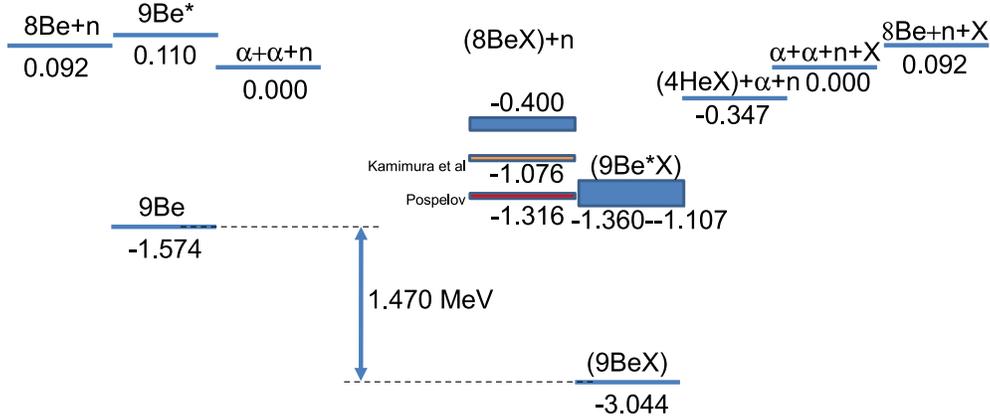}
\end{center}
\caption{\it Level schemes calculated for unbound and bound $X^-$ states in the \be9 system.  
State labels appear above the corresponding level.  Level values shown are based on Table 1, and
all levels are shifted such that zero energy corresponds to the free-particle state.  
In the case of $(\be8X^-)$, we show the result of our 3-body calculation (a blue band whose 
the width corresponds to the uncertainty),
as well as the 2-body calculations of Kamimura {\it et al.}~\cite{Kamimura09} and 
of Pospelov~\cite{Pospelov07} (labelled).
In the case of $(\be9^*X^-)$, we consider a range of level values including our and Pospelov's estimates.
\label{fig:be9}
}
\end{figure}

Fig.~\ref{fig:be9} shows the level structure for the \be9 system when
either in the ordinary unbound state 
(left, \cite{tunl}) or bound to an $X^-$ particle
(right, our calculations).  The zero-point of energy is taken to be
that of unbound, free particles.   
Thus, for example, unbound $\be8+n$ lies at $-Q_{\rm 8,free} = +0.092$ MeV.
The level positions for each of the bound states
are shifted relative to the corresponding 
unbound case due to the binding energy:
level $i$ lies at energy $E_{(iX)} = E_{i,\rm free} - B_i$.
Thus, level {\em spacings} are shifted by {\em differences}
in binding:  $E_{(jX)}-E_{(iX)} = E_{j,\rm free}-E_{i,\rm free} - B_j + B_i$.
For example, the $(\be9^* X)$ excited state lies
above the ground state by an amount that is
the sum of the unbound \be9 level spacing $1.684$ MeV,
minus the difference $B_{9^*}-B_9$.  If the bindings are the same,
then the first excited state spacing is also the same as in the
ordinary case.  But if the excited state is more weakly bound,
then the level spacing is larger than in ordinary \be9.

The three strips for the $(\be8 X^-)$ bound
state represent the three values as determined by this work (highest strip), 
as well that
of~\cite{Pospelov07,Kamimura09} (the lower two strips, as labelled).  
The thick box for the excited
$(\be9^* X^-)$ state corresponds to the range of possible charge radii
$R_{\rm c,9^*} \in (2.519, 3.39)$~fm. Even with the broad range of
possible rms charge radii for the first excited state in \be9, one can
see that the excited state probably lies substantially below the
$(\be8 X^-)$ entrance level.  If this is so, this drives the reaction
away from resonance, substantially reducing or even eliminating the
mechanism discussed in~\cite{Pospelov07}, and thus suppressing \be9
production via bound states.  Ultimately, however, a true determination of
the first exited state position would require a four-body calculation
of $(\alpha \alpha n X^-)$.  This
challenging work has not been done,
and lies beyond the scope of our paper.
Note also that our 2-body and 3-body calculations of $(\be8 X^-)$
gave significantly different binding energies, underscoring
the importance of detailed calculations for these states
with relatives high $Z$.
Consequently,  we cannot exclude the possibility that
the $(\be8 X^-)$ entrance channel will be near resonance.

Clearly, the situation regarding \be9 is uncertain, reflecting 
the poor state of knowledge regarding the $(\be8 X^-)$ and $(\be9 X^-)$ nuclear
properties.   In this paper, our approach is to illustrate the ability
of \be9 to constrain supersymmetric models, within the most optimistic
scenario in which the resonant production discussed
in~\cite{Pospelov07} occurs.  Thus, for most
of our calculations we adopt this (constant) resonant rate for $(\be8
X^-) + n \rightarrow \be9 + X^-$~\cite{Pospelov07}.  However, the
reader should bear in mind that the resulting \be9 abundances and
resulting constraint therefore represent a `most optimistic'
scenario.  Consequently, we also make comparisons with calculations in which
this production channel is suppressed, considering the
cases:
\begin{enumerate}
\item{}Bound state structure and resonant rate from~\cite{Pospelov07}.
\item{}$(\be8 X^-)$ is more tightly and more weakly bound. 
\item{} The resonant rate set to zero for $(\be8 X^-) + n \rightarrow \be9 + X^-$.
\end{enumerate}
It goes without saying that there is a pressing need for precise and
accurate calculations of $(\be8 X^-)$ and $(\be9 X^-)$ properties.

\section{Bound-State Reactions and Abundances}

\subsection{Formalism}

In looking at the effects of bound states of $X^-$,
we must track the abundances of its
bound states with various nuclei, e.g., $(pX^-)$,
$(\he4X^-)$ and $(\be7X^-)$.
For this purpose, we need to incorporate reactions that affect
these bound-state abundances, namely:
(1) recombination and photodissociation processes
on each nuclear species $i$,
such as
\beq
\label{eq:rec}
i + X^- \leftrightarrow (iX^-) + \gamma \ \ ,
\eeq
and (2) charge-exchange processes between $(iX^-$) bound states
and other nuclides $j$, such as
\beq
\label{eq:ce}
(i X^-) + j \leftrightarrow i + (j X^-)    \ \ .
\eeq
We refer to these processes collectively as `bound-state chemistry',
and solve the rate equations for the processes (\ref{eq:rec}) and (\ref{eq:ce}) 
to determine the corresponding chemical abundances.

Denoting the total $X^{-}$ number density by $n_{X^-} = Y_{X^-} n_{\rm B}$, 
we decompose it into the free and bound abundances,
respectively  $n_{X^-,\rm{free}}$ and $n_{(jX^-)}$, with $j\in p,d,\ldots$.
To remove the effect of cosmic expansion, 
as usual we follow the evolution
of the `mole fractions'
\beq
Y_i \equiv \frac{n_i}{n_{\rm B}} ,
\eeq
where the states $i$ include:
(a) ordinary {\em unbound}, free nuclei, 
(b) $X^-$ bound states,
and (c) free $X^-$. Thus, we treat these in a manner completely parallel to the
usual BBN accounting for ordinary (unbound) nuclides. 
Note that the total abundance of a nuclear species $i$ sums its
unbound and bound states
$Y_{i,\rm tot} = Y_{i,\rm free} + Y_{(iX^-)}$,
while the total abundance of $X^-$ is 
$Y_{X^-,\rm tot} = Y_{X^-,\rm free} + \sum_i Y_{(iX^-)}$.

For two-to-two reactions of the form $a b \rightarrow c d$,
the reaction rate per unit volume is $n_a n_b \avg{\sigma_{ab \rightarrow cd} v}$,
with $\avg{\sigma_{ab \rightarrow cd} v}$ the appropriate thermally-averaged rate coefficient.
The reaction rate per particle $a$ is thus
\beq
\Gamma_{ab \rightarrow cd} = \avg{\sigma_{ab \rightarrow cd} v} n_b
  = N_{\rm Avo} \avg{\sigma_{ab \rightarrow cd} v} \, \rho_{\rm B} Y_b
  \equiv \lambda_{ab \rightarrow cd} \, \rho_{\rm B} Y_b ,
\eeq
where $N_{\rm Avo} = 1/m_u$ is Avogadro's number,
and $\lambda_{ab \rightarrow cd} = N_{\rm Avo} \avg{\sigma_{ab \rightarrow cd} v}$
is the form in which thermonuclear rates are normally tabulated.
Thus the {\em total} rate per target $b$ nucleus is 
\beq
\Gamma_{ab \rightarrow cd}^{\rm tot}
 = \Gamma_{ab \rightarrow cd} + \Gamma_{(aX)b \rightarrow cd X}
 = \avg{\sigma_{ab \rightarrow cd} v} n_{\rm B} Y_{a,\rm free} 
  +  \avg{\sigma_{(aX)b \rightarrow cdX} v} n_{\rm B} Y_{(aX)} \ ,
\eeq
which can be substantially larger than in the ordinary
case if the catalyzed rate coefficient has a large
enhancement and if there is a substantial $(aX)$ abundance.

With these definitions, 
the evolution of bound state $(iX)$, with $i\in p,d,\ldots$
can be expressed as a sum over several kinds of processes:
\beq
\label{eq:bsev}
\frac{\partial}{\partial t} Y_{(iX)}
    =  
  \left. \frac{\partial}{\partial t} Y_{(iX)} \right|_{\rm chem}
  + \left. \frac{\partial}{\partial t} Y_{(iX)} \right|_{\rm nuc}
  + \left. \frac{\partial}{\partial t} Y_{(iX)} \right|_{\rm decay} .
\eeq
The bound-state chemistry reactions do not change the type of nuclides
in the initial state, 
and are 
\beq
  \left. \frac{\partial}{\partial t} Y_{(iX)} \right|_{\rm chem} 
    =   - \left( \Gamma_{\gamma(iX) \rightarrow iX} 
	+ \sum_j \Gamma_{j(iX) \rightarrow i(jX)} \right) Y_{(iX)}
   + \Gamma_{iX \rightarrow \gamma(iX)}  Y_{X,\rm free}
   +  \sum_j \Gamma_{i(jX) \rightarrow j(iX)} Y_{(jX)} \ .
\eeq
Bound-state nuclear reactions have final-state nuclides
different than those in the initial state, and take the form
\beq
\left. \frac{\partial}{\partial t} Y_{(iX)} \right|_{\rm nuc}
    =   - \left (\sum_{k\ell} \Gamma_{j(iX) \rightarrow k\ell X} \right) Y_{(iX)}
	+ \sum_{k\ell} \Gamma_{\ell(kX) \rightarrow j(iX)} Y_{(kX)} \ ,
\eeq
where the last term includes only those reactions that produce
bound $(iX^-)$ rather than free $i$.
Finally, the decays of $X^-$ with lifetime $\tau_X$ destroy bound states:
\beq
\left. \frac{\partial}{\partial t} Y_{(iX)} \right|_{\rm decay}  =    - \Gamma_X Y_{(iX)} \ ,
\eeq
where the decay rate $\Gamma_X  = 1/\tau_X$.

Turning to free $X^-$, we have
\beq
\frac{\partial}{\partial t} Y_{X,\rm free}
  = - \sum_i \left. \frac{\partial}{\partial t} Y_{(iX)} \right|_{\rm chem}
  - \sum_i \left. \frac{\partial}{\partial t} Y_{(iX)} \right|_{\rm nuc}
  - \Gamma_X Y_{X,\rm free}  \ \ .
\eeq
The bound-state 
contribution to the evolution of
a species of unbound, free nuclei $i$ is given by
\beqar
\left. \frac{\partial}{\partial t} Y_{i,\rm free} \right|_{\rm BS}
  & = &
  - \left. \frac{\partial}{\partial t} Y_{(iX)} \right|_{\rm chem} 
  - \left. \frac{\partial}{\partial t} Y_{(iX)} \right|_{\rm decay}
\nonumber \\
  & &  - \left (\sum_{j} \Gamma_{i(jX) \rightarrow k\ell X} \right) Y_{i,\rm free}
  +  \sum_{k\ell} \Gamma_{\ell(kX) \rightarrow ijX} Y_{(kX)} \ ,
\eeqar
where the last term includes only bound-state reactions 
that produce free $i$ in the final state.

\subsection{Reaction Rates}

Tables~\ref{tab:rxns-chem} and \ref{tab:rxns-nuke} summarize the treatments of bound-state chemistry and
nuclear rates in our work and in the recent literature.
Entries for recent literature are given to the best of our
knowledge; in some cases full details were not given
in the published papers.
Blank entries mean that to the best of our knowledge no rate was 
assigned, effectively setting the rate to zero.

Wherever possible, we adopted the most up-to-date chemistry and
nuclear rates
from the literature.  In many cases, rates were not available
for channels we wished to examine.  Thus we adopted simple
rules to estimate the needed rates from published ones;
these cases are identified in the Tables.
In the case of bound-state chemistry, we adopted 
recombination rates for nuclide $i$ using the
scaling $\sigma_{\rm rec} \propto Z_i^2 B_{(iX)}$.

As we see below, bound state chemistry strongly favors
$(\he4 X^-)$ production, which 
essentially locks up all of the $X^-$,
for the cases of physically interesting abundances where $Y_{X^-} < Y_\alpha$.
Consequently, nuclear reactions involving $(\he4 X^-)$ are the most
important. On the other hand, $(pX^-)$ reactions are relatively unimportant
due to the small abundance of this state, and $(dX^-)$ and $(tX^-)$ have
negligible effects. 

In general, bound states enhance nuclear rates.
This is in part because they reduce the Coulomb barrier, to which the
rates are exponentially sensitive.
For nuclides and channels for which bound-state nuclear
rates were not available in the literature,
we estimated the rates assuming this is the only
source of perturbation.  In these cases, we
adopted the ordinary thermonuclear rates, but with
a Gamow penetration factor appropriate for a nucleus
of effective charge $Z_{(iX)}^{\rm eff} = Z_i - 1$.

In many cases, bound states also enhance nuclear channels through
catalysis effects, which may be described as follows \cite{Maxim}.
Consider the important example of catalyzed \li6 formation,
\beq
(\he4X^-) + d \rightarrow \li6 + X^-  \ \ .
\eeq
The corresponding ordinary process is the
$\he4 + d \rightarrow \li6 + \gamma$ radiative capture reaction,
which is suppressed because it must proceed through 
the E2 mode.  The bound-state rate does not require
the emission of a photon and is
substantially larger than the ordinary rate
in the typical situation in which the $(\he4 X^-)$ abundance is large.

We have included rates for the formation and processing of $(\be8X^-)$
which, as we will see, can lead to substantial \be9 production under
optimistic nuclear physics assumptions.
We also have included rates for \bor{10} and \bor{11} production
via rates involving $(\alpha X^-)$.
We find that boron production is indeed increased over the (very small)
standard BBN level.  However, the B/H abundance always remains many orders of magnitude
below the levels seen in halo stars.  Thus we find
that boron is not a promising signature of decaying particle effects.

We have also studied whether reionization
by the emitted \xlya\ photons could inhibit the net rates for the NLSP recombination reactions
$A + X^- \rightarrow (AX^-) + \gamma$, as is the case in ordinary hydrogen
recombination. As discussed in Appendix~A, we find that,
whereas the optical depth for reionization by \xlya\ photons emitted by NLSP recombination is much smaller
than that for ordinary hydrogen and helium recombination, it is still very large, so that the net rate of
recombination might be very suppressed. However, as we also show in Appendix~A, 
\xlya\ photons Compton scatter rapidly
off free electrons.  This rapidly degrades the energies
off resonance, so that they are ineffective
for reionization. We conclude that NLSP recombination to the ground state proceeds unimpeded,
unlike the case of ordinary hydrogen and helium recombination.

\section{Bound-State Chemical Effects}

We present later results from a code that treats self-consistently
the bound states as separate nuclei, which
then can have their own set of bound-state chemical and nuclear
reactions with other species.
As a warm-up exercise, 
we first present some results with catalysis effects turned off,
and so only incorporate bound-state chemistry, i.e., recombination onto bound
states.  We include decays as part of the chemistry, i.e., decays remove free and bound $X^-$,
but we turn off nonthermal decay effects.    This exercise tests our code and illustrates the 
interplay between recombination and charge transfer.

For this purpose, we choose an initial $X^-$ abundance $Y_{X^-,\rm tot}^{\rm init} = 10^{-2}$,
which is typical for interesting supersymmetric models, and we vary the lifetime $\tau_X$,
to show the sensitivity to this parameter.



\begin{table}[htb]
\footnotesize
\caption{\it Summaries of the treatments of bound-state chemistry 
rates assumed in our work and in the recent literature (I)
 \label{tab:rxns-chem}}
\bigskip
\hspace{-2cm}
\begin{tabular}{c|cccc|c}
\hline\hline
 & CEFOS & Bailly 
  & Pospelov & Kamimura & This  \\
Reaction & 2006~\cite{CEFOS} & et al 2009~\cite{Bailly} 
  & et al 2008~\cite{Pospelov08} & et al 2009~\cite{Kamimura09} & Work\\
\hline
$p + X^- \rightarrow (pX^-) + \gamma $ &
  simple scaling &  Kamimura 09~\cite{Kamimura09} & estimated & 
  & Pospelov~\cite{Pospelov08} \\
$d + X^- \rightarrow (dX^-) + \gamma $ &
  simple scaling &  & & 
  & scaled Pospelov~\cite{Pospelov08} \\
$t + X^- \rightarrow (tX^-) + \gamma $ &
  simple scaling &  &  & 
  & scaled Pospelov~\cite{Pospelov08} \\
$\he3 + X^- \rightarrow (\he3X^-) + \gamma $ &
  simple scaling & &  & 
  & scaled Pospelov~\cite{Pospelov08} \\
$\alpha + X^- \rightarrow (\alpha X^-) + \gamma $ &
  simple scaling & \verb+"+ & Pospelov 07~\cite{Maxim} &  
  & Pospelov~\cite{Pospelov08} \\
$\li6 + X^- \rightarrow (\li6X^-) + \gamma $ &
  simple scaling & \verb+"+ & & & $\sigma_{\rm rec} \propto Z^2 B$ scaling \\
$\li7 + X^- \rightarrow (\li7X^-) + \gamma $ &
  simple scaling & \verb+"+ & & & $\sigma_{\rm rec} \propto Z^2 B$ scaling \\
$\be7 + X^- \rightarrow (\be7 X^-) + \gamma $ &
  simple scaling & \verb+"+ & Bird 08~\cite{Bird08} & & Bird 08~\cite{Bird08} \\
$(\be8 X^-) +\gamma \rightarrow \be8 + X^-$ &
  simple scaling & & & & $\sigma_{\rm rec} \propto Z^2 B$ scaling \\
$\be9 + X^- \rightarrow (\be9 X^-) + \gamma $ &
   & & & & $\sigma_{\rm rec} \propto Z^2 B$ scaling \\
$\bor8 + X^- \rightarrow (\bor8 X^-) + \gamma $ &
   & & & & $\sigma_{\rm rec} \propto Z^2 B$ scaling \\
\hline
$(pX^-) + \alpha \rightarrow (\alpha X^-) + p$ &
  & \verb+"+ & estimated & QM 3-body & Kamimura 09~\cite{Kamimura09} \\
$(dX^-) + \alpha \rightarrow (\alpha X^-) + d$ &
  & \verb+"+ &  & QM 3-body & Kamimura 09~\cite{Kamimura09}  \\
$(tX^-) + \alpha \rightarrow (\alpha X^-) + t$ &
  & \verb+"+ &  & QM 3-body & Kamimura 09~\cite{Kamimura09}  \\
\hline\hline
\end{tabular}
\end{table}




\begin{table}[htb]
\footnotesize
\caption{\it Summaries of the treatments of bound-state nuclear
rates assumed in our work and in the recent literature (II)
 \label{tab:rxns-nuke}}
\bigskip
\hspace{-2cm}
\begin{tabular}{c|cccc|c}
\hline
\multicolumn{6}{c}{\it Bound State Nuclear} \\
\hline\hline
 & CEFOS & Bailly 
  & Pospelov & Kamimura & This \\
Reaction & 2006~\cite{CEFOS} & et al 2009~\cite{Bailly} 
  & et al 2008~\cite{Pospelov08} & et al 2009~\cite{Kamimura09} & Work \\
\hline
$(d X^-)+ \alpha \rightarrow \li6+X^-$ &
   &  Kamimura 09 &  
  & QM 3-body &  Kamimura 09~\cite{Kamimura09} \\
$(\alpha X^-)+ d \rightarrow \li6+X^-$ &
  simple scaling & \verb+"+ &  
  & QM 3-body &  Kamimura 09~\cite{Kamimura09} \\
$(t X^-)+ \alpha \rightarrow \li7+X^-$ &
   & \verb+"+ & & QM 3-body &  Kamimura 09~\cite{Kamimura09} \\
$(\alpha X^-)+ t \rightarrow \li7+X^-$ &
  simple scaling & \verb+"+ & & QM 3-body &  Kamimura 09~\cite{Kamimura09} \\
$(\alpha X^-)+ \he3 \rightarrow \be7 + X^-$ &
  simple scaling & \verb+"+ & & QM 3-body &  Kamimura 09~\cite{Kamimura09} \\
$(\alpha X^-)+ \he4 \rightarrow (\be8 X^-) + \gamma$ &
   & & & & Pospelov 07~\cite{Pospelov07} \\
$(\alpha X^-)+ \li6 \rightarrow (\bor{10} X^-) + \gamma$ &
   & & & & scaling from Caughlan 88~\cite{CF88} \\
$(\alpha X^-)+ \li7 \rightarrow (\bor{11} X^-) + \gamma$ &
   & & & & scaling from Angulo99~\cite{nacre} \\
$(\alpha X^-)+ \be7 \rightarrow (\iso{C}{11} X^-) + \gamma$ &
   & & & & scaling from Angulo99~\cite{nacre} \\
$(\li6X^-)+ p \rightarrow \alpha + \he3 + X^-$ &
  simple scaling & \verb+"+ & & QM 3-body &  Kamimura 09~\cite{Kamimura09} \\
$(\li6X^-)+ n \rightarrow t + \alpha + X^-$ &
   & & &  & Caughlan 88~\cite{CF88} \\
$(\li6X^-)+ d \rightarrow \li7 + p + X^-$ &
   & & &  & scaling Malaney 89~\cite{Malaney89} \\
$(\li6X^-)+ d \rightarrow \be7 + n + X^-$ &
   & & &  & scaling Malaney 89~\cite{Malaney89} \\
$(pX^-)+ \li6 \rightarrow \he4 + \he3 + X^-$ &
  & \verb+"+ & estimated & & Pospelov~\cite{Pospelov08} \\
$(\li7X^-)+ p \rightarrow \alpha + \alpha + X^-$ &
  simple scaling & & & QM 3-body &  Kamimura 09~\cite{Kamimura09} \\
$(pX^-)+ \li7 \rightarrow \be8 + X^-$ &
  & \verb+"+ & & QM 3-body &  Kamimura 09~\cite{Kamimura09} \\
$(\be7X^-)+ n \rightarrow \li7 + p + X^-$ &
  & &  & & scaling from Cyburt04~\cite{Cyburt04} \\
$(\be7X^-)+ p \rightarrow \bor8 + X^-$ &
  & \verb+"+ & & QM 3-body${}^{\dagger}$ &  Kamimura 09~\cite{Kamimura09} \\
$(pX^-)+ \be7 \rightarrow \bor8 + X^-$ &
  & \verb+"+ & & QM 3-body &  Kamimura 09~\cite{Kamimura09} \\
$(\be7X^-)+ d \rightarrow p + 2\alpha + X^-$ &
  & &  & & Caughlan 88~\cite{CF88} \\
$(\be8X^-)+ n \rightarrow \be9 + X^-$ &
   & & estimated & ${}^{\star}$ &  Pospelov~\cite{Pospelov08} \\
$(\be8X^-) + d \rightarrow \bor{10} + X^-$ & 
   & & &  & scaled from Coc 12~\cite{Coc12} \\
$(\be8X^-) + d \rightarrow \li6 + \alpha + X^-$ & 
   & & &  & scaled from Coc 12~\cite{Coc12} \\
$(\bor8X^-) \rightarrow \be8 + X^-$ &
   & & &  & $\beta$ lifetime Matt 64~\cite{Matt64} \\
\hline\hline
\end{tabular}
{${}^{\dagger}$ \it This rate is $m_X$-dependent.} \\
{${}^{\star}$ \it The reaction $(\be8X^-)+n$ is argued in \cite{Kamimura09}
to be non-resonant, which would reduce the \be9 production \\
from the levels given by this rate.} \\
\end{table}


\normalsize

\begin{figure}[ht!]
\centering
\includegraphics[width=9.5cm,angle=+90]{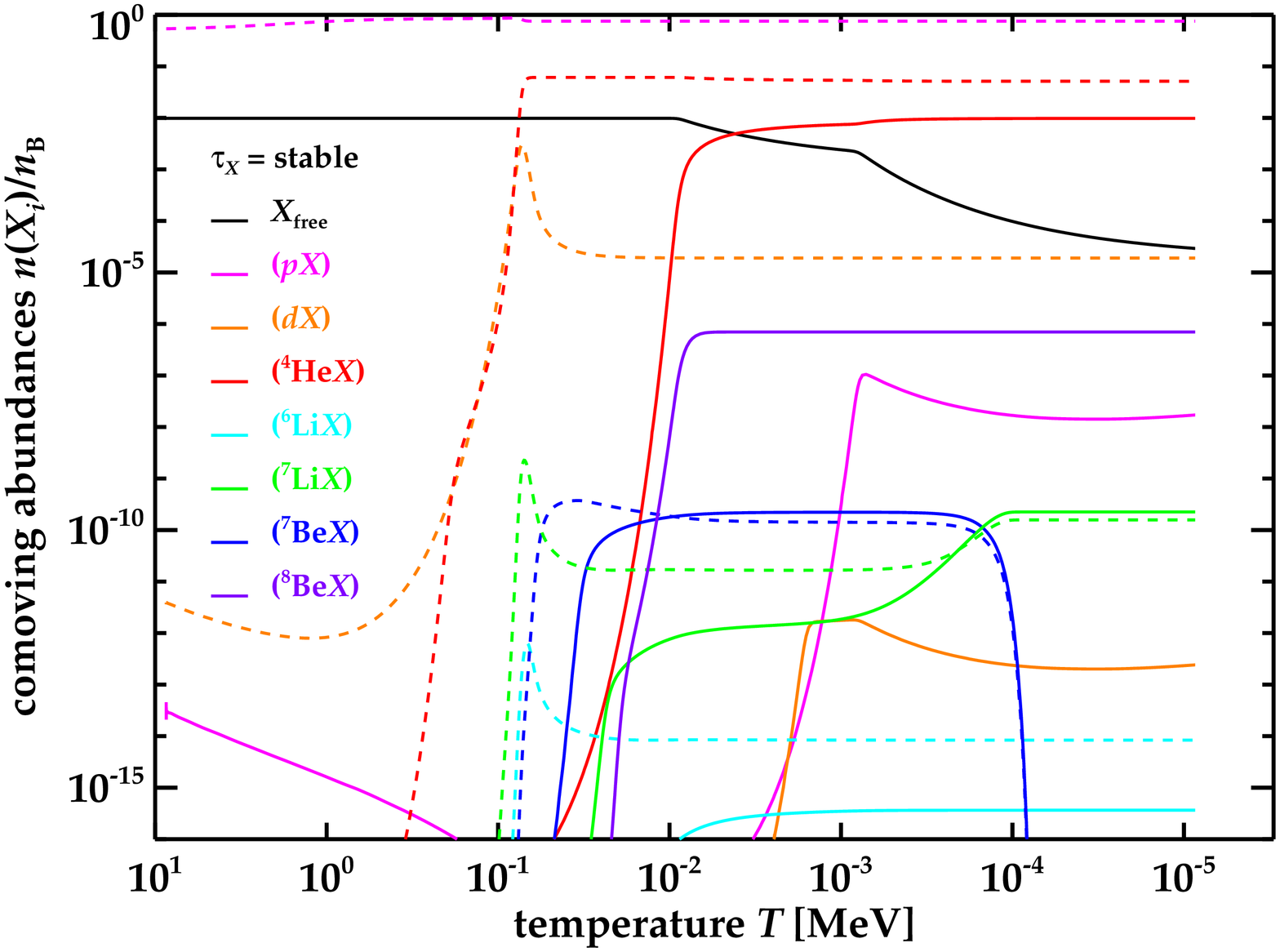}
\caption{
\it The abundances of free nuclei (dashed lines) and nuclear bound states (solid lines) as
functions of temperature.
The black line corresponding to the abundance of free $X^-$ particles, which is assumed to be $10^{-2}$ initially.
In this case the $X$ lifetime $\tau_{X}$ is assumed to be infinite.
\label{fig:chemistry1}}
\end{figure}

\begin{figure}[ht!]
\centering
\includegraphics[width=9.5cm,angle=+90]{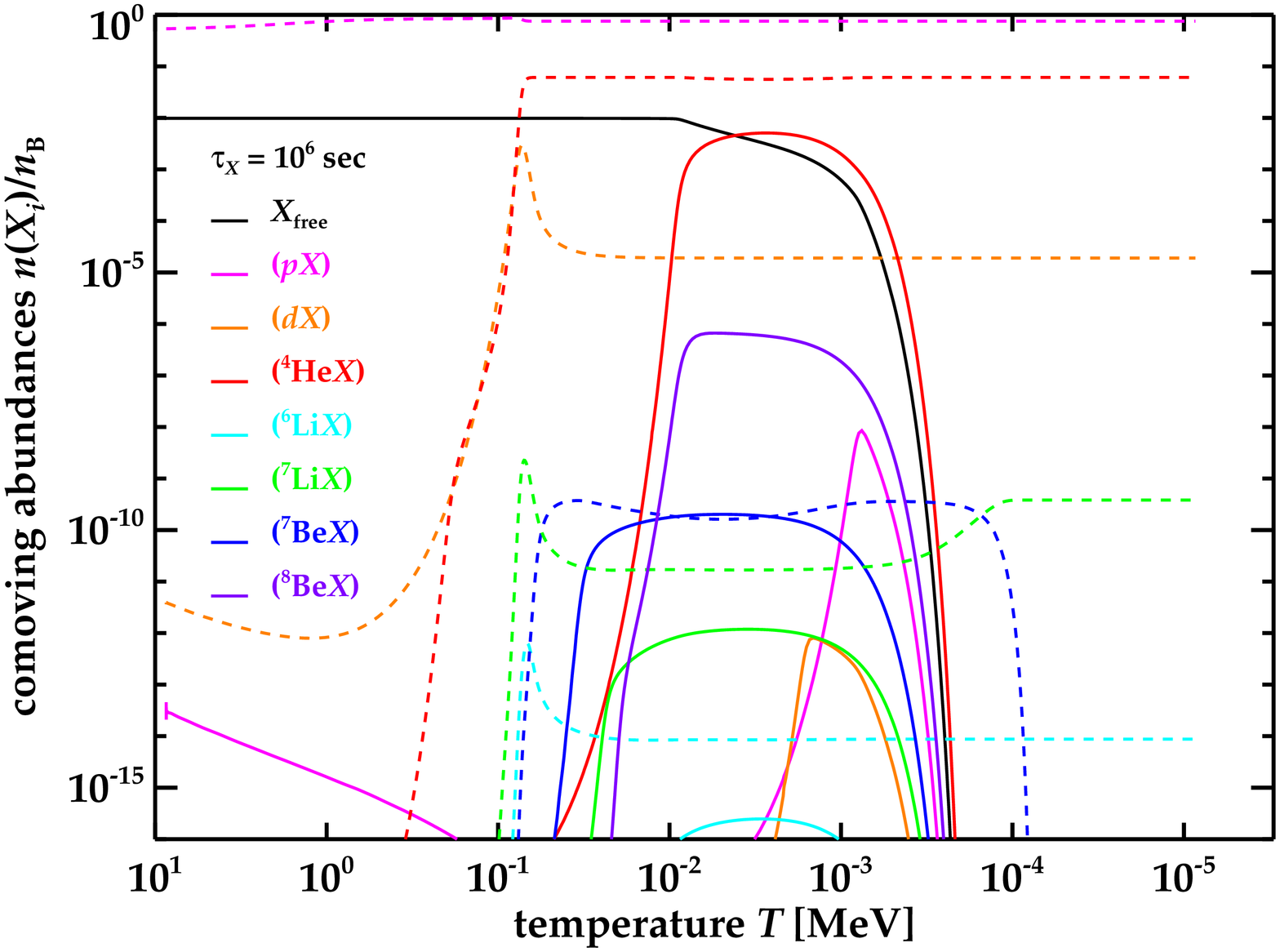} 
\caption{
\it As in Fig.~\ref{fig:chemistry1}, but with an assumed lifetime  $\tau_{X} = 10^6$~s.
\label{fig:chemistry2}}
\end{figure}

\begin{figure}[ht!]
\centering
\includegraphics[width=9.5cm,angle=+90]{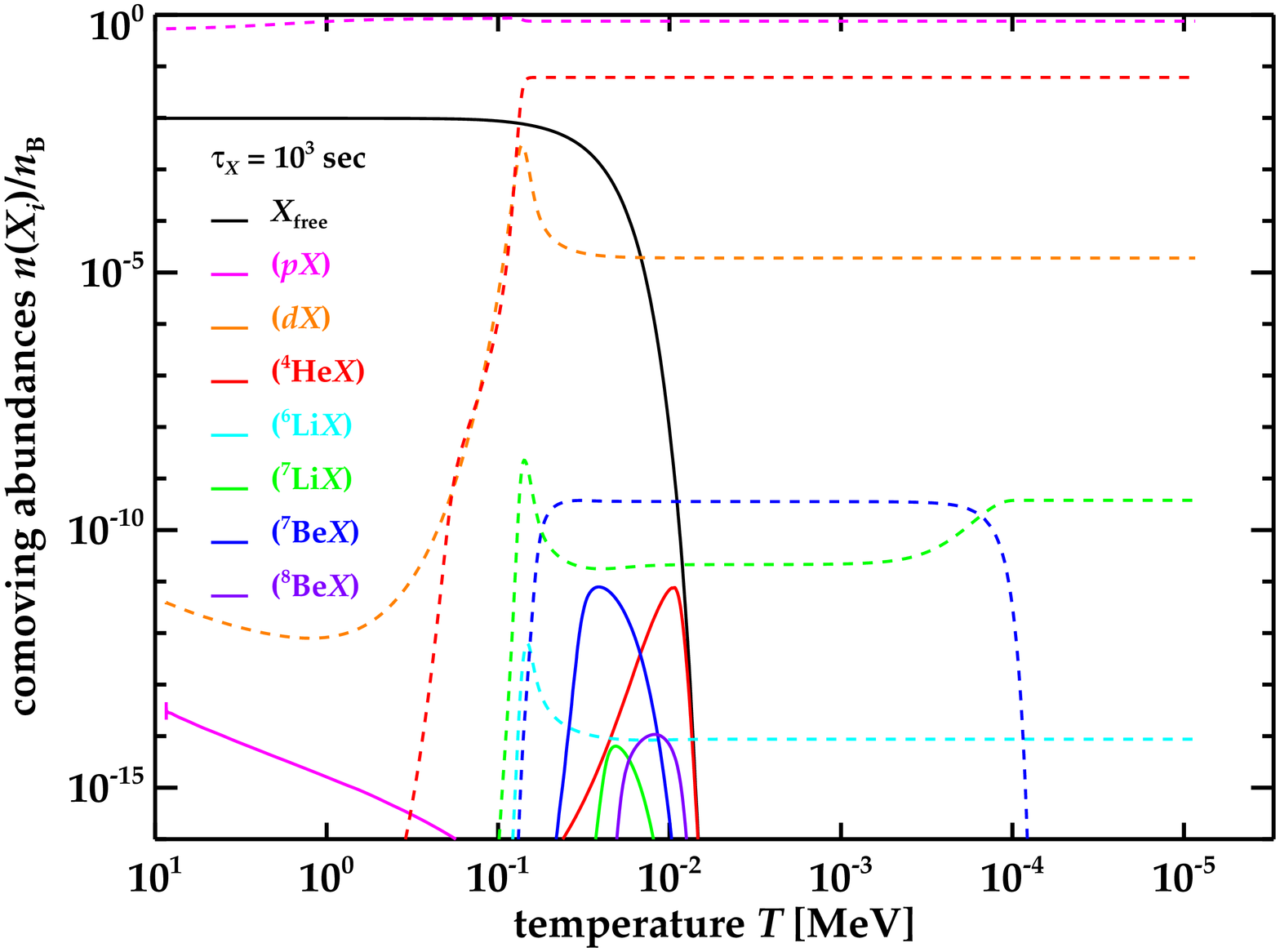}
\caption{
\it As in Fig.~\ref{fig:chemistry1}, but with an assumed lifetime  $\tau_{X} = 10^3$~s.
\label{fig:chemistry3}}
\end{figure}

In Figs.~\ref{fig:chemistry1} - \ref{fig:chemistry3}, we show the abundances $Y_i \equiv n_i/n_{\rm B}$ for both bound and
free species, as functions of the temperature $T$.  The solid black line corresponds to the abundance of free $X^-$'s,
whereas the other solid lines are the abundances of the
bound states, as labeled by colour, and dashed lines of the same color are the corresponding
free states, e.g., the solid red line represents the (\he4$X^-$) abundance, while the dashed red represents that of free \he4. 
Fig.~\ref{fig:chemistry1} illustrates the case of a very long-lived $X^-$.
Focusing first on the bound states (solid colored lines) we see that 
\be7 recombines first, followed by \li7, then \he4, and finally protons. These results
were to be expected, given that recombination 
occurs at $T_{Rec} \sim B/|{\rm ln}~\eta| \sim B/25$, where $B \sim Z^2$,
with $\eta \equiv n_{\rm B}/n_\gamma$ the baryon-to-photon ratio.

Comparing the solid lines, we see that free $X^-$ particles
dominate until \he4 recombines, leading to the first kink
in the black curve, after which most $X^-$ are in (\he4$X^-$) bound states~\footnote{Note that
 it is important for this analysis that $Y_{X^-} < Y_{\he4}$, so that all $X^-$ particles
can find \he4 partners.}.  Subsequently
the protons recombine, and this leads to the second kink in
the black free $X^-$ curve, as well as a small rise in
the \he4 curve.  This is because, immediately after the ($pX^-$) states recombine,
the important ($pX^-$) + \he4 $\to$ p + (\he4$X^-$) charge-exchange process
converts ($pX^-$) states into more (\he4$X^-$) bound states.
The $(dX^-)$ state actually form earlier than
$(pX^-)$ states, because the deuteron states are more tightly bound.
However, deuterons are much rarer, and thus the $(dX^-)$ abundance
is always quite small, and ultimately the $(dX^-)/(pX^-)$ ratio
is comparable to the ordinary $D/H$ ratio. 

Turning attention now to the $A = 7$ states (blue and green curves), we see that
recombination into these nuclei occurs mostly after they are formed.
The (\be7$X^-$) state has almost the same abundance as the free \be7 state,
whereas the abundance of (\li7$X^-$) is smaller than that of free \li7.  At late times $T \sim 10^{-5}$~MeV,
the $(\be7 X^-)$ captures an electron and converts to $(\li7 X^-)$, which
remains bound because here $Q$ is smaller than the difference
in binding energies.  
Looking at the \li6 abundances, we see that the bound state has a much smaller
abundance than the free nucleus.

Finally, we turn to the special case of $(\be8 X^-)$.
Because this state has no analogue nuclide in standard BBN, 
there is no $\be8 + X^-$ recombination,
and thus $(\be8 X^-)$ 
does not emerge when the temperature drops below its binding.
Rather, production occurs via $(\he4 X^-) + \he4 \rightarrow (\be8X^-) + \gamma$,
and thus we see that the abundances rises after that of $(\he4 X^-)$.

Our results are similar to those shown in~\cite{Pospelov08},  apart from \li6,
where we have purposely removed catalysis effects for this exercise only.
One difference is that we use the charge-exchange rates of~\cite{Kamimura09}, which are larger than those
used in~\cite{Pospelov08}, and thus give a {\it smaller} ($pX^-$) abundance due to the more efficient ($pX^-$) conversion.

Figs.~\ref{fig:chemistry2} and \ref{fig:chemistry3} are for $\tau_X = 10^{6}$ and $10^{3}$~s,
respectively, and reveal few surprises. They are similar to the case of long-lived $X$
until the time $t \sim \tau_X$, where $t \sim 1/T_{\rm MeV}^2$, at which stage the $X$ particles decay away.
A corollary of this exercise is that if $\tau_X < 10^3$~s, recombination cannot form bound states
before the $X$ particles decay, and hence bound-state chemical effects are negligible.


\section{Decays  of the NLSP in Gravitino Dark Matter Scenarios}

As in~\cite{CEFOS}, we base our discussion here on the CMSSM, as described in
Appendix B. 
The mass of the gravitino is a free parameter in the CMSSM, and is the LSP and constitutes the dark matter if its
mass, $m_{3/2}$ is chosen to be less than $\min(m_\chi,m_{\tilde \tau})$. 
The abundances of the light elements provide some of the most important
constraints on such a gravitino dark matter (GDM) scenario~\cite{SFT,0404231,eoss5,vcmssm,EOV,Jed1,Jed2,stef,CEFOS,PS,Kawasaki08,Bailly}.  
Their abundances also impose important constraints on neutralino LSP scenarios,
since a gravitino NLSP could decay sufficiently slowly to affect them.  Here, however, 
we restrict our attention to GDM scenarios with either a stau or neutralino NLSP,
later focusing more closely on the stau NLSP case.

In the process of calculating the lifetime of the NLSP, we
calculate the partial widths of the dominant relevant decay channels of the NLSP
and hence the various NLSP decay branching ratios. We also calculate the resulting electromagnetic (EM) and hadronic spectra,
which impact the light-element abundances.
The decay products that yield EM energy obviously include directly-produced photons,
and also indirectly-produced photons (e.g., via the decays of neutral pions, $\pi^0$), and charged leptons (electrons and muons)
that may be produced via the secondary decays of gauge and Higgs bosons.
Hadrons (nucleons and  mesons such as the $K_L^0$, $K^\pm$ and $\pi^\pm$) 
are produced through quark-antiquark pairs and via the secondary decays of gauge and Higgs bosons,
as well as (in the case of mesons) via the decays of $\tau$ leptons.
We note that mesons decay before interacting with the 
hadronic background~\cite{kohri,SFT,0404231}. They are therefore irrelevant 
for BBN processes and to our analysis, except via their decays into photons
and charged leptons. Thus the hadronic injections on which we focus our attention are those that 
produce nucleons, namely the decays via gauge and Higgs bosons and quark-antiquark pairs.

In the case of the  neutralino NLSP, we include the two-body decay
channels $\chi \to \grav \, H_i$ and $\chi \to \grav \, V$, where $H_i=h,H,A$ and $V=\gamma,Z$,
and also the dominant 
three-body decays  $\chi \to \grav \, \gamma^*  \to \grav \, q  \qbar$ and 
  $\chi \to \grav \, W^+ W^-$.
  In the case of $\chi \to \grav \, W^+ W^-$ we have included all the contributing tree-level amplitudes,
  as was done in~\cite{CEFLOS}, thus treating correctly the longitudinal components of  the $W$ bosons.
In general, the two-body channel $\chi \to \grav \, \gamma$ dominates the 
$\chi$ NLSP decays and yields the bulk of the injected EM energy.
When the $\chi$ is heavy enough to produce a real $Z$ boson, the next most important 
channel is $\chi \to \grav \, Z$, which is also the dominant channel for producing hadronic 
injections in this case. The Higgs boson channels are smaller by a few orders
of magnitude, and those to heavy Higgs bosons ($H,A$), in particular,
become kinematically accessible only for heavy $\chi$ in the large-$m_{1/2}$ region. 
 
Turning to the three-body channels, the decay through the virtual photon to a $q \qbar$ pair 
can become comparable to the subdominant channel $\chi \to \grav \, Z$, 
injecting nucleons even in the kinematical region $\mchi < \mgrav + M_Z$,
where direct on-shell $Z$-boson production is not possible. In principle, one 
should also include $q \qbar$ pair production through the virtual $Z$-boson
channel $\chi \to \grav \, Z^*  \to \grav \, q  \qbar$~\cite{kmy} 
and the corresponding interference term. However,
this process is suppressed by a factor of $M_Z^4/m({\bar q} q)^4$ with respect to
$\chi \to \grav \, \gamma^*  \to \grav \, q  \qbar$, and the interference term is
suppressed by $M_Z^2/m({\bar q} q)^2$. Numerically, these contributions are
unimportant, and we drop them in our calculation. 
Finally, we note that the three-body decays to $W^+W^-$ pairs and a gravitino are usually 
at least five  orders of magnitude smaller. Analytical results for the amplitudes for these  gravitational decays of a neutralino NLSP
have been presented in~\cite{CEFLOS}. There,  they were calculated for the inverse processes 
$ \grav \to \chi + X $, but the decay amplitudes are the same, and the only adjustment needed is
to interchange the neutralino and gravitino mass in the phase space. 

We presented in~\cite{CEFLOS} our method of estimating  the EM and
the hadronic decays of the direct products of the $\chi$ decays
using  the  {\tt PYTHIA} event generator~\cite{pythia}. 
We first generate a sufficient number of spectra for the secondary decays of the
gauge and Higgs bosons and the quark pairs.  Then, we perform fits to
obtain the relation between the energy of the decaying particle and the
quantity that characterizes the hadronic spectrum, namely $dN_h/dE_h$, the
number of produced nucleons as a function of the nucleon energy.  These
spectra and the fraction of the energy of the decaying particle that is
injected as EM energy are then used to calculate the light-element
abundances. We use the same approach here. 

An analogous procedure is followed for the $\stau$ NLSP case. 
In~\cite{CEFOS}, we assumed that the lighter stau was right-handed,
so we ignored the stau mixing effects and the stau interactions with the $W^\pm$.
However, in this analysis here we include the full effects  of stau mixing.
The decay rate for the dominant two-body decay channel, namely
$\stau \to \grav \, \tau$, was given in~\cite{eoss5}. However, this
decay channel does not yield any nucleons.  Therefore, one must
calculate some three-body decays of the $\stau$ to obtain any protons or
neutrons. The most relevant channels are  $\stau \to \grav \, \tau^* \to
\grav \, Z \, \tau $, $\stau \to Z \, \stau^* \to \grav \, Z \, \tau $,
$\stau \to \tau \widetilde{\chi}_i^{\,0*} \to \grav \, Z \, \tau $ and $\stau \to \grav \, Z
\, \tau $. In addition, so as to include the full effects of stau mixing,  we included the processes
$\stau \to \grav \, \tau^* \to \grav \, W^- \, \nu_\tau $,
 $\stau \to W^- \, \snutau^* \to \grav \, W^- \, \nu_\tau $,
$\stau \to \nu_\tau  \widetilde{\chi}_i^{-*} \to \grav \, W^- \, \nu_\tau  $ 
and $\stau \to \grav \, W^- \, \nu_\tau  $. 

Analytical results for three-body stau decays can be found 
in Appendix~C. We then use {\tt PYTHIA} to obtain the
hadronic spectra and the EM energy injected by the secondary $W$, $Z$-boson and
$\tau$-lepton decays. As in the case of the $\chi$ NLSP, this information
is then used for the BBN calculation.

\section{NLSP Lifetimes in the CMSSM with Gravitino LSP}

As discussed above, we study the constraints from the cosmological light element abundances in the context of 
the CMSSM. The recent discovery of a new boson with mass $\sim 125$ to 126~GeV with properties that resemble
those of the Standard Model Higgs boson~\cite{LHCH}
motivates us to concentrate on regions of the CMSSM parameter space where
the lightest neutral Higgs boson has a mass close to this range, taking into account the theoretical
uncertainty in the calculation of its mass for any fixed values of the CMSSM parameters~\cite{FeynHiggs}.
As discussed in~\cite{Ellis:2012aa}, this mass range favours large values of $A_0$ and $\tan \beta$:
see also~\cite{MC8,post-mh}. On
the other hand, the constraint from $B_s \to \mu^+ \mu^-$~\cite{bmmexp} disfavours very large $\tan \beta$
\cite{bmmth,bmmth2}. Accordingly,
in this paper we discuss one example of a $(m_{1/2}, m_0)$ plane with $\tan \beta = 10$ and two examples
with $\tan \beta = 40$~\cite{Ellis:2012aa}. In many models of supersymmetry breaking, the
soft trilinear supersymmetry-breaking parameter $A_0 \propto m_0$.
For $\tan \beta = 10$ we consider the single value $A_0 = 2.5 \, m_0$, and in the $\tan \beta = 40$
case we consider the two options $A_0 = 2 \, m_0, 2.5 \, m_0$. In the absence of clear indications on the
gravitino mass $m_{3/2}$, in each case we consider two options: 
fixed $m_{3/2} = 100$~GeV and $m_{3/2} = 0.1 m_0$. We also consider  in less detail two examples of
$(m_{1/2}, A_0)$ planes with fixed $\tan \beta = 40$ and $m_{3/2} = 100$~GeV, namely with
$m_0 = 1000, 3000$~GeV.

\begin{figure}[ht!]
\centering
\includegraphics[width=6cm]{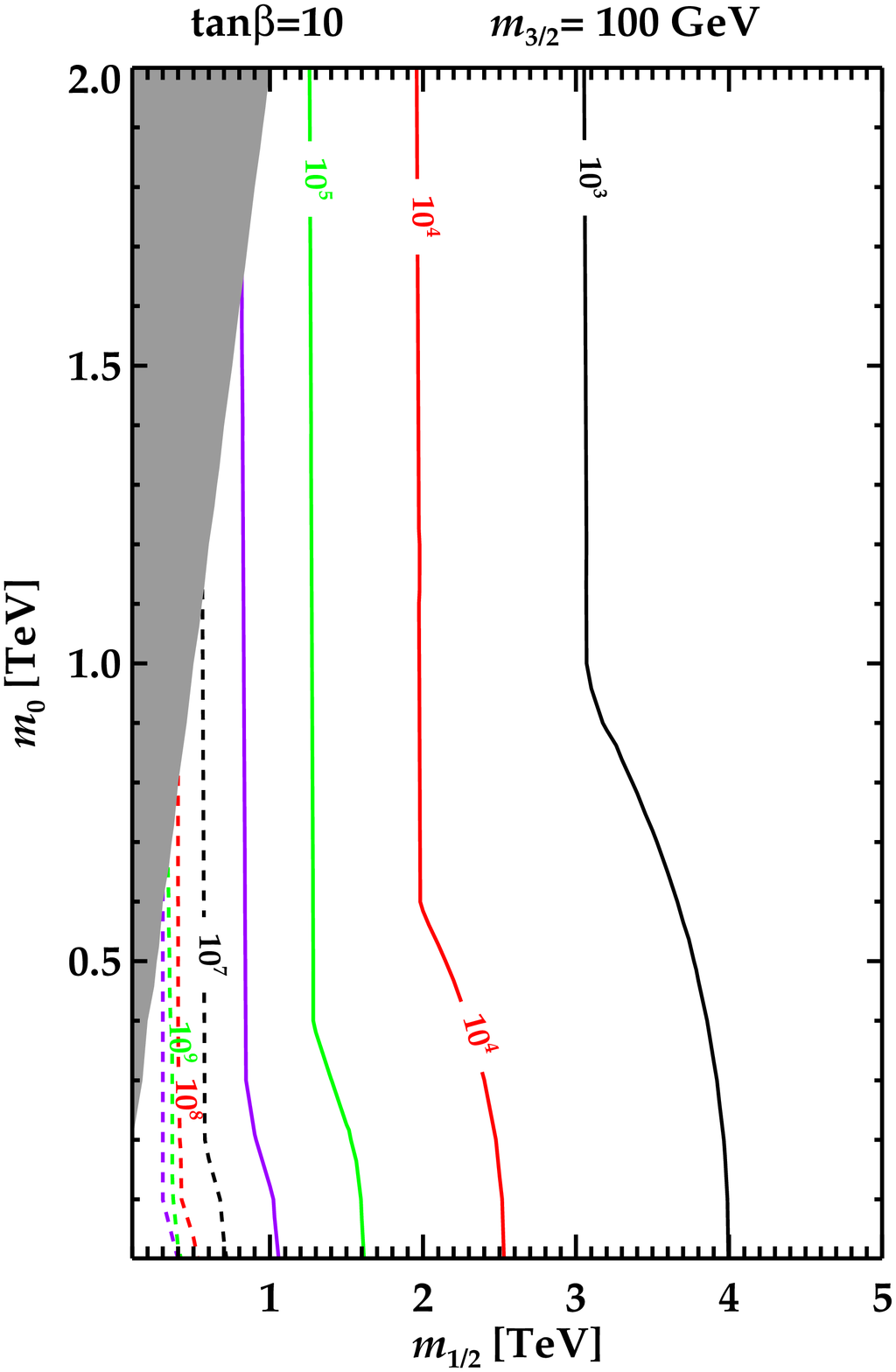}
\includegraphics[width=6cm]{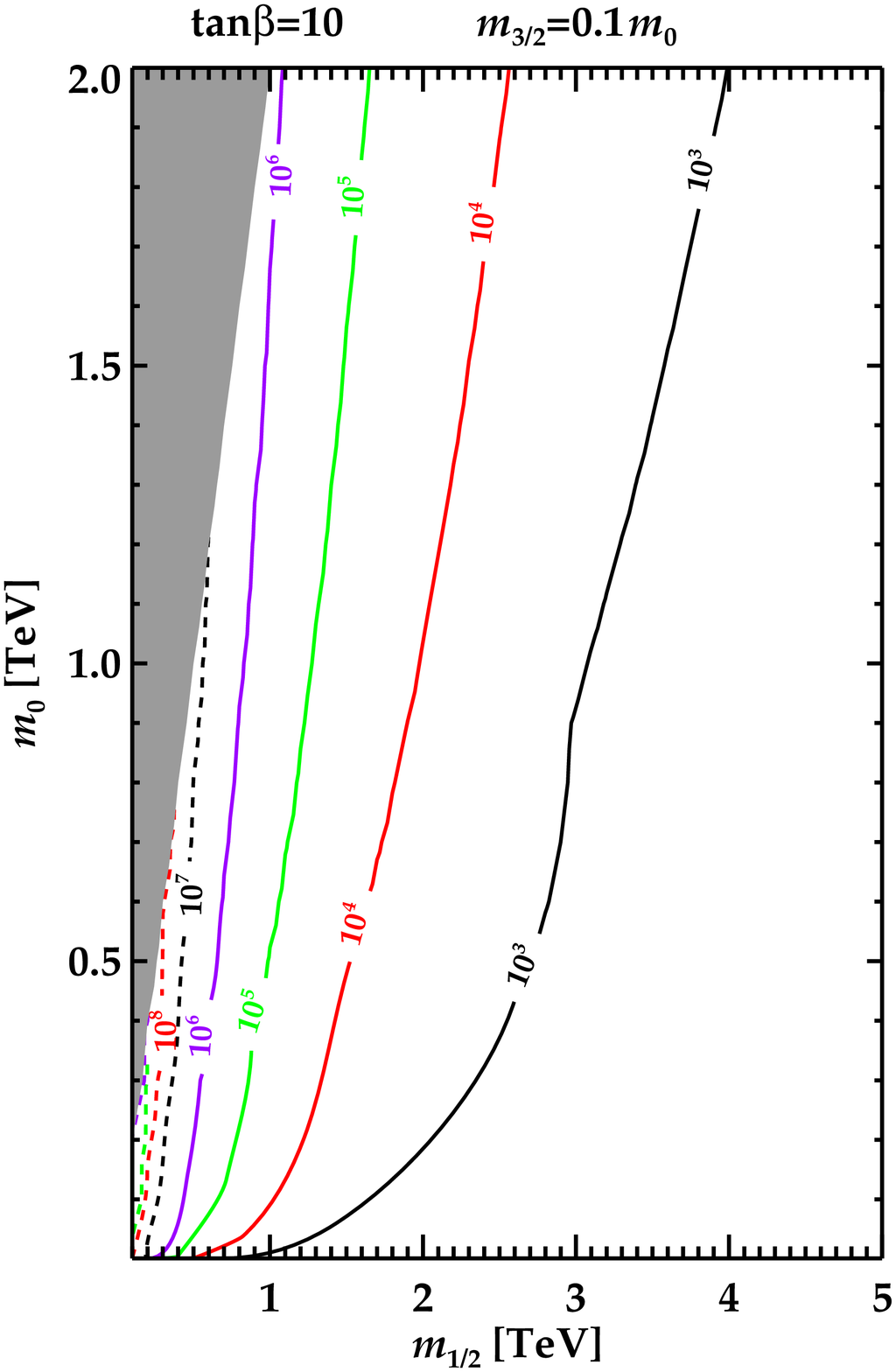}
\caption{
\it The NLSP lifetime $\tau_{\rm NLSP}$ in the $(m_{1/2}, m_0)$ plane for $A_0=2.5 \, m_0$, $\tanb=10$ and $m_{3/2}= 100$~GeV (left) and
$m_{3/2} = 0.1 \, m_0$ (right).}
\label{fig:tau_10_2.5}
\end{figure}

An important ingredient in understanding the morphology of our results in the
$(m_{1/2}, m_0)$ planes is provided by the NLSP lifetime $\tau_{\rm NLSP}$. Fig.~\ref{fig:tau_10_2.5}
displays contours of $\tau_{\rm NLSP}$ in the first cases studied above, namely the $(m_{1/2}, m_0)$
planes for $\tan \beta = 10$, $A_0 = 2.5 \, m_0$ and $m_{3/2} = 100$~GeV (left) and $m_{3/2} = 0.1 \, m_0$ (right).
In the upper part of the left panel, where the lightest neutralino is the NLSP,
we see that the lifetime contours are essentially vertical, since they
depend mainly on the relationship of $m_\chi$ (and hence $m_{1/2}$) to $m_{3/2}$. These contours
appear only when the gravitino is the LSP, i.e., for $m_\chi > m_{3/2}$, and there is a vertical band
at small $m_{1/2}$ where this condition is not satisfied. Also, we note that the lighter
stop squark is either the NLSP or tachyonic in the grey shaded triangular regions in the small-$m_{1/2}$, large-$m_0$ corners of these panels.  
In the lower part of the left panel of Fig.~\ref{fig:tau_10_2.5}
where the lighter stau is the NLSP,
the contours of constant NLSP lifetime curve, track the relationship between $m_{1/2}$ and $m_{\tilde \tau_1}$.
The lifetime contours in the right panel of Fig.~\ref{fig:tau_10_2.5}, for $m_{3/2} = 0.1 \, m_0$, are
everywhere sloping up from left to right.

\begin{figure}[ht!]
\centering
\includegraphics[width=6cm]{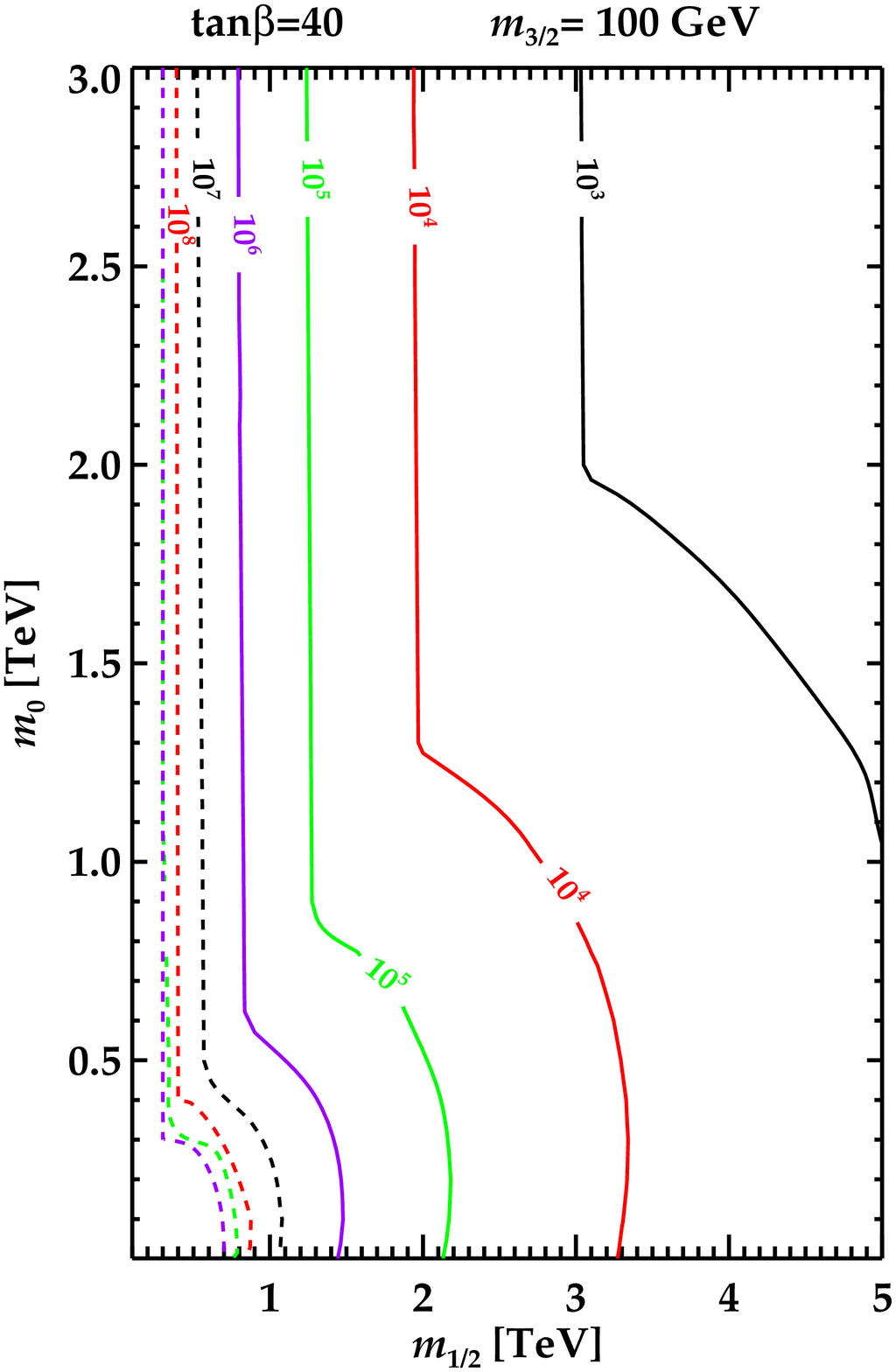}
\includegraphics[width=6cm]{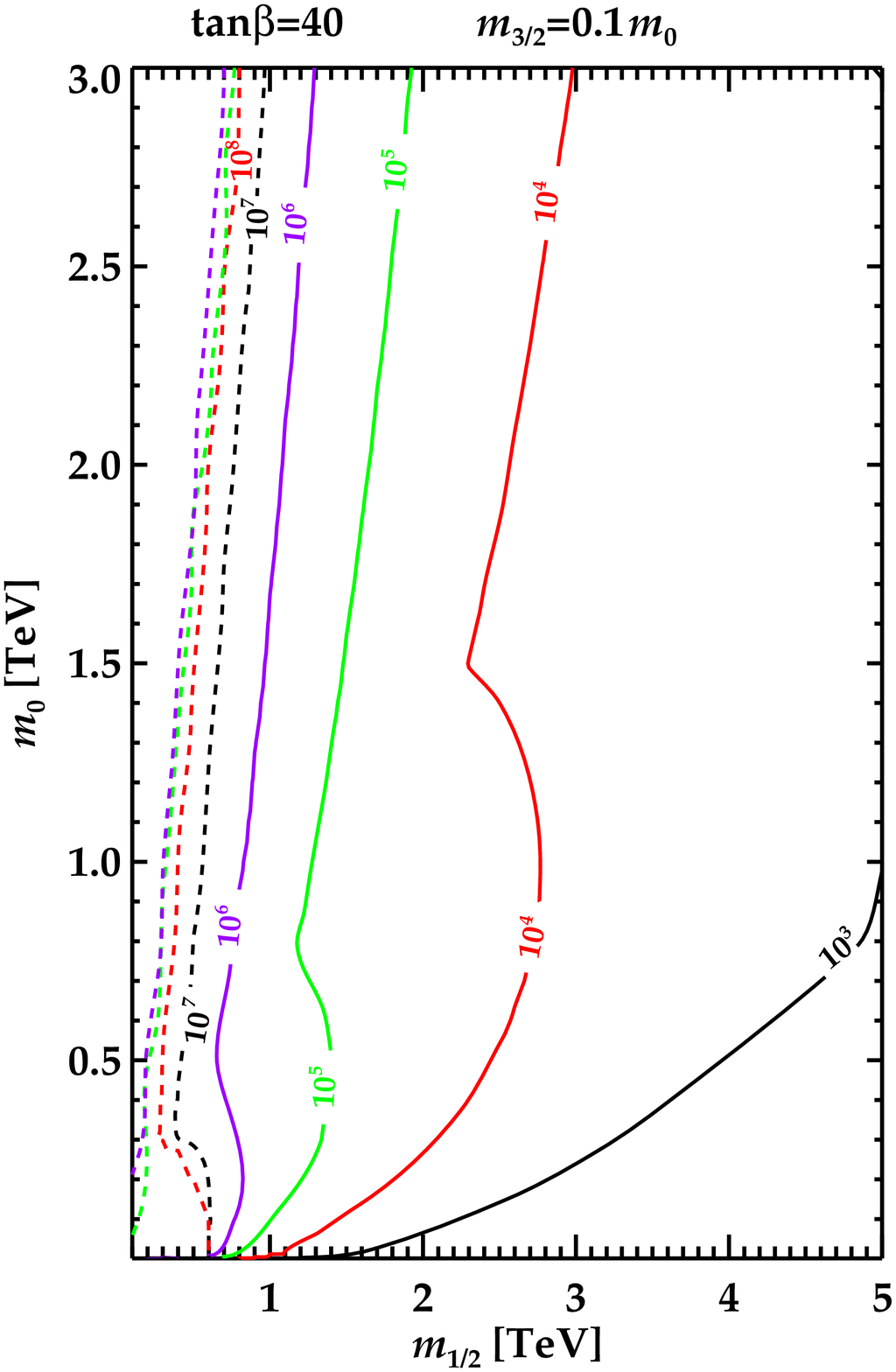}
\caption{
\it The NLSP lifetime in the $(m_{1/2}, m_0)$ plane for $A_0=2 \, m_0$, $\tanb=40$ and $m_{3/2}= 100$~GeV (left) and
$m_{3/2} = 0.1 \, m_0$ (right).}
\label{fig:tau_40_2}
\end{figure}

Fig.~\ref{fig:tau_40_2} displays the corresponding contours of $\tau_{\rm NLSP}$ for the cases
$\tan \beta = 40$, $A_0 = 2 \, m_0$ and $m_{3/2} = 100$~GeV (left) and $m_{3/2} = 0.1 \, m_0$ (right).
These exhibit similar features to the previous case, except that the stau NLSP region is now larger,
as a result of the larger value of $\tan \beta$, and now we see a difference in the behaviours of
the lifetime contours in the stau and neutralino NLSP regions. The vertical band
at small $m_{1/2}$ where the gravitino is not the LSP is now fully visible. 
In this case, there is a triangular region at small $m_{1/2}$ and large $m_0$
where the gravitino is no longer the LSP.

\begin{figure}[ht!]
\centering
\includegraphics[width=6cm]{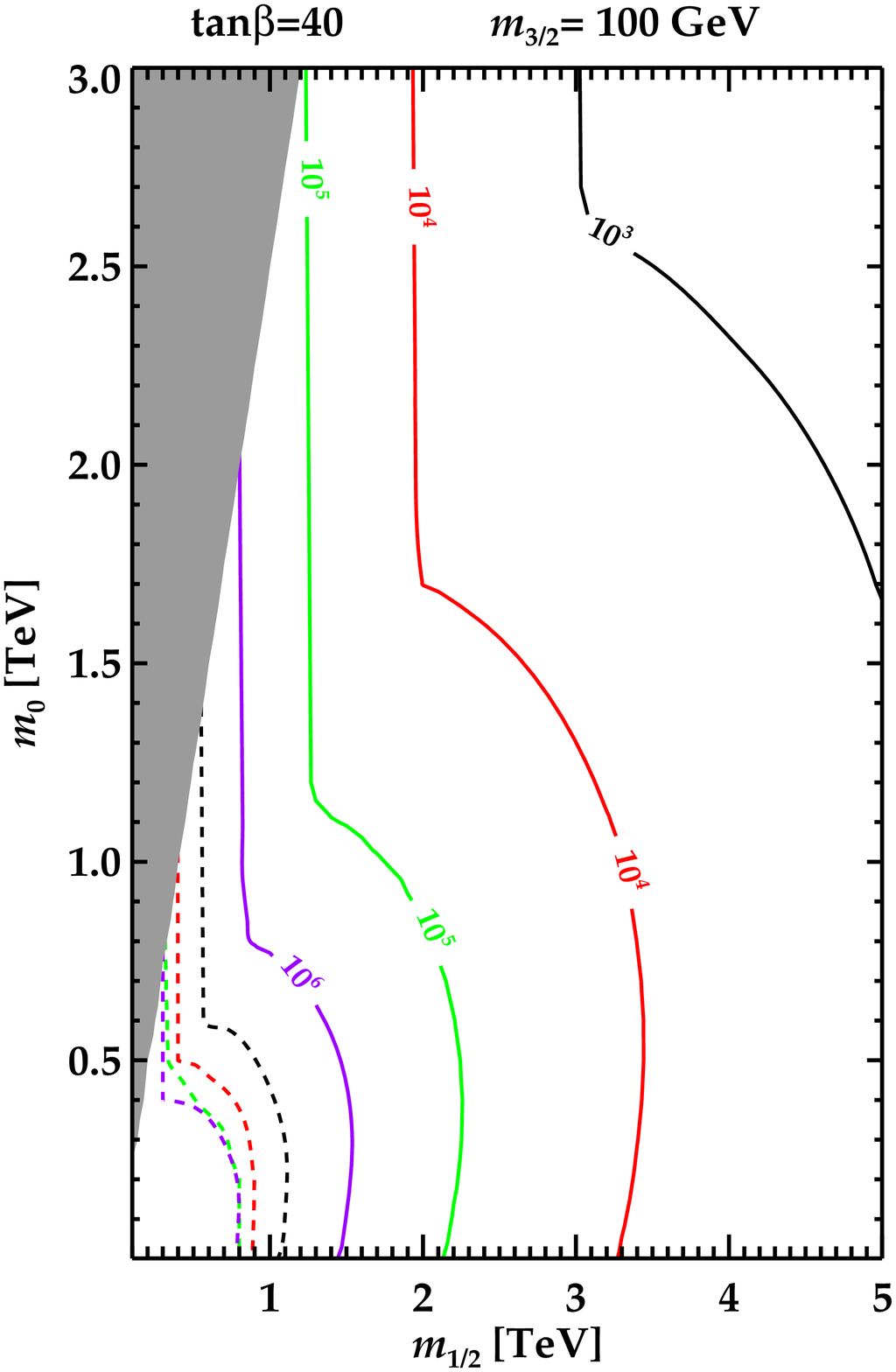}
\includegraphics[width=6cm]{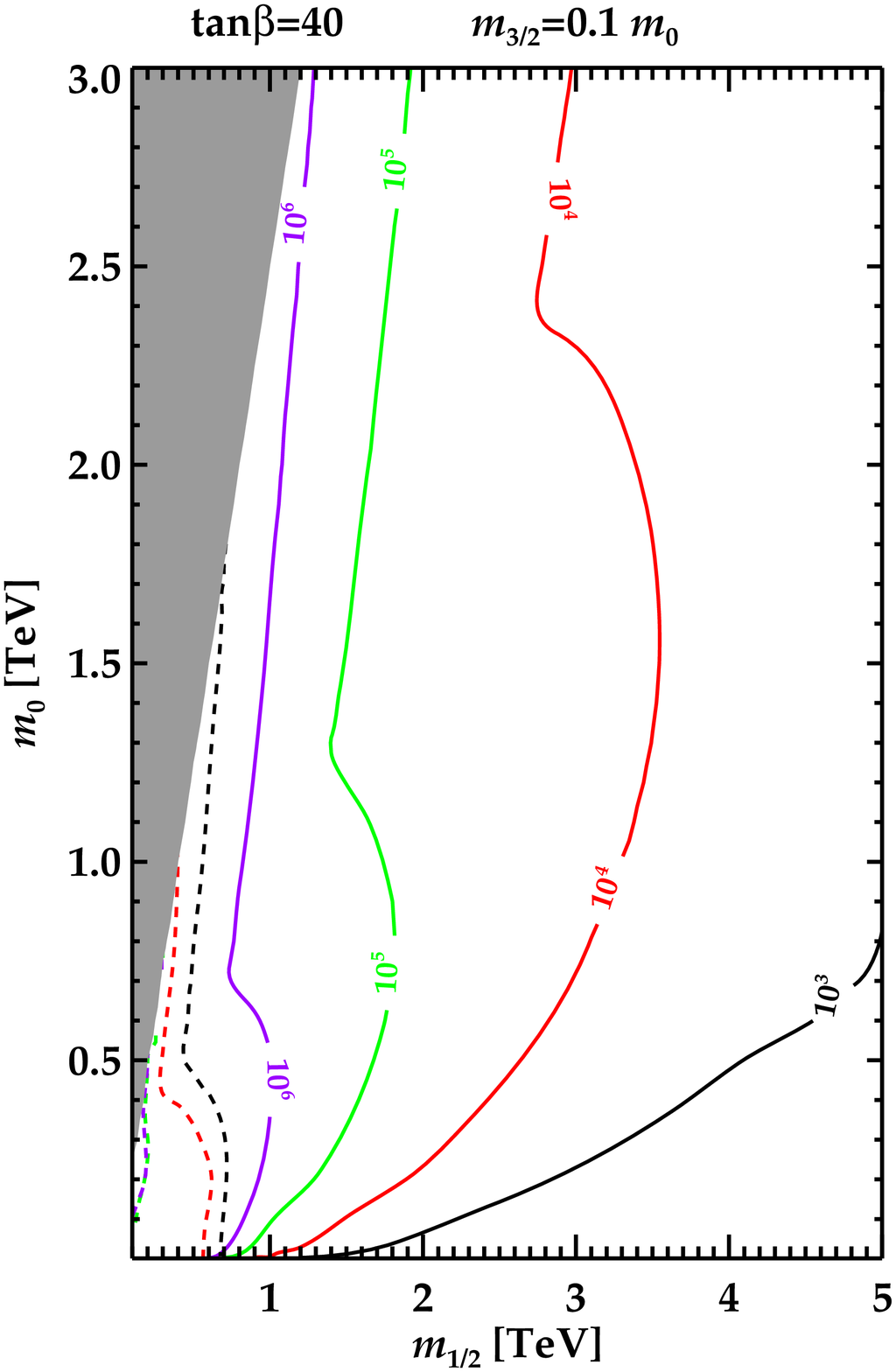}
\caption{
\it The NLSP lifetime in the $(m_{1/2}, m_0)$ plane for $A_0=2.5 \, m_0$, $\tanb=40$ and $m_{3/2}= 100$~GeV (left) and
$m_{3/2} = 0.1 \, m_0$ (right).}
\label{fig:tau_40_2.5}
\end{figure}

Finally, Fig.~\ref{fig:tau_40_2.5} displays the corresponding contours of $\tau_{\rm NLSP}$ for the cases
$\tan \beta = 40$, $A_0 = 2.5 \, m_0$ and $m_{3/2} = 100$~GeV (left) and $m_{3/2} = 0.1 \, m_0$ (right).
This is qualitatively similar to Fig.~\ref{fig:tau_40_2}, though we note that the stau NLSP region
has expanded again, this time as a result of the larger value of $A_0$. Note also that the 
triangular region where the stop is light (or tachyonic) has reappeared at the larger value of $A_0$.

\section{Light-Element Constraints in the CMSSM with a Metastable Stau NLSP}

\begin{figure}[ht!]
\centering
\includegraphics[width=13cm,angle=+90]{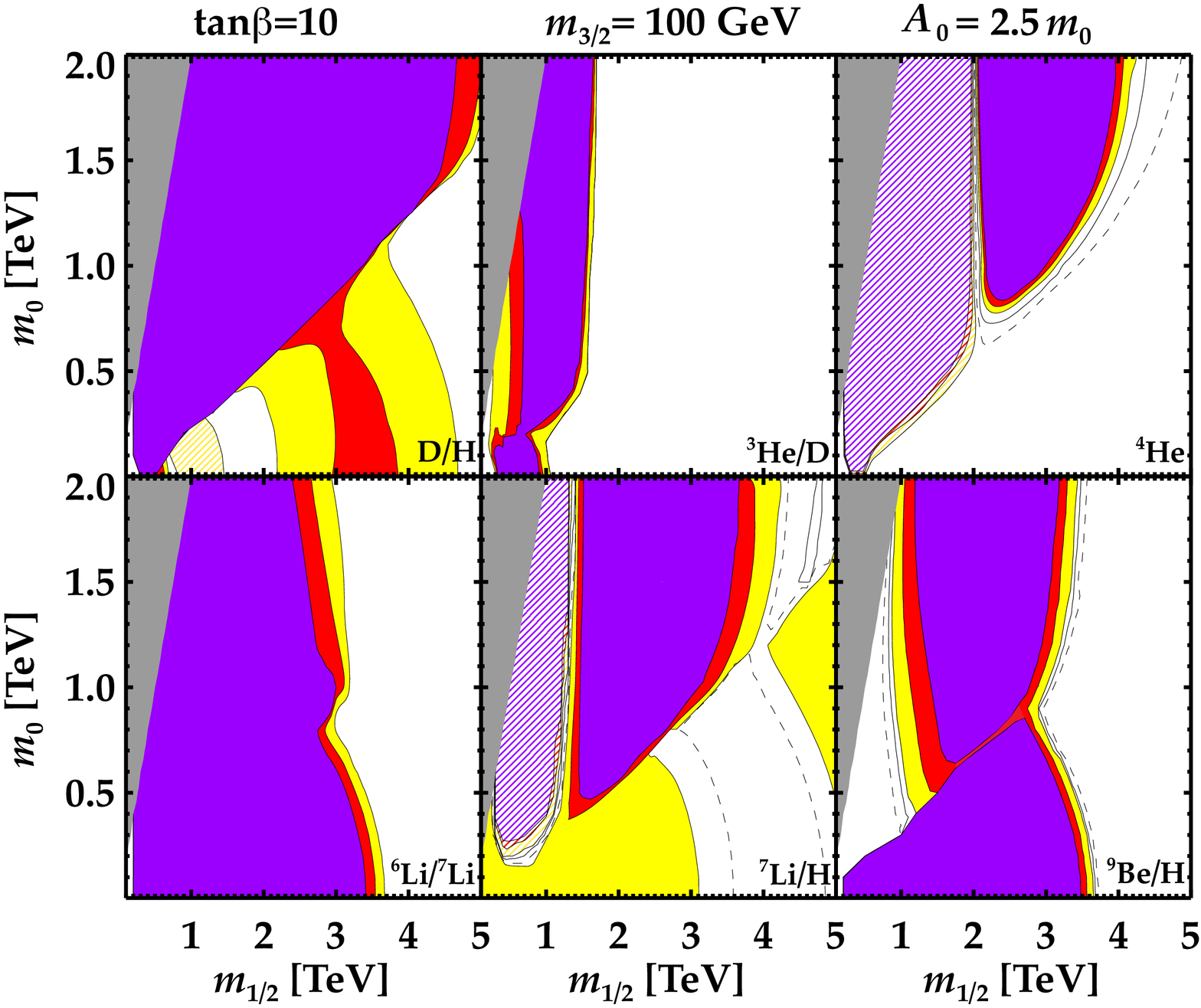}
\caption{
\it Light-element abundances in the $(m_{1/2}, m_0)$ plane for $A_0=2.5 m_0$, $\tanb=10$ and $m_{3/2}=100$ GeV.}
\label{fig:Aeq2.5m0_10_100m0}
\end{figure}

We display in Fig.~\ref{fig:Aeq2.5m0_10_100m0} the light-element abundances
we calculate in the $(m_{1/2}, m_0)$ plane for the first example introduced
above, namely $A_0=2.5 \, m_0$, $\tanb=10$ and $m_{3/2}=100$ GeV.
In this and subsequent figures, the stau is the NLSP in a wedge of each
plane at low $m_0$ and large $m_{1/2}$. (The outline of this wedge can be seen in each of the preceding 
lifetime plots by connecting the points where the lifetime vs. $m_{1/2}$ changes from a curve to a straight 
line.) In most of the planes at larger
$m_0$ the lightest neutralino is the NLSP. However, when $A_0 = 2.5 m_0$,  
there are also wedges at large $m_0$
and small $m_{1/2}$, shaded grey, in which the NLSP is the lighter stop squark.
Indeed for very low $m_{1/2}$, the stop mass squared is negative and hence for parameter
choices inside this grey wedge the sparticle spectrum is not physical.
There is only a very narrow strip along the wedge where the stop is actually the NLSP.
We do not consider this case in the present work (see however \cite{yudi}), discussing only the
neutralino and stau NLSP cases.

\subsection{Summary of Light-Element Abundance Constraints}

As in subsequent similar figures, the upper left panel of Fig.~\ref{fig:Aeq2.5m0_10_100m0}
displays the D/H ratio, the upper
middle panel displays the \he3/D ratio, the upper right panel displays the \he4 abundance,
the lower left panel the \li6/\li7 ratio, the lower middle panel the \li7/H ratio, and the lower
right panel the \be9/H ratio. As a general rule, we consider the regions left unshaded to be
compatible with observation, whereas the yellow regions are problematic, and the red and magenta regions
are progressively more strongly excluded. Solid shadings are used for regions with excess
abundances, and hashed shadings for regions with low abundances. 
The criteria adopted for the light-element abundances
are similar to those used in our previous work, and are summarized in Table~\ref{tab:abundances}~\footnote{The values
corresponding to `strong exclusion' are somewhat arbitrary, but serve to indicate how rapidly the
abundances are varying in relevant regions of parameter space.}. 

\begin{table}[htb]
\begin{center}
\vspace{0.5cm}
\begin{tabular}{|c||c|c|c|c|c|c|}
\hline\hline
Comparison with & D/H & \he3/D & \he4 & \li6/\li7 & \li7/H & \be9/H \\
observation & $(\times 10^{-5})$ &  &  &  & $\times 10^{-10}$ & $\times 10^{-13}$ \\
\hline\hline
{Strongly excluded} & $< 0.5$ &  $-$ & $< 0.22$ & $-$ & $< 0.1$ & $-$ \\
{Excluded} & $< 1.0$ &  $-$ & $< 0.23$ & $-$ & $< 0.2$ & $-$ \\
{Problematic} & $< 2.3$ &  $-$ & $< 0.24$ & $-$ & $< 0.5 $ & $-$ \\
\hline
Acceptable & [2.3, 3.7] &  [0.3, 1.0] & [0.24, 0.27] & ${< 0.05}$ & [0.5, 2.75] & ${< 0.3}$ \\
\hline
{Problematic} & $> 3.7$ &  $> 1.0$ & $> 0.27$ & $> 0.05$ & $> 2.75$ & $> 0.3$ \\
{Excluded} & $> 5.0$ &  $> 3.0$ & $> 0.28$ & $> 0.1$ & $> 10$ & $> 1.0$ \\
{Strongly excluded} & $> 10$ &  $> 5.0$ & $> 0.29$ & $> 0.2$ & $> 30$ & $> 3.0$ \\
\hline\hline
\end{tabular}
\end{center}
\caption{\it The ranges of light-element abundances whose
comparisons with observation we
consider in this work to be acceptable, problematic and (strongly) excluded, as shown
in the unshaded/yellow/red/magenta regions in the Figures.}
\label{tab:abundances}
\end{table}

\vspace{1cm}
\noindent
\underline{D/H} \\
We assume the mean value given in~\cite{opvs}
\beq
\label{eq:D}
\left( \frac{\rm D}{\rm H} \right)_p = \left( 3.0 \pm 0.7 \right)
  \times10^{-5} \, ,
\eeq
corresponding to the deuterium abundance measured in 10 quasar absorption systems~\cite{deut},
and the quoted uncertainty is given by the sample variance in the data. This is considerably larger
than the error in the mean, which is only 0.2. Therefore, we consider any value outside the range
$(2.3 - 3.7) \times 10^{-5}$ as problematic, as indicated in Table~\ref{tab:abundances}, which also
includes ranges that we consider to be (strongly) excluded.

\vspace{0.5cm}
\noindent
\underline{\he3/D} \\
Whilst it is difficult to use \he3 to constrain BBN, it is possible to use the ratio
\he3/D \cite{sigl}.  Although \he3 may be created or destroyed in stars, D is always
destroyed in the pre-main sequence of stellar evolution and, as a result,
the ratio \he3/D is a monotonically increasing function of time.
Thus one can use the solar ratio of about 1~\cite{geiss} to constrain the BBN ratio.
Because \he3 can be produced and/or D can be destroyed, we do not assume
a lower bound to the ratio.

\vspace{0.5cm}
\noindent
\underline{\he4} \\
Although the determination of the \he4 abundance in extragalactic HII
regions is dominated by systematic uncertainties \cite{osk},
using the Markov Chain-Monte Carlo methods described in \cite{aos2} and data
compiled in \cite{its}, one finds~\cite{aos3} 
\beq
Y_p = 0.2534 \pm 0.0083
\eeq
based on a regression of $Y$ vs. O/H and
\beq
\langle Y \rangle  = 0.2574 \pm 0.0036
\eeq
based on a weighted mean.  
As we will see, once the standard BBN value of $Y_p$ is affected by NLSP decays,
it varies very rapidly and it suffices to consider values outside the range
[0.24, 0.27] to be problematic.

\vspace{0.5cm}
\noindent
\underline{\li6/\li7} \\
Some observations of Li absorption lines in halo dwarf stars have claimed evidence for a relatively
large ratio of \li6/\li7 $\simeq 0.05$ \cite{asp06} over a broad range of metallicities, though it remains 
possible that these observations are also dominated by systematic uncertainties \cite{cayrel}.
There are a few reliable observations of stars with a similar ratio of \li6/\li7 in a very narrow range
of metallicity~\cite{li6obs} consistent with galactic cosmic-ray nucleosynthesis
\cite{gcr}. However, no observations
indicate a ratio greater than 0.05 which we set as our lower boundary of the problematic range.

\vspace{0.5cm}
\noindent
\underline{\li7/H} \\
The cosmological \li7 problem \cite{CFO} is now well established.  There are many observations of \li7
in halo dwarf stars \cite{li7obs} that indicate a far lower \li7/H abundance than predicted in 
standard BBN. We adopt the range found in the plateau of
Lithium versus metallicity~\cite{rbofn}, namely 
\begin{equation}
\left( \frac{\li7}{\rm H} \right)_{\rm halo*} \; = \; (1.23^{+0.34}_{-0.16}) \times 10^{-10} ,
\label{Li7Hhalo}
\end{equation}
although the lithium abundance observed in globular cluster stars may be
a factor $\sim 2$ higher \cite{liglob}. Although the preferred range in (\ref{Li7Hhalo}) is rather narrow,
we deem that any reduction from the BBN value of~\footnote{This corresponds to the BBN value at 
a baryon-to-photon ratio of  $\eta = 6.16 \times 10^{-10}$~\cite{wmap7}. A similar value of 
$(5.24\pm 0.5) \times 10^{-10}$ was found in another recent analysis~\cite{Coc12}.} 
 \li7/H $ = (5.07^{+0.71}_{-0.62}) \times  10^{-10}$~\cite{CFO}
to $< 2.75 \times 10^{-10}$ represents a significant improvement in the \li7 problem, and we take this to be 
the lower bound of our problematic region. NLSP decays can also destroy too much \li7
and we will consider any value below $0.5 \times 10^{-10}$ as similarly problematic. 

\vspace{0.5cm}
\noindent
\underline{\be9/H} \\
Finally, \be9 is also observed in halo dwarf stars, and is found
to vary strongly with metallicity, as seen in a recent set of observations~\cite{boes}.
These observations extend down to [O/H] of about -2.5 with a \be9/H abundance of 
$3 \times 10^{-14}$. Though there is a single observation \cite{ito} of \be9 with an abundance about 3 times lower, conservatively we will consider problematic any `primordial' abundance in excess of the highest
value seen at the lowest metallicity.

\vspace{0.5cm}
In the cases of \he3/D, \li6/\li7 and \be9/H, there are no observational lower limits, so
we do not quote ranges of abundances that we consider too low. Within the
unshaded regions, we also display extra contours for the \he4 abundance = 0.25 and 0.26 (dashed and solid 
lines, respectively), 
for \li7/H $= 0.91, 1.91 \times 10^{-10}$ (dashed and solid) and \be9/H $=1, 2 \times 10^{-14}$
(dashed and solid). In the case of \li7/H, as already discussed, it is well known that standard BBN gives a ratio significantly higher
than that indicated by observations. Therefore, large parts of the regions coloured
yellow in the \li7 panels yield an abundance that is no worse than in standard cosmology,
and may even be in somewhat better agreement with observation. Depending how
seriously one takes the cosmological \li7 problem, the favoured (unshaded) regions
in subsequent plots could be expanded. In general,
we see discontinuities in the colouring along a rising diagonal line: above it, the lightest neutralino is the NLSP,
and below it the lighter stau is the NLSP, which is the case of main interest here.

We recall from previous analyses that hadronic processes are mostly
relevant for lifetimes $\lappeq 10^4$~s, whereas electromagnetic processes
are generally dominant for longer lifetimes, i.e., at smaller $m_{1/2}$ for any
fixed value of $m_0$. We also note that, for any fixed $m_{1/2}$, the abundance of
metastable relic particles (before decay) is generally largest at large $m_0$.
We therefore expect hadronic processes to be most important when both
$m_{1/2}$ and $m_0$ are large. Indeed, in the upper right parts of the \he4 panels
in Fig.~\ref{fig:Aeq2.5m0_10_100m0} and later figures we see triangular regions
where the \he4 abundance is enhanced unacceptably by hadroproduction. This is
generally accompanied by hadronic depletion of the \he3/D ratio and enhancements
in D/H, \li6/\li7, \li7/H and \be9/H. On the other hand, staying at large $m_0$, the
dominant electromagnetic processes  at smaller $m_{1/2}$ include photodestruction
of \he4 and \li7, accompanied by photoproduction of \he3/D and D/H.

\subsection{Application to the CMSSM with a Metastable Stau NLSP}

We see in the upper panels of Fig.~\ref{fig:Aeq2.5m0_10_100m0}
 that for $A_0=2.5 \, m_0$, $\tanb=10$ and $m_{3/2}=100$~GeV the D/H ratio
is acceptable in arcs with $m_{1/2} \sim 2$~TeV and $> 4$~TeV, whereas the \he3/D ratio is
generally acceptable for $m_{1/2} > 1.6$~TeV and the \he4 abundance is acceptable throughout
the stau NLSP wedge of the $(m_{1/2}, m_0)$ plane (This demarkation is displayed in the summary plot below). 
In the lower panels of
Fig.~\ref{fig:Aeq2.5m0_10_100m0} we see that the \li6/\li7 ratio is unacceptable for
$m_{1/2} < 3$~TeV, that the \li7/H ratio is acceptable in an arc with $m_{1/2} > 2.5$~TeV, as well as in the neutralino NLSP  region
with $m_{1/2} > 4$~TeV at large $m_0$. 
The \be9/H ratio favours either $m_{1/2} > 3$~TeV or a triangular region with a neutralino NLSP with
$m_{1/2} \la 1$ TeV. The overall conclusion is that all
the light-element abundances are acceptable in a narrow arc starting on the stau/neutralino
NLSP boundary where $(m_{1/2}, m_0) \sim (4.0, 1.1)$~TeV and extending to larger $m_{1/2}$ at lower $m_0$~\footnote{We
note in passing that there is no region of the neutralino NLSP wedge where all the light-element
abundances are acceptable.}.
Note that there is excessive photo-destruction of both \he4 and \li7 in the low mass neutralino NLSP 
region. While there is a narrow strip along $m_{1/2} \approx 1.4$ TeV where \li7 is just right, this strip
is excluded by many of the other light elements. This behaviour is also seen in the subsequent parameter
choices.

\begin{figure}[ht!]
\centering
\includegraphics[width=13cm,angle=+90]{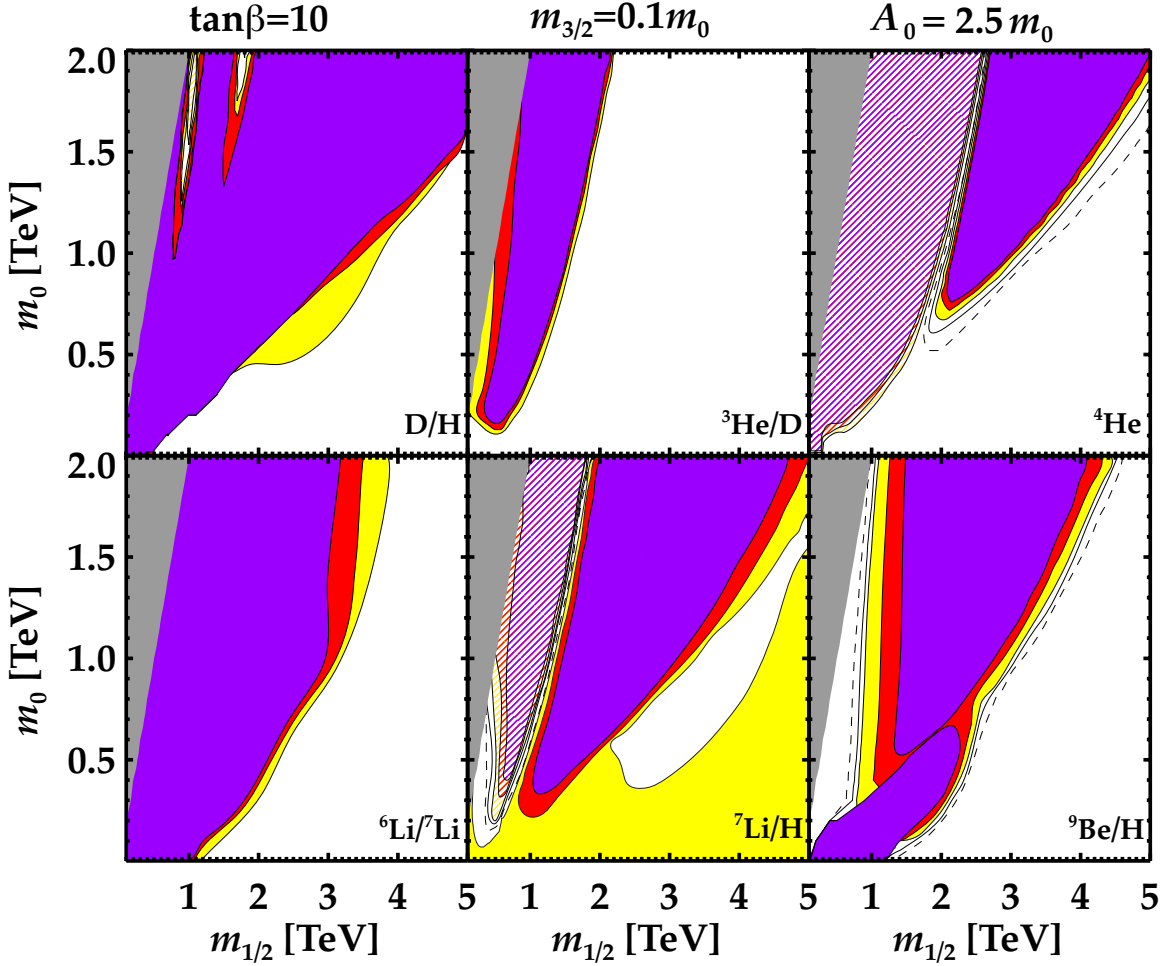}
\caption{
\it Light-element abundances in the $(m_{1/2}, m_0)$ plane for $A_0=2.5 \, m_0$, $\tanb=10$ and $m_{3/2}=0.1\,  m_0$.}
\label{fig:Aeq2.5m0_10_m32eq0.1m0}
\end{figure}

Fig.~\ref{fig:Aeq2.5m0_10_m32eq0.1m0} displays a similar analysis for the same CMSSM parameters
$A_0=2.5 \, m_0$, $\tanb=10$, but with $m_{3/2} = 0.1\,  m_0$. In this case, we see that the light-element
abundances are all acceptable in a narrow arc through the stau NLSP region between $(m_{1/2}, m_0) \sim
(2.3, 0.4)$~TeV and $\sim (4.8, 1.5)$~TeV, that is defined essentially by the D/H and \li7/H
constraints~\footnote{Again, there is no acceptable region where the neutralino is the NLSP.}.

\begin{figure}[ht!]
\centering
\includegraphics[width=6cm]{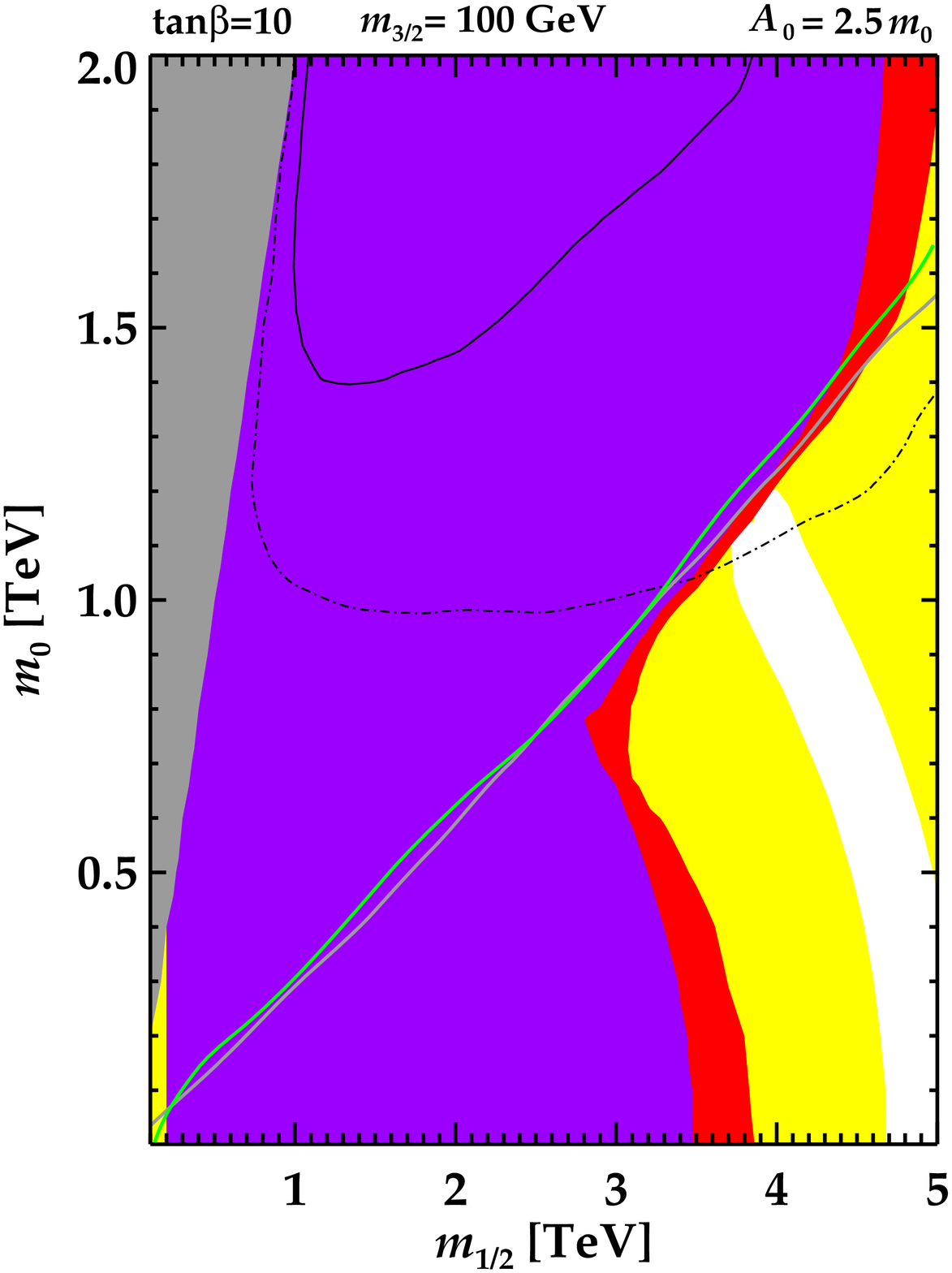}
\includegraphics[width=6cm]{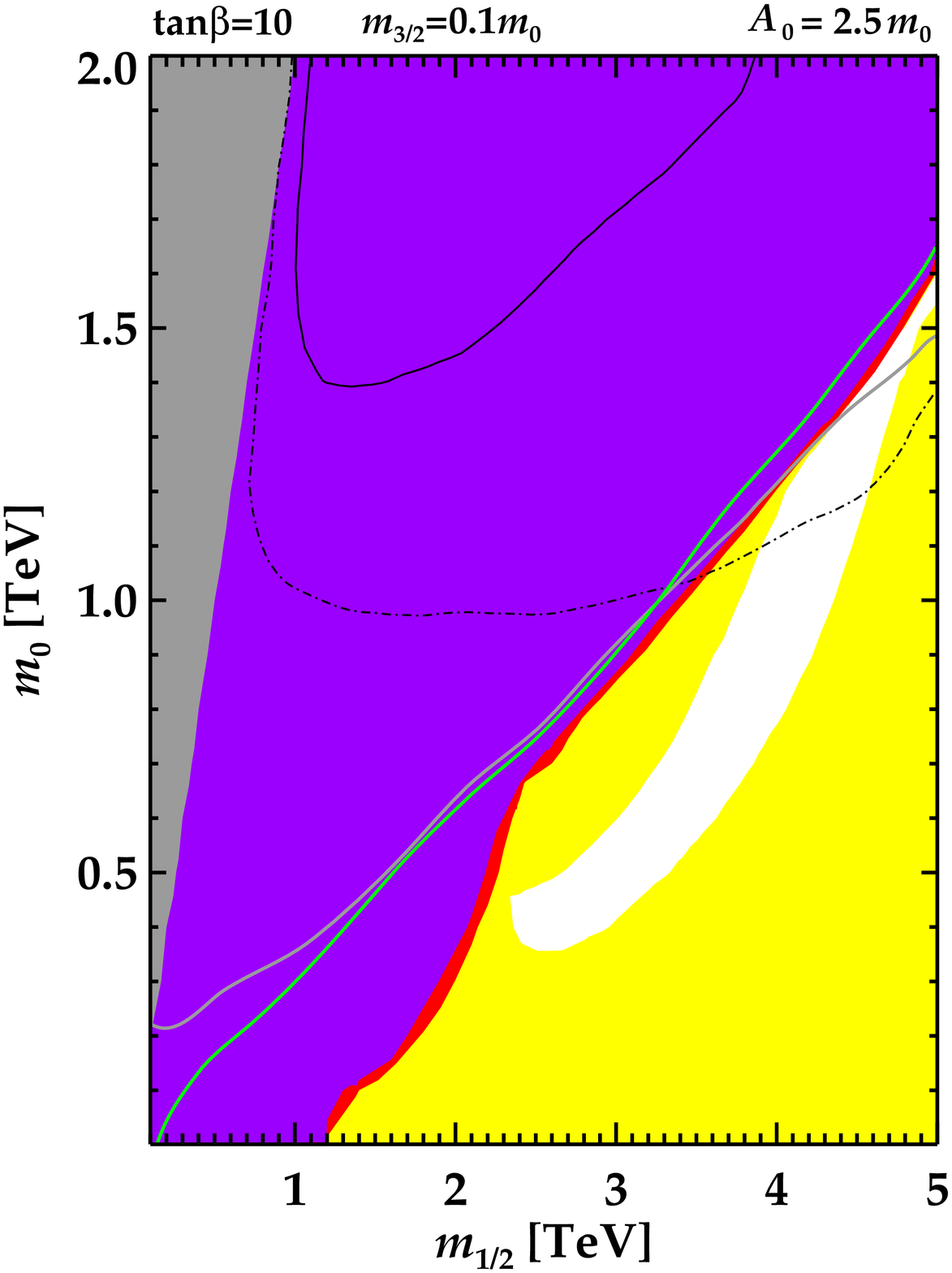}
\caption{
\it Summary of the light-element-abundance constraints in the $(m_{1/2}, m_0)$ plane for 
$A_0=2.5\, m_0$, $\tanb=10$ and $m_{3/2}=100$ GeV (left) and $m_{3/2}=0.1 \, m_0$ (right).}
\label{fig:Aeq2.5m0_10_sum}
\end{figure}

Fig.~\ref{fig:Aeq2.5m0_10_sum} summarizes our results for the CMSSM $(m_{1/2}, m_0)$ planes
for $A_0=2.5 m_0$, $\tanb=10$, with $m_{3/2} = 100$~GeV (left) and $m_{3/2} = 0.1\, m_0$ (right).
Here and in subsequent similar figures, the magenta regions are strongly excluded by one or
more constraints, the red regions are excluded by one or more constraints, and the yellow regions are problematic for at least
one constraint. We see explicitly the unshaded narrow arcs where all the constraints are satisfied. These are the
regions where the cosmological \li7 problem is solved in the presence of metastable stau NLSPs:
they are all below the grey line that marks the boundary between the neutralino
and stau NLSP wedges. 

Also shown is a green line, above which the gravitinos arising from NLSP
decays have a density greater than the range allowed by WMAP and other observations~\cite{wmap7}. 
This line corresponds to the gravitino relic abundance determined from NLSP decays
\beq
\Omega_{3/2} h^2 = \Omega_{\rm NLSP} h^2 \left( \frac{m_{3/2}}{m_{\rm NLSP}} \right) \, ,
\eeq
where $\Omega_{\rm NLSP}$ is the thermal relic density of NLSPs
left over after annihilations.  We note that there may be other sources
of gravitinos such as reheating after inflation which would
further strengthen this bound.
This constraint excludes almost completely the neutralino NLSP regions in
Fig.~\ref{fig:Aeq2.5m0_10_sum} and the subsequent analogous figures, but does not
impact the white regions compatible with all the light-element constraints. 
Also shown in this and subsequent summary figures
are some contours of calculated values of $M_h = 124$~GeV (dash-dotted),
125~GeV (solid), 126~GeV (dotted) and 127~GeV (dashed). The present experimental and
theoretical uncertainties are such that no calculated value of $M_h \in [124, 127]$~GeV can currently be
excluded, and an even larger range may be permitted at large $m_{1/2}$, where the {\tt FeynHiggs} code~\cite{FeynHiggs}
warns of theoretical uncertainties considerably exceeding 1.5~GeV.

Looking back at the contours of constant $\tau_{\rm NLSP}$ in described arcs in
Fig.~\ref{fig:tau_10_2.5}, we see that in the stau NLSP segment
of the $(m_{1/2}, m_0)$ plane they parallel the contours in the corresponding regions of the $(m_{1/2}, m_0)$
planes in Figs.~\ref{fig:Aeq2.5m0_10_100m0} and \ref{fig:Aeq2.5m0_10_m32eq0.1m0} for the different light-element abundances. This confirms the
important influence of $\tau_{\rm NLSP}$. Comparing with the summary of this case displayed in the left panel of
Fig.~\ref{fig:Aeq2.5m0_10_sum}, we see that in this case the optimal lifetime for solving the cosmological Lithium
problem is $\tau_{\rm NLSP} \sim {\rm few} \times 10^{2}$~s. In the case when $m_{3/2} = 0.1 \, m_0$ (right panel
of Fig.~\ref{fig:tau_10_2.5}), we see that in the stau NLSP segment the contours of constant $\tau_{\rm NLSP}$ parallel 
those of constant \li6/\li7 and \be9/H ratios, though the shapes of the D/H and \li7/H contours are quite different.
Looking at the right panel of Fig.~\ref{fig:Aeq2.5m0_10_sum}, we see that in this case the optimal lifetime for 
solving the cosmological Lithium problem is $\tau_{\rm NLSP} \sim {\rm few} \times 10^{3}$~s.

Though we do not show the results here, we have studied other choices for the
gravitino mass for the values of $\tan \beta = 10$ and $A_0 = 2.5 \, m_0$.
For example, for a larger fixed gravitino mass of 500 GeV, one must consider 
larger $m_{1/2} \ga 1$ TeV to ensure a gravitino LSP.  For a given gaugino mass, 
the NLSP lifetime is longer. As a result, the acceptable arc of D/H moves to larger $m_{1/2}$.
More importantly, the \li6 constraint would now exclude all values of $m_{1/2}$ between 1 and 5 TeV.
The \be9 constraint similarly would exclude the entire stau NLSP region displayed.
Had we chosen instead $m_{3/2} = m_0$, a gravitino LSP would be present only in the lower
right half of the plane. Once again lifetimes would typically be longer, affecting the 
light element abundances. In this case, only a small corner of the parameter space at very large
$m_{1/2}$ and very small $m_0$ would survive all constraints.

\begin{figure}[ht!]
\centering
\includegraphics[width=13cm,angle=+90]{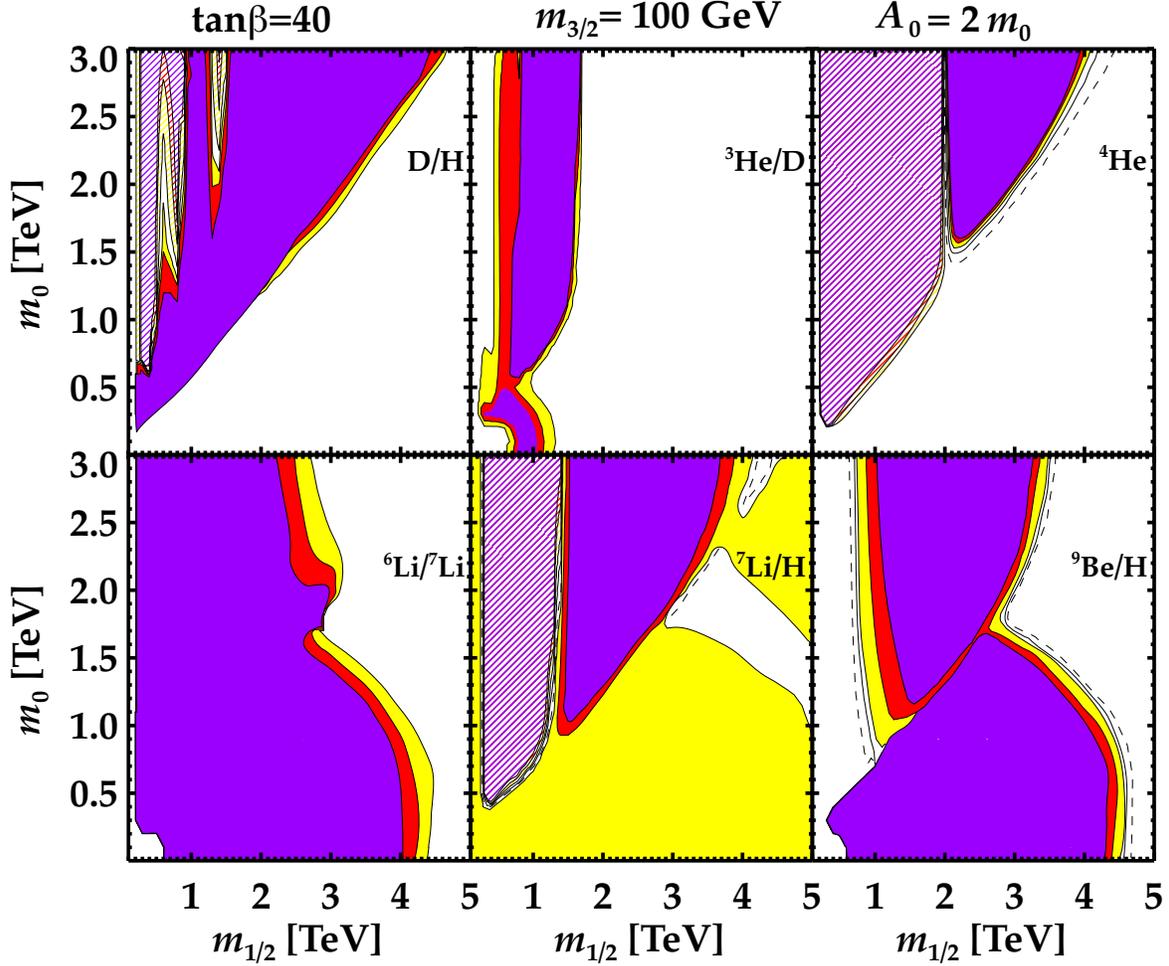}
\caption{
\it Light-element abundances in the $(m_{1/2}, m_0)$ plane for $A_0=2 \, m_0$, $\tanb=40$ and $m_{3/2}= 100$~GeV.}
\label{fig:Aeq2m0_40_100m0}
\end{figure}

We now describe an analogous analysis for the CMSSM $(m_{1/2}, m_0)$ planes
for $A_0=2 \, m_0$, $\tanb=40$. Fig.~\ref{fig:Aeq2m0_40_100m0} displays our results for the option 
$m_{3/2} = 100$~GeV. In this case, the D/H constraint would allow most of the lower
half of the parameter plane. This regions would be allowed by both
the \he3/D (except for a small area with low $m_{1/2}$ and $m_0$) and \he4 constraints, 
but much of it is excluded by the \li6/\li7 ratio, and more strongly excluded by the \be9/H ratio.
Improvement in the \li7/H ratio only occurs around an arc starting at $(m_{1/2},m_0) = (3.2,2)$ TeV.
This arc is for the most part allowed by the other constraints.

\begin{figure}[ht!]
\centering
\includegraphics[width=13cm,angle=+90]{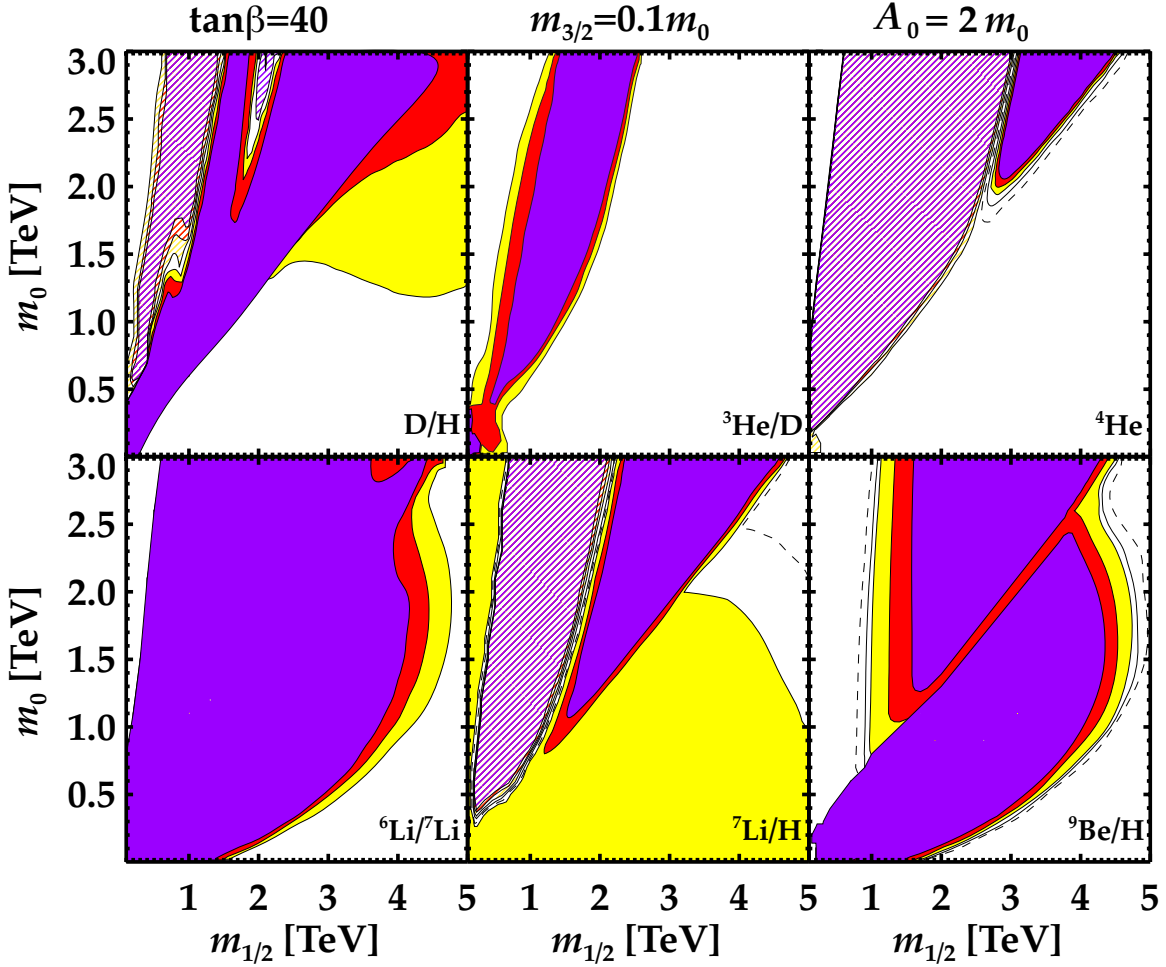}
\caption{
\it Light-element abundances in the $(m_{1/2}, m_0)$ plane for $A_0=2 \, m_0$, $\tanb=40$ and $m_{3/2}=0.1 \, m_0$.}
\label{fig:Aeq2m0_40_m32eq0.1m0}
\end{figure}

Fig.~\ref{fig:Aeq2m0_40_m32eq0.1m0} displays the results of a similar analysis for $m_{3/2}=0.1\, m_0$,
but with the same values of the CMSSM parameters. Once again, the neutralino NLSP region is
excluded by the D/H ratio, which is also problematic for a large area with $m_0 > 1$~TeV. The \he3/D and \he4
constraints are qualitatively similar to the previous case. However, the effect of the \li6/\li7
constraint is somewhat different: it excludes a bulbous region of the stau NLSP segment extending almost
to $m_{1/2} \sim 5$~TeV as does the \be9 constraint. 
In this case, the arc allowed by the \li7/H ratio is wider and has shifted to larger masses. 
As a result, the only region
that has a chance of being compatible with all the light-element constraints has $m_{1/2} \sim 5$~TeV
and $m_0 \sim 1$~TeV.

\begin{figure}[ht!]
\centering
\includegraphics[width=6cm]{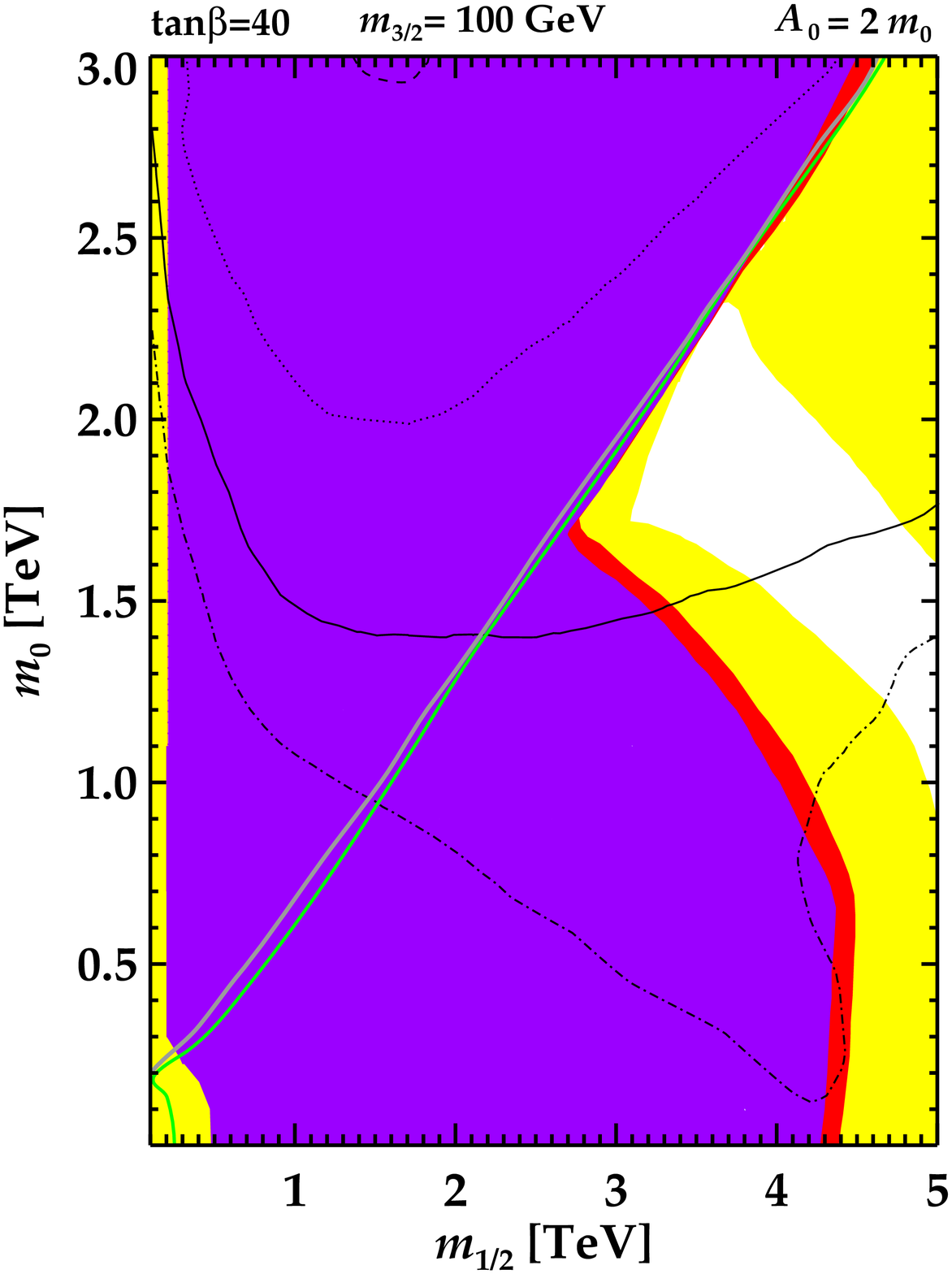}
\includegraphics[width=6cm]{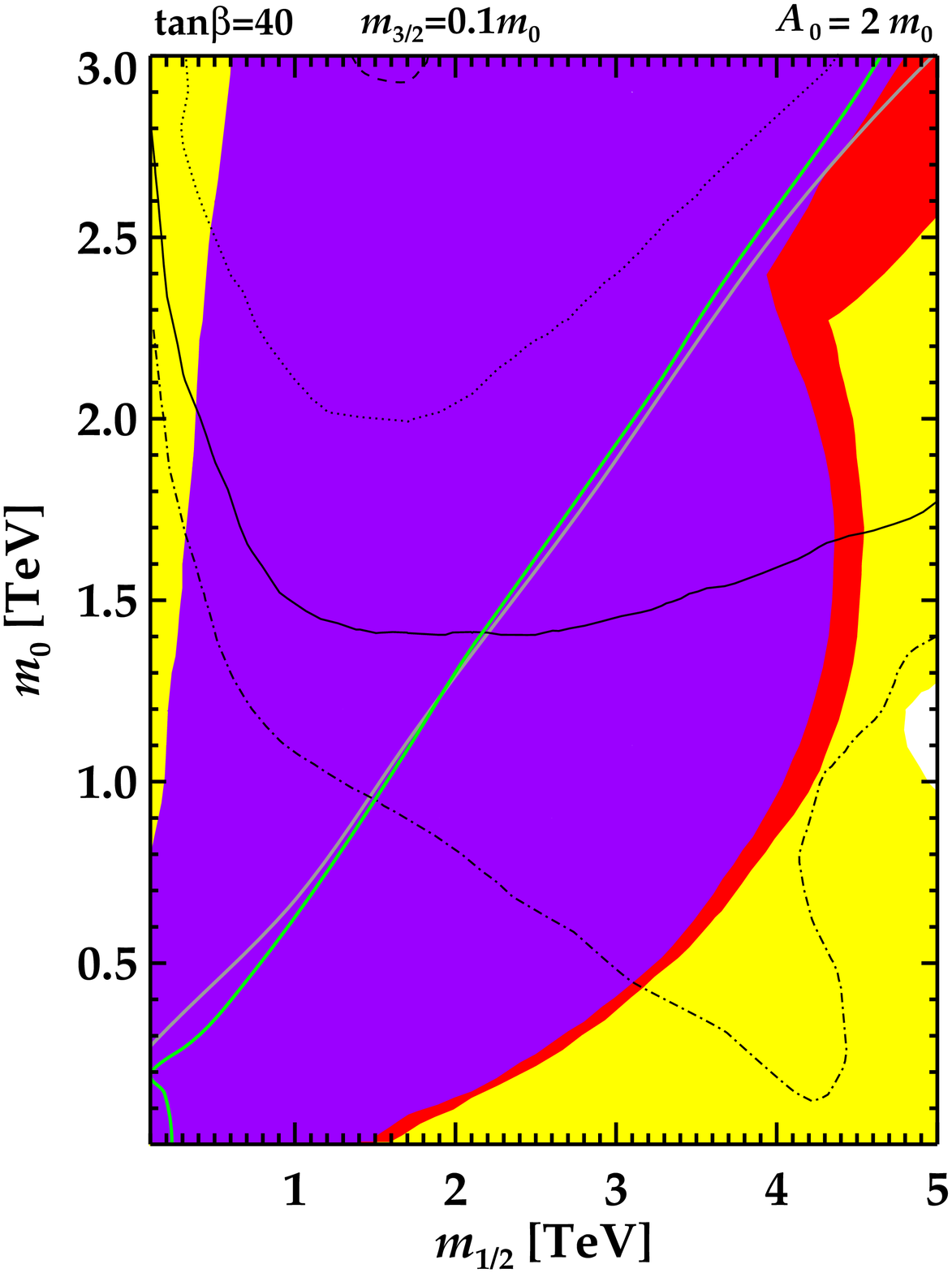}
\caption{
\it Summary of the light-element-abundance constraints in the $(m_{1/2}, m_0)$ plane for 
$A_0=2\, m_0$, $\tanb=40$ and $m_{3/2}=100$ GeV (left) and $m_{3/2}=0.1 \, m_0$  (right).}
\label{fig:Aeq2m0_40_sum}
\end{figure}

The results in the $(m_{1/2}, m_0)$ planes
for $A_0=2\, m_0$ and $\tanb=40$ are summarized in Fig.~\ref{fig:Aeq2m0_40_sum}.  In the case of
$m_{3/2}=100$ GeV (left panel) we see an allowed arc across the stau NLSP region
extending from $(m_{1/2}, m_0) \sim (3.2, 2)$~TeV to $\sim (5, 1.5)$~TeV. In the case of
$m_{3/2}=0.1\, m_0$ (right panel), there is only a very small region of marginal consistency
close to $(m_{1/2}, m_0) \sim (5, 1)$~TeV. 
In the stau NLSP region the contours of the light-element
constraints displayed in Figs.~\ref{fig:Aeq2m0_40_100m0} and \ref{fig:Aeq2m0_40_m32eq0.1m0}
(except for the D/H ratio in the latter case) again parallel the contours of
constant $\tau_{\rm NLSP}$ in Fig.~\ref{fig:tau_40_2}, and we see in Fig.~\ref{fig:Aeq2m0_40_sum} that the
preferred ranges of $\tau_{\rm NLSP}$ are somewhat below and above $10^3$~s, respectively.

For the same choice of $\tanb = 40$ and $A_0 = 2.0 m_0$, had we taken $m_{3/2} = 500$ GeV
we would have found that, due to the increased lifetimes, virtually the entire parameter plane with
a gravitino LSP would be strongly excluded by the \li6/\li7 ratio.  The \be9 constraint also
would strongly exclude the stau NLSP region shown. For $m_{3/2} = m_0$, we would once again
be forced into a tiny area in the lower right corner of the $(m_{1/2}, m_0)$ plane.

\begin{figure}[ht!]
\centering
\includegraphics[width=13cm,angle=+90]{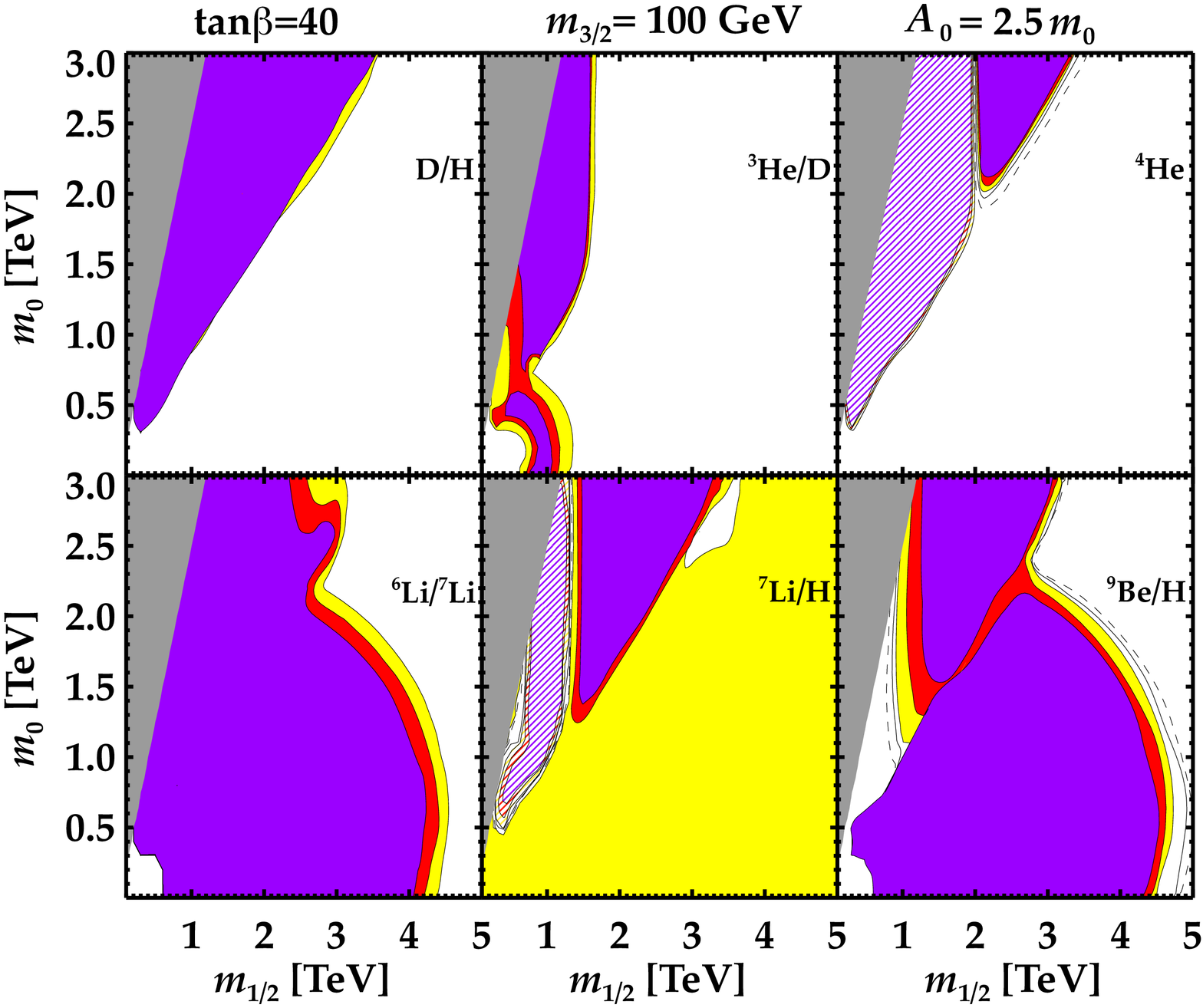}
\caption{
\it Light-element abundances in the $(m_{1/2}, m_0)$ plane for $A_0=2.5 \, m_0$, $\tanb=40$ and $m_{3/2}= 100$~GeV.}
\label{fig:Aeq2.5m0_40_100m0}
\end{figure}

Turning now to the case $A_0=2.5\, m_0$, $\tanb=40$ and $m_{3/2} = 100$~GeV shown
in Fig.~\ref{fig:Aeq2.5m0_40_100m0}, we note in particular that there is virtually no
improvement over standard BBN in the  \li7/H abundance 
throughout almost all the stau NLSP region. Only a small region close to the
stau-neutralino NLSP boundary extending to higher masses from $m_{1/2} \sim 3$~TeV
is consistent with this constraint. The \li6/\li7 and \be9 constraints once again dominate in the 
stau NLSP region.

\begin{figure}[ht!]
\centering
\includegraphics[width=13cm,angle=+90]{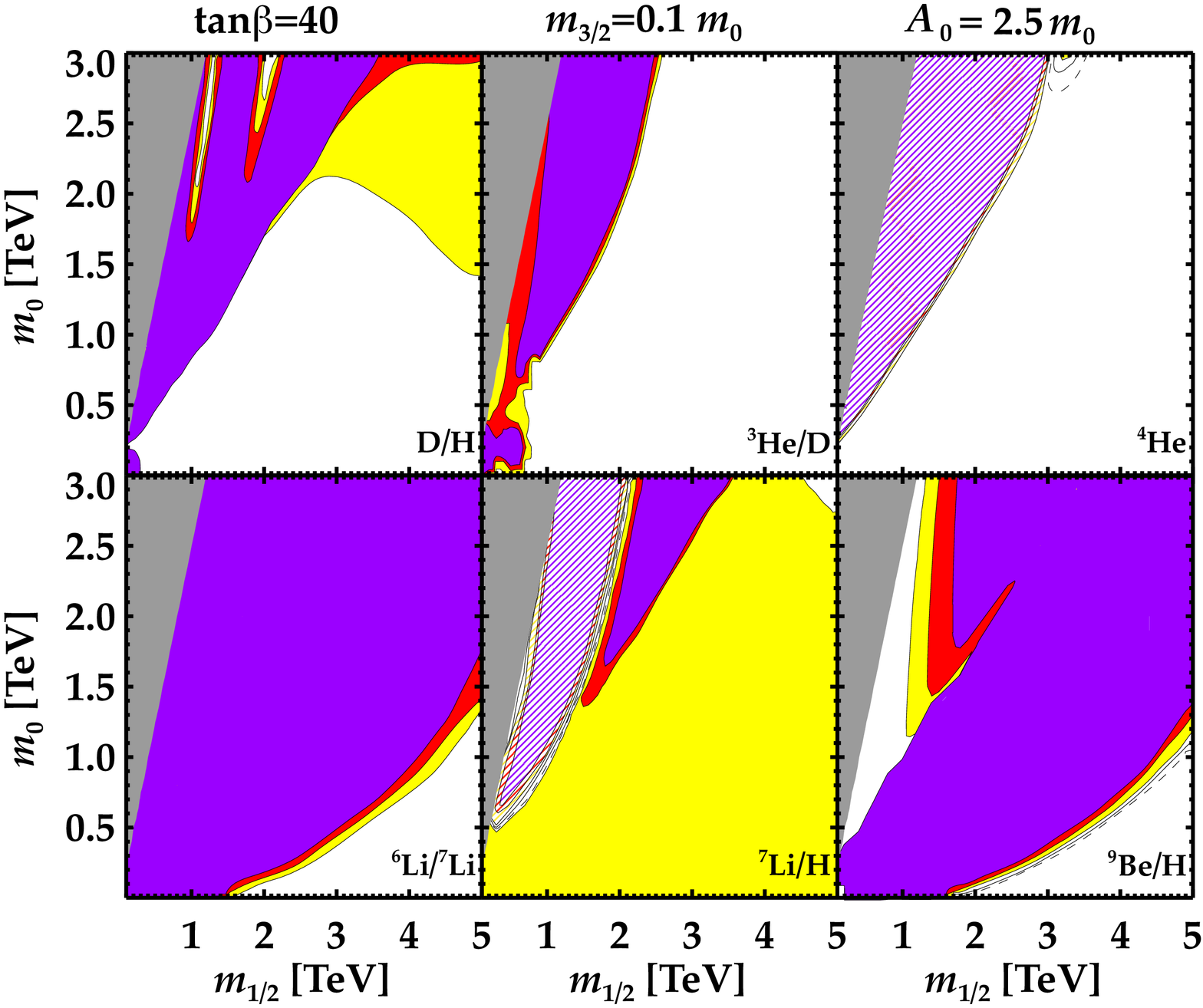}
\caption{
\it Light-element abundances in the $(m_{1/2}, m_0)$ plane for $A_0=2.5 \, m_0$, $\tanb=40$ and $m_{3/2}=0.1\, m_0$.}
\label{fig:Aeq2.5m0_40_m32eq0.1m0}
\end{figure}

Even this small region of consistency is eradicated in the case $A_0=2.5\, m_0$, $\tanb=40$ and 
$m_{3/2} = 0.1 \, m_0$ shown in Fig.~\ref{fig:Aeq2.5m0_40_m32eq0.1m0}. Now, the improvement in \li7/H
is limited to a small region at very large masses $(m_{1/2}, m_0) \sim
(5, 3)$~TeV, and this corner of parameter space is excluded by both the \li6/\li7 and \be9/H
ratios.

\begin{figure}[ht!]
\centering
\includegraphics[width=6cm]{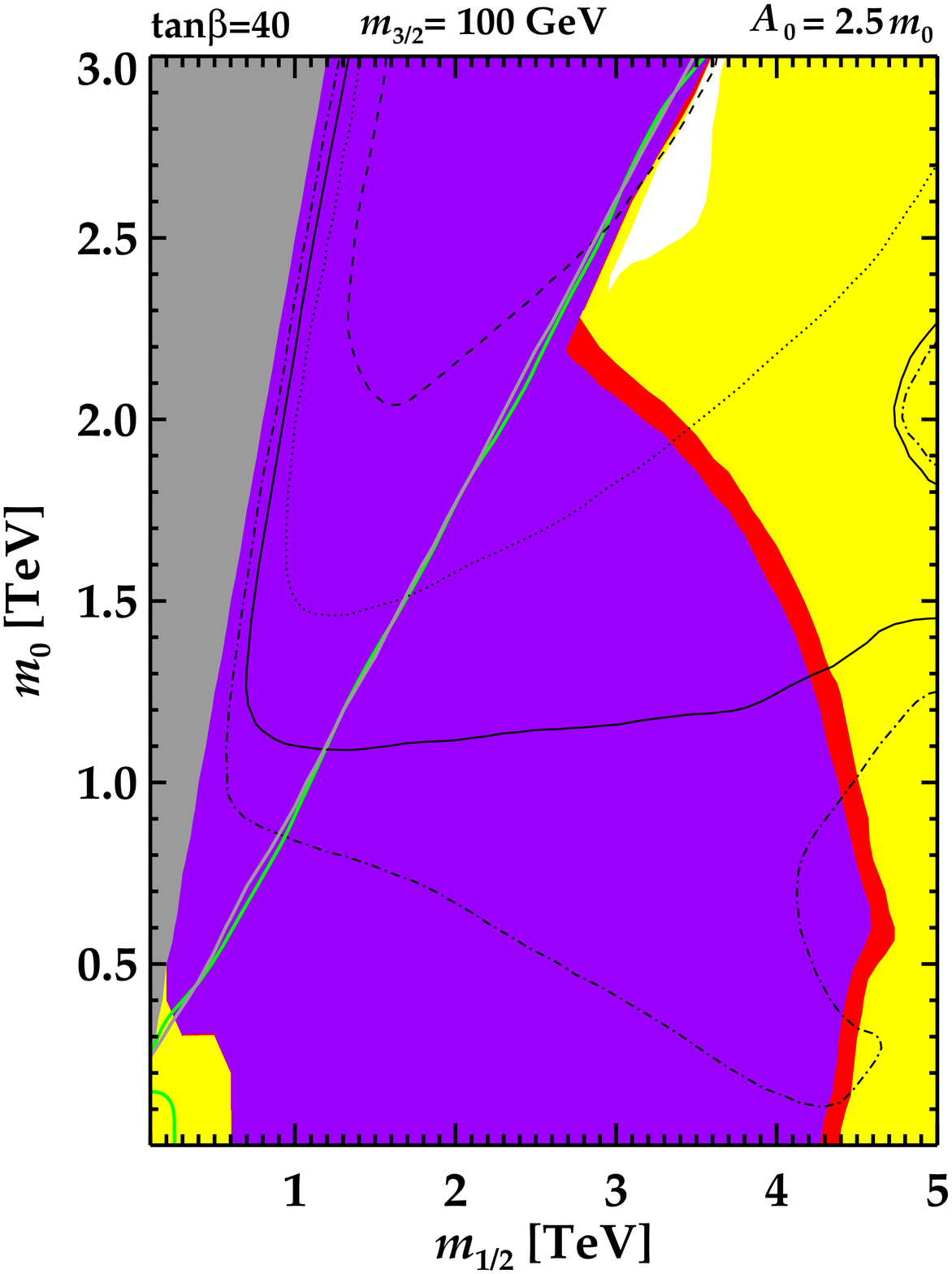}
\includegraphics[width=6cm]{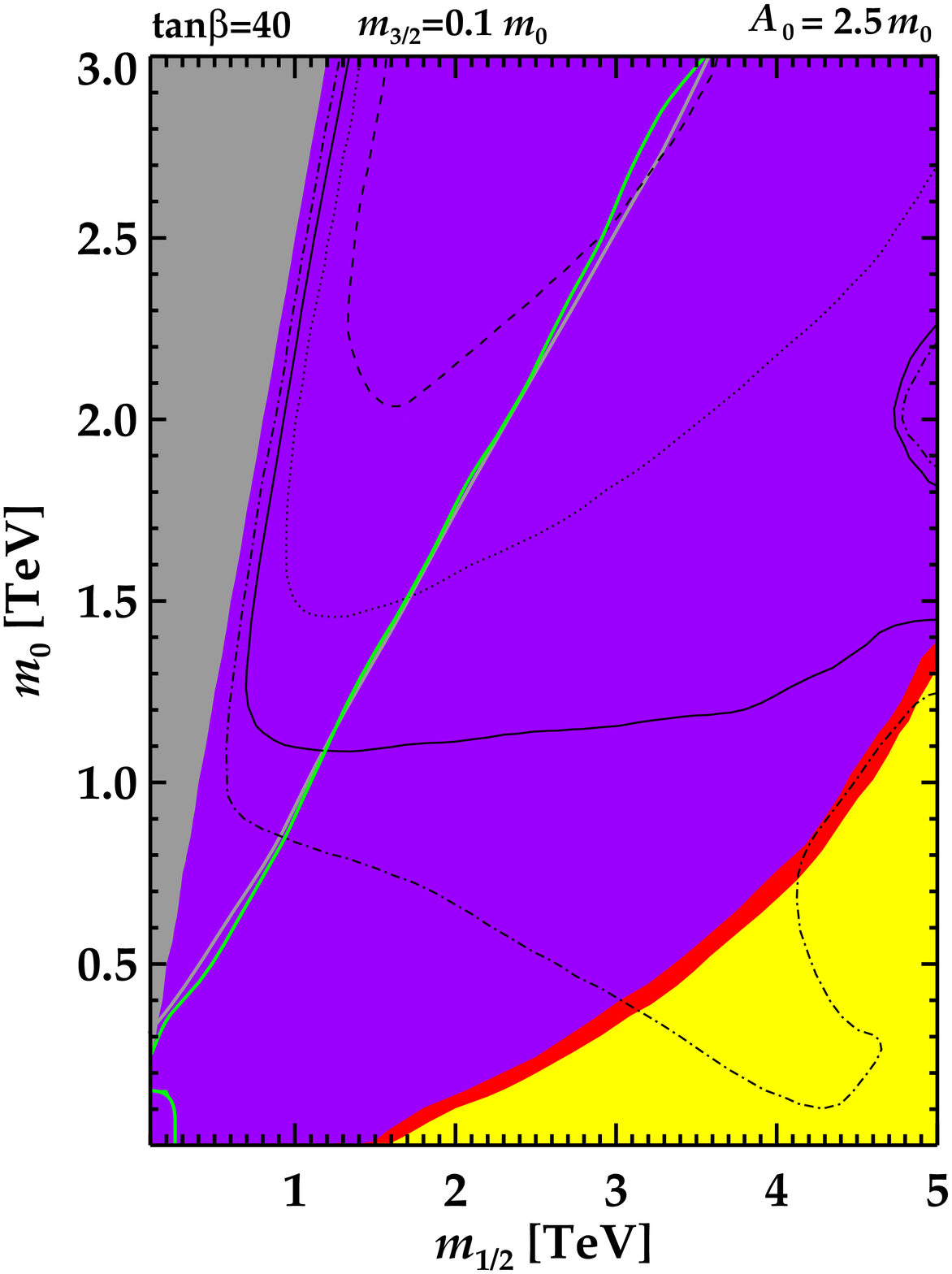}
\caption{
\it Summary of the light-element-abundance constraints in the $(m_{1/2}, m_0)$ plane for 
$A_0=2.5 \, m_0$, $\tanb=40$ and $m_{3/2}=100$ GeV (left) and $m_{3/2}=0.1\, m_0$ (right).}
\label{fig:Aeq2.5m0_40_sum}
\end{figure}

These results are summarized in Fig.~\ref{fig:Aeq2.5m0_40_sum}. We see in the left panel for
$m_{3/2} = 100$~GeV that only a very small region with $(m_{1/2}, m_0) \sim (3.5, 2.8)$~TeV
is compatible with all the constraints, whereas we see no allowed region in the 
right panel for $m_{3/2} = 0.1 \, m_0$.
Comparing the lifetime contours in Fig.~\ref{fig:tau_40_2.5}
with Figs.~\ref{fig:Aeq2.5m0_40_100m0}, \ref{fig:Aeq2.5m0_40_m32eq0.1m0}
and \ref{fig:Aeq2.5m0_40_sum}, we again see that the
preferred ranges of $\tau_{\rm NLSP}$ are $\sim 10^3$~s.
Choosing the gravitino masses $m_{3/2} = 500$ GeV or $= m_0$ would leave us with results
very similar to those described for $A_0 = 2 m_0$.

\begin{figure}[ht!]
\centering
\includegraphics[width=6cm]{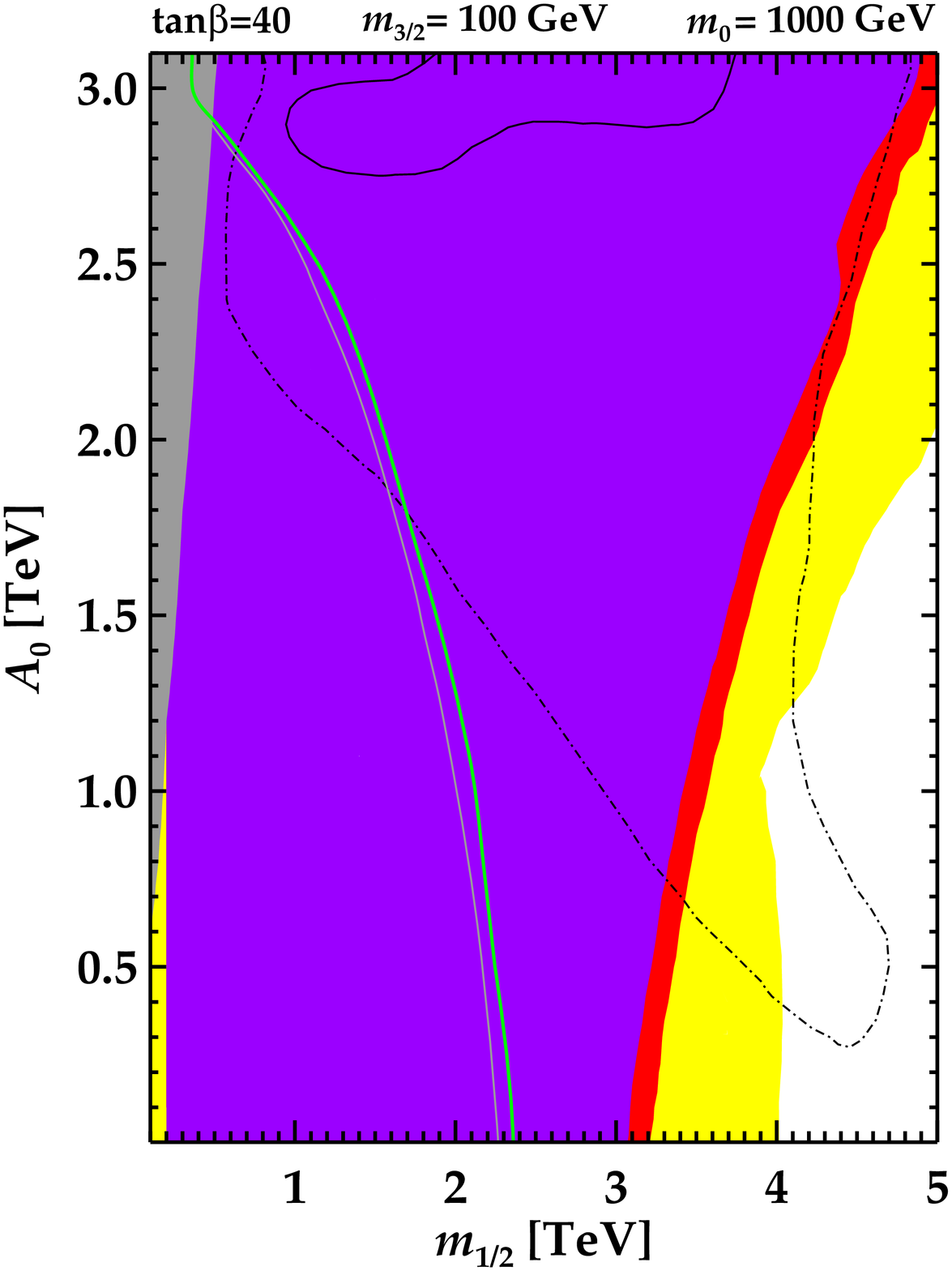}
\includegraphics[width=6cm]{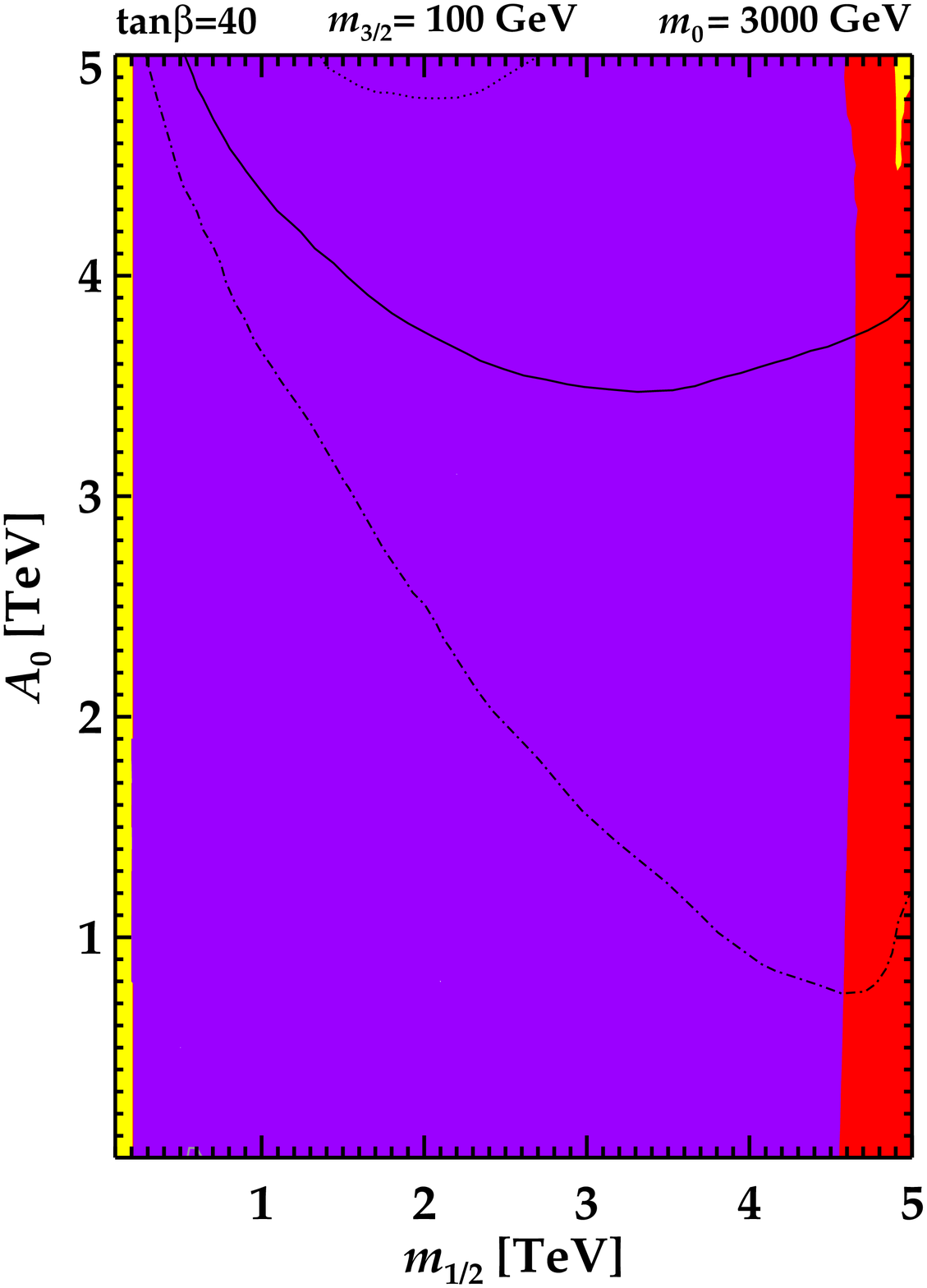}
\caption{
\it Summary of the light-element-abundance constraints in the $(m_{1/2}, A_0)$ plane for 
$\tanb=40$ and $m_{3/2}=100$ GeV with $m_0 = 1000$~GeV (left) and $m_0 = 3000$~GeV (right).}
\label{fig:varyA0}
\end{figure}

The results shown above have been for slices through the CMSSM parameter space
corresponding to $(m_{1/2}, m_0)$ planes for fixed $\tan \beta$ and $A_0$. We have also
explored how the results for $\tan \beta = 40$ vary as functions of $A_0$ for a couple of
values of $m_0 = 1000, 3000$~GeV, with the results summarized in Fig.~\ref{fig:varyA0}.
The left panel is for $m_0 = 1000$~GeV, which is typical of the range of $m_0$ in the
unshaded regions in the cases studied above. We see that a large region
with $m_{1/2} > 4$~TeV and $A_0 < 2$~TeV is unshaded and hence \li7-compatible.
On the other hand, we see no unshaded region in the right panel for $m_0 = 3000$~GeV,
which is less typical of the values of $m_0$ found in the unshaded regions of previous
summary plots. Therefore, we expect that the features found earlier are quite generic.

Also shown in Figs.~\ref{fig:Aeq2.5m0_10_sum}, \ref{fig:Aeq2m0_40_sum}, \ref{fig:Aeq2.5m0_40_sum} and \ref{fig:varyA0}
are some representative contours of the lightest MSSM Higgs boson $M_h$, as
calculated using the {\tt FeynHiggs} code~\cite{FeynHiggs}. This code is generally thought to
have an uncertainty $\sim 1.5$~GeV for generic sets of CMSSM parameters, but warns of
larger uncertainties at the large values of $m_{1/2}$ of interest here~\footnote{This may be linked
with the irregular behaviours of some calculated contours of $M_h$ in 
Figs.~\ref{fig:Aeq2m0_40_sum}, \ref{fig:Aeq2.5m0_40_sum} and \ref{fig:varyA0}.}.
Accordingly, we consider calculated values of $M_h \in [124, 127]$~GeV to be
compatible with the observed range of 125 to 126~GeV~\cite{LHCH}, and an even larger range
of calculated values of $M_h$ may be acceptable at large $m_{1/2}$. In the cases displayed
in Fig.~\ref{fig:Aeq2.5m0_10_sum}, we see that the ends of the BBN-compatible arcs with higher $m_0$
have $M_h \sim 124$~GeV, i.e., within the acceptable range, and hence may be preferred.
In Figs.~\ref{fig:Aeq2m0_40_sum} we see that the preferred arc for $m_{3/2} = 100$~GeV
corresponds to $M_h \sim 124$ to 126~GeV, all within the range suggested by the LHC,
whereas in the case $m_{3/2} = 0.1 \, m_0$ the BBN-compatible region has $M_h \sim 124$~GeV.
In Fig.~\ref{fig:Aeq2.5m0_40_sum} we see that the small BBN-compatible region for
$m_{3/2} = 100$~GeV corresponds to a nominal value of $M_h \sim 127$~GeV, at the
upper end of the LHC-compatible range. Finally, in Fig.~\ref{fig:varyA0} we see that the
unshaded region in the left panel corresponds generally to $M_h \sim 124$~GeV,
which is compatible within theoretical uncertainties with the LHC discovery.

\subsection{Bound-State Effects and Uncertainties}

We conclude this Section with a brief discussion of the importance of
bound-state effects and their uncertainties.

\begin{figure}[ht!]
\centering
\includegraphics[width=13cm,angle=+90]{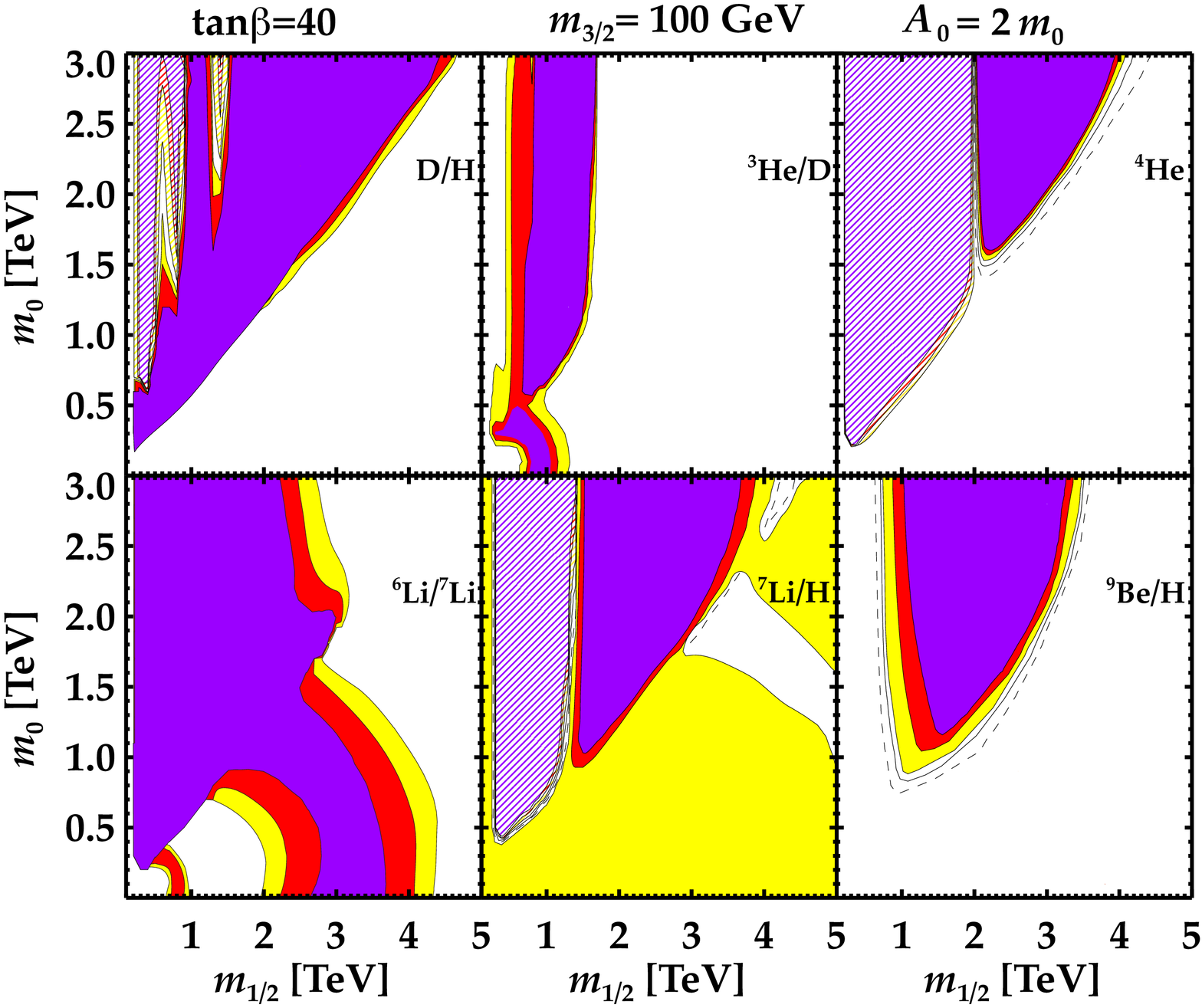}
\caption{
\it Light-element abundances in the $(m_{1/2}, m_0)$ plane for $A_0=2 \, m_0$, $\tanb=40$ and $m_{3/2}= 100$~GeV,
with decay effects retained but all bound-state effects switched off.}
\label{fig:noBS}
\end{figure}

\begin{figure}[ht!]
\centering
\includegraphics[width=5cm]{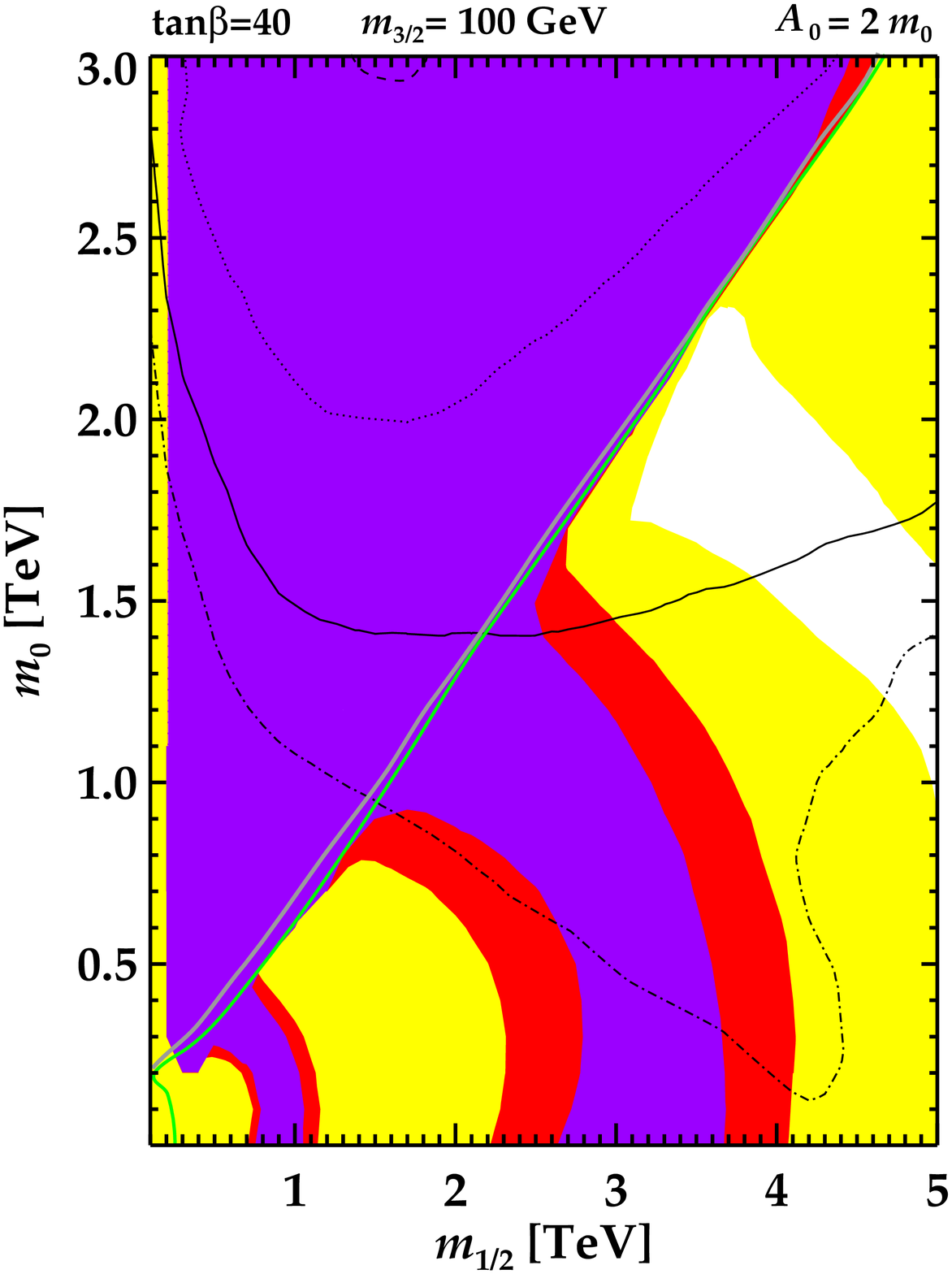}
\includegraphics[width=5cm]{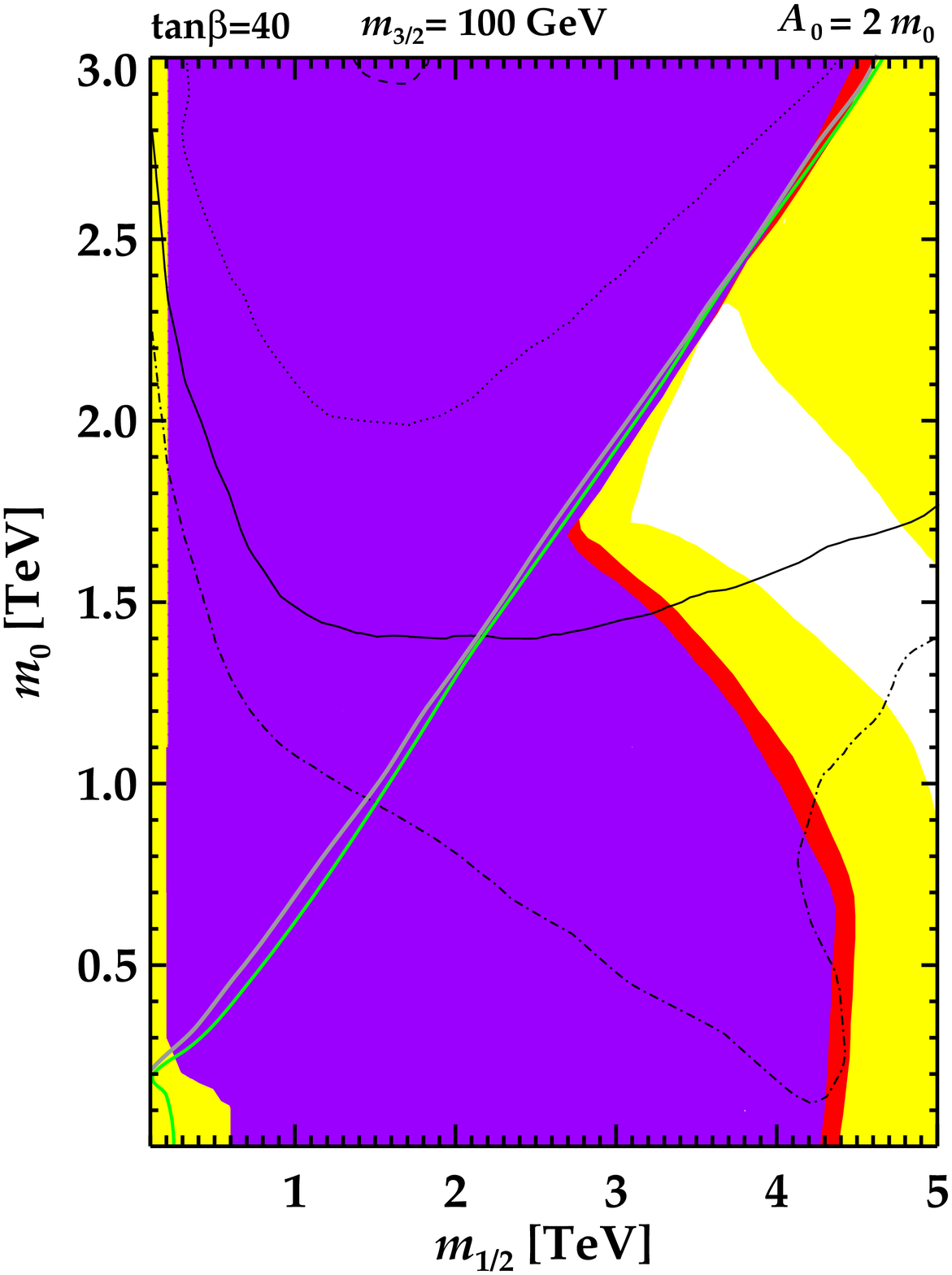}
\includegraphics[width=5cm]{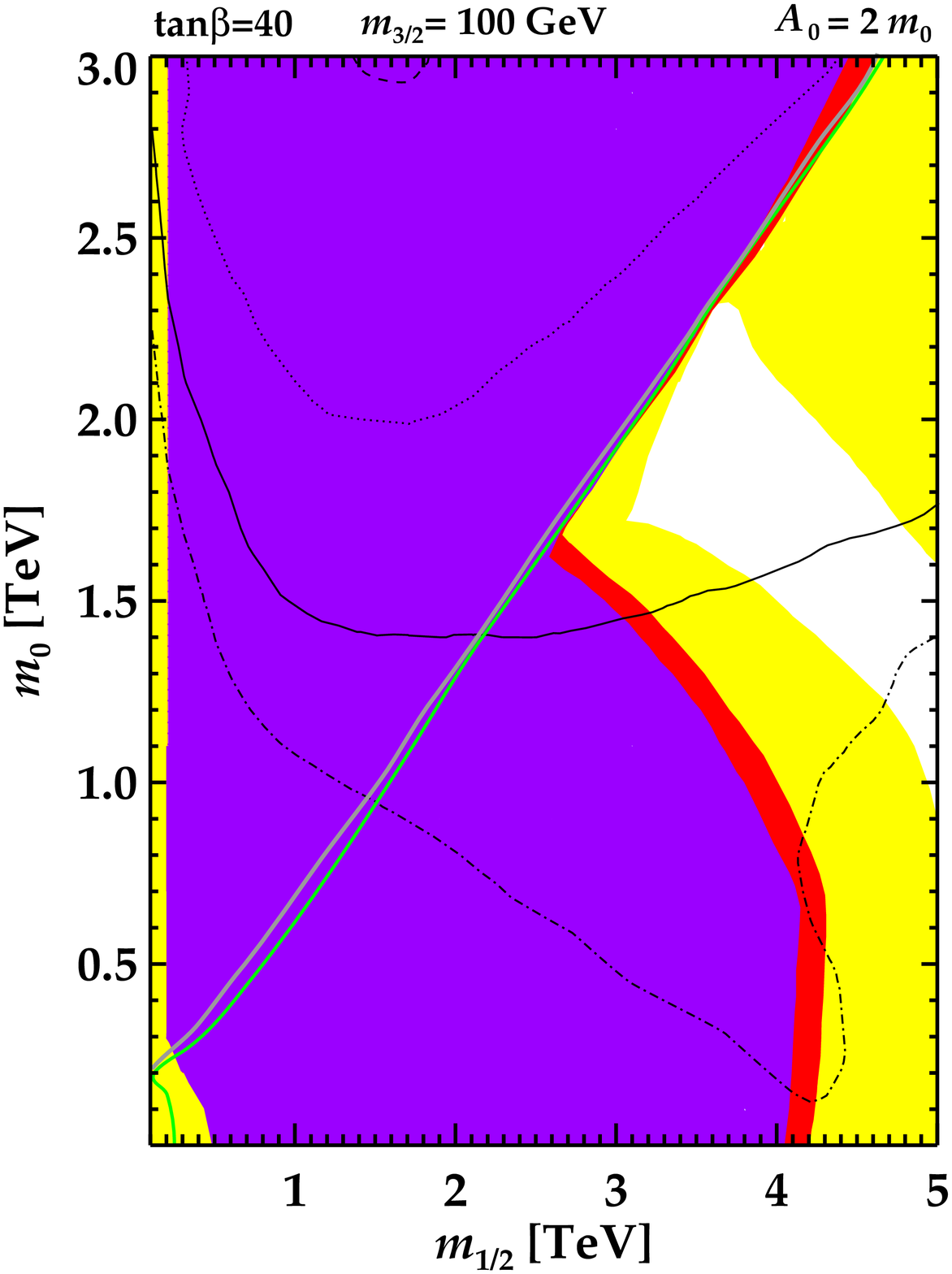}
\caption{
\it Summary of the light-element-abundance constraints in the $(m_{1/2}, m_0)$ plane for 
$A_0=2.5 \, m_0$, $\tanb=40$ and $m_{3/2}=100$ GeV, if all bound-state effects
were switched off (left), with greater \be8-X binding energy than our default choice
(centre) and with \be9 bound-state production suppressed (right).}
\label{fig:motherofallsummaries}
\end{figure}

Fig.~\ref{fig:noBS} shows how our results in the $(m_{1/2}, m_0)$ plane
for $A_0=2 \, m_0$, $\tanb=40$ and $m_{3/2}= 100$~GeV
would change if all bound-state effects
were to be switched off, but with decay effects retained.  Comparing with
Fig.~\ref{fig:Aeq2m0_40_100m0}, we see that the D/H, \he3/D and \li7/H
ratios are unaffected, as is the \he4 abundance. However, there are major changes in the
\li6/\li7 ratio and \be9 abundance. In particular, arcs at small $m_{1/2}$ and $m_0$ that
would have been permitted (modulo the lack of improvement in the  \li7/H abundance) in the 
absence of bound-state effects are robustly excluded
by both the \li6/\li7 and \be9/H ratios once bound-state effects are included.
On the other hand, the allowed arc at larger $m_{1/2}$ and $m_0$ is
quite unaffected by bound-state effects, as seen by comparing the left panel of
Fig.~\ref{fig:motherofallsummaries} with the left panel of Fig.~\ref{fig:Aeq2m0_40_sum}.
Even more dramatically, we see that the stau NLSP region excluded by \be9
was entirely due to bound state effects.

As discussed in Section~2, bound-state \be9 production hinges on
two principal uncertainties in our bound-state analysis.
One of these uncertainties 
is the $(\be8 X)$ binding energy, must be high enough
to allow for $(\he4 X^-) + \he4 \rightarrow (\be8X^-)+\gamma$ to be
exothermic, i.e., $Q > 0$ (cf.~eq.~\ref{eq:8be_lim}). 
Our analysis has assumed as default the 
$B_8 = 1.1679$ value of ref~.\cite{Kamimura09}, which implies
the formation reaction is strongly exothermic, and thus the reverse
photodissociation of $(\be8 X^-)$ is strongly suppressed.
We have also considered both greater and smaller values of
the binding energy.
Using the larger value $B_8 = 1.408$ MeV~\cite{Pospelov07}
makes the reaction even more exothermic;
this gives results that are almost identical to our
default analysis, since the bound-state formation rate remains very similar. 
The central panel of Fig.~\ref{fig:motherofallsummaries} summarizes the overall
effect on the allowed region of $(m_{1/2}, m_0)$ plane for 
$A_0=2 \, m_0$, $\tanb=40$ and $m_{3/2}=100$~GeV, which is almost indistinguishable from the
default result shown in the left panel of Fig.~\ref{fig:Aeq2m0_40_sum}.

\begin{figure}[ht!]
\centering
\includegraphics[width=13cm,angle=+90]{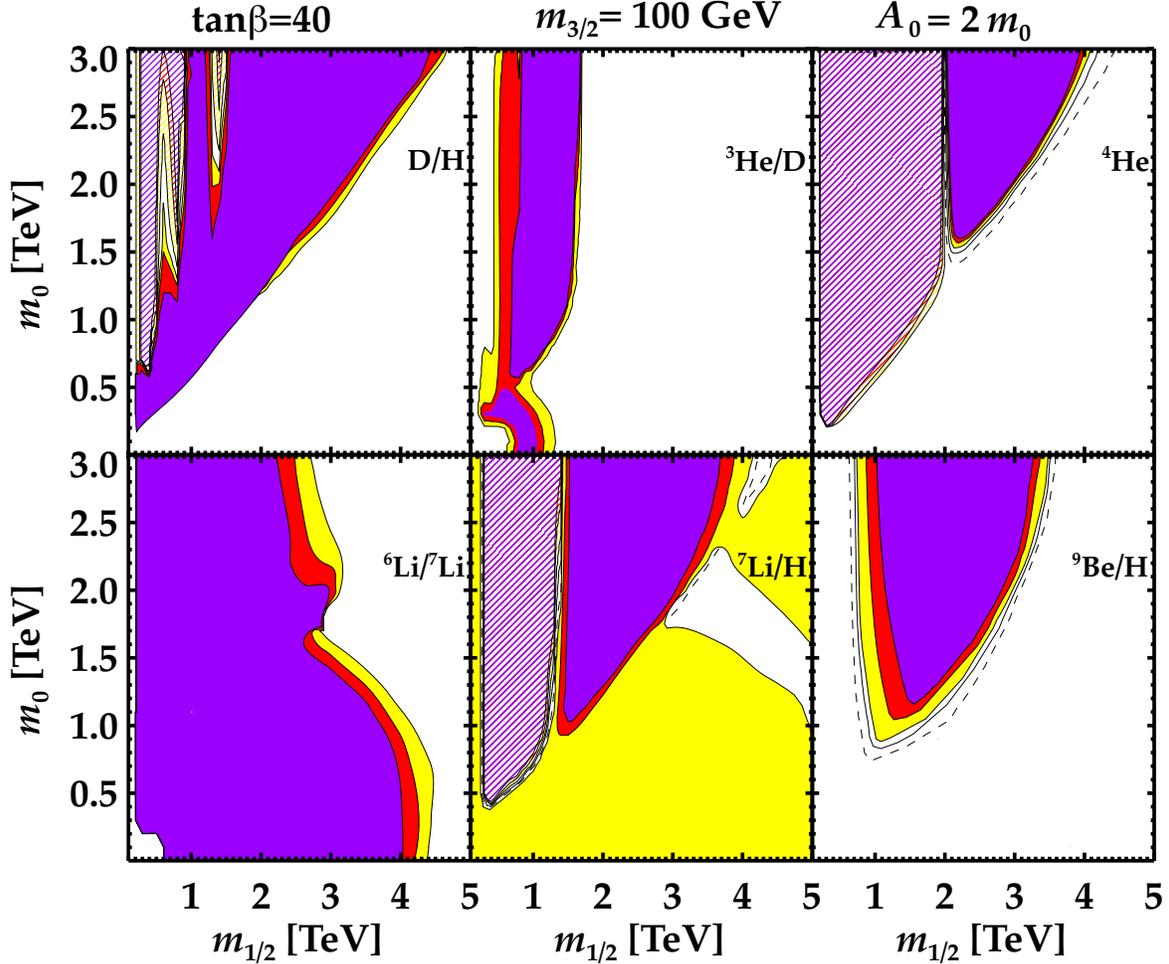}
\caption{
\it Light-element abundances in the $(m_{1/2}, m_0)$ plane for $A_0=2 \, m_0$, $\tanb=40$ and $m_{3/2}= 100$~GeV,
with bound-state \be9 production suppressed.}
\label{fig:RC}
\end{figure}

Our own three-body estimate of the $(\be8 X^-)$ binding energy in Section~2
gives
$B_8 = 492 \pm 50$ keV,
a value  that exceeds the effective `no-go' limit in 
(\ref{eq:8be_lim})only by 
$Q = B_8 - B_8^{\rm min} = 53$ keV.
In this situation, $(\be8 X^-)$ production is weakly exothermic
but remains highly vulnerable to photodissociation
back to $(\he4 X^-) + \he4$.  This reverse reaction 
suppresses $(\be8 X^-)$ formation
until the temperature drops
to $\sim Q/| \ln \eta | \sim 2$ keV. But at this late time,
no free neutrons are available, and the result is effectively that
no bound-state \be9 production occurs.  
Results for this case appear in Fig.~\ref{fig:RC}. 
We see that in the lower-right region, where
bound-state effects are important, the \be9 production is now missing.
Indeed, we have checked that the results are unchanged if 
bound-state \be9 production is switched off entirely,
as would be the case if the 
if $(\be8 X)$ binding energy 
drops below the limit in eq.~(\ref{eq:8be_lim}). 
Note also that even in the absence of bound-state production,
\be9 contours do remain in the upper-left region in Fig.~\ref{fig:RC}. 
In this regine
the thermalized \he4 fragments deuterium,
tritium and \he3 are overproduced, and these can still make \be9 
via reactions with background \be7.

The right panel of Fig.~\ref{fig:motherofallsummaries} 
summarizes the overall effect of suppressed \be9 bound-state production
on the allowed region of the $(m_{1/2}, m_0)$ plane for 
$A_0=2 \, m_0$, $\tanb=40$ and $m_{3/2}=100$~GeV. Perhaps surprisingly, there
is little visible effect, and specifically none on the allowed unshaded arc at large
$m_{1/2}$ and $m_0$. This is because the \be9 and \li6/\li7 constraint contours
very closely 
shadow each other in the lower part of the $(m_{1/2}, m_0)$ plane.

A second uncertainty in \be9 production comes from the requirement
that the $(\be8 X^-) + n \rightarrow (\be9 X^-) + \gamma$ reaction
is on resonance with the first excited state of $(\be9^* X^-)$.
As noted in section~2, this requires that the 
$(\be8 X^-)$ and $(\be9^* X^-)$ bindings conspire 
in such a way that the entrance channel is on resonance,
as would be the case for the larger binding energy
$B_8 = 1.408$ MeV.
Because the $(\be9^* X^-)$ binding is quite uncertain, 
this possibility remains viable.  However, if the excited
state level turns out to fall far ($\ga 100$ keV) 
from the $(\be8 X^-)+n$ entrance, then the reaction will
be non-resonant and suppressed.  And here again,
the bound-state \be9 production would become unimportant,
similar to the results in 
Fig.~\ref{fig:RC}
and in the right panel of Fig.~\ref{fig:motherofallsummaries}.

We conclude that, whereas the overall bound-state effects are very important,
the principal uncertainties associated with the $(\be8 X)$ binding energy have little
effect on our final results. This comes about because the regions excluded by
bound-state \be9 production overlap almost completely with those also excluded
by bound-state \li6 production.
In particular, though our analysis incorporated the
resonant $(\be9^\ast X^-)$ reaction rate postulated in~\cite{Pospelov07}, our final results are
not very sensitive to this assumption.

\section{Summary}

We have presented in this paper a new treatment of the possible effects of
bound states of metastable charged particles in the light-element
abundances yielded by Big-Bang Nucleosynthesis (BBN), including in
our analysis calculations of the abundances of D, \he3, \he4, \li6, \li7 and \be9.
We have applied our code to the case of metastable ${\tilde \tau_1}$ NLSPs
in the framework of the CMSSM with a gravitino LSP. Motivated by the
discovery of (apparently) a Higgs boson weighing $\sim 125$ to 126~GeV,
we have concentrated on regions of the CMSSM in which this may be
interpreted as the lightest neutral Higgs boson, specifically $(m_{1/2}, m_0)$
planes with $A_0 \ge 2 \, m_0$. 

We find interesting examples in which the light-element
abundances are as consistent with observations as are calculations of standard
homogeneous BBN with no metastable relic particles. Indeed, we find
generic strips of the CMSSM parameter space in which the cosmological
\li7 problem may be solved without altering unacceptably the abundances
of the other light elements. Examples are given for $\tan \beta = 10$ and 40
with Higgs masses compatible with the LHC discovery.

Characteristics of these models include relatively large values of the
soft supersymmetry-breaking parameters $m_{1/2}$ and $m_0$, with
heavy supersymmetric particles that could not be detected directly at the LHC.
Another characteristic of these models is that the ${\tilde \tau_1}$ lifetime is 
${\cal O}(10^3)$~s. Avenues for future research include a more complete
examination of the CMSSM parameter space, possible extensions to more
general supersymmetric models, as well as to non-supersymmetric scenarios.
In addition, we reiterate the need
for careful study of the nuclear physics behind
bound-state \be8 and \be9 production.

The observational situation with the \li6 and \li7 abundances is still
evolving, and the fat lady has not yet sung the final aria in the cosmological
lithium saga. It is perhaps still possible that the current discrepancy with
standard homogeneous BBN will eventually dissipate. However, we have
shown that, should it survive, it could have a plausible supersymmetric
solution.

\section*{Acknowledgments}
B.D.F.~is pleased to acknowledge illuminating discussions with 
Chris Hirata regarding bound-state recombination, and we thank
Sven Heinemeyer for discussions about {\tt FeynHiggs} calculations.
The work of R.H.C.~was supported by the U.S. National Science
Foundation Grants PHY-02-016783 and PHY-08-22648 (JINA).  
The work of
J.E. and F.L. was supported in part by the London Centre for Terauniverse
Studies (LCTS), using funding from the European Research Council via
the Advanced Investigator Grant 267352: this also supported
visits by K.A.O. and V.C.S. to the CERN TH Division, which they thank
for its hospitality. 
The work of B.D.F.~was partially supported by the
U.S. National Science
Foundation Grant PHY-1214082. The work of F.L. and K.A.O. was supported in part
by DOE grant DE--FG02--94ER--40823 at the University of Minnesota, and 
the work of F.L. was also supported in part by the Doctoral Dissertation Fellowship at the University of Minnesota. 
The work of V.C.S. was supported by Marie Curie International
Reintegration grant SUSYDM-PHEN, MIRG-CT-2007-203189.

\appendix

\section{Recombination of $X$ Bound States}

A comparison of  $(AX^-)$ recombination with 
that of ordinary hydrogen recombination reveals important similarities
but also crucial differences.
The recombination of ordinary cosmic hydrogen and helium
does not proceed primarily through 
$pe \rightarrow (pe)_1$ transitions
directly to the $n = 1$ ground state.
This is because such recombinations
emit Lyman limit photons with energy $E_\gamma = B_{(pe)} = 13.6$ eV,
which have a large cross section.  Thus, during the
era of (ordinary) recombination, these photons
have a short mean free path  against absorption
by neighboring ground-state hydrogen atoms.
Thus, at this epoch the universe is optically thick to 
Lyman limit photons.
Consequently, the overwhelming majority of recombinations to the ground
state in one atom lead to a reionization of a neighboring
atom, and there is no net change in the number of atoms.

Ordinary recombination therefore proceeds via transitions initially to 
excited states, particularly the
$n = 2$ first excited state, and then to the ground state.
However, the cosmic plasma is also optically
thick to $2P \rightarrow 1S$ Lyman-$\alpha$ photons, 
and so the transition to
the ground state is dominated by the much slower
two-photon $2S \rightarrow 1S$ transition.
Because of these effects, ordinary hydrogen recombination
is not instantaneous, but delayed due to the
`bottleneck' of the large optical depth for 
Lyman series photons.

Our NLSP case of $pX^- \rightarrow (pX)$ and $\he4 X^- \rightarrow (\he4 X)$
recombination is controlled by the same underlying atomic physics,
and under conditions of a similar matter-to-photon ratio.
It thus is worthwhile to check whether we expect
similar effects.

For the hydrogen and NLSP recombination, we are interested in the 
absorption of \lya\ and `\xlya' photons, respectively,
by ground-state atoms:
\beq
\gamma_{1 \rightarrow 2} + (\anion p)_{1} \rightarrow (\anion p)_{2} \ \ ,
\eeq
where the `anion' $\anion \in (e,X)$ corresponds
to ordinary and NLSP recombination, respectively.
Note, however, that in the NLSP case the
proton, being lighter, plays the role of the electron
in setting the relevant reduced mass $\mu$,
so $\mu = m_e$ for hydrogen and $\mu = m_p$ for NLSP.
On the other hand, the atomic mass $m_p + m_{\cal A} - B_{\cal A}$
is well approximated by
$m = m_p$ for hydrogen and $m = m_X$ for NLSP.

The \xlya\ photon optical depth against absorption by $(pX)$
atoms is
\beq
\tau_\alpha(\anion p) 
 = \sigma_\alpha(\anion p) \ n_\anion \ d_{\rm hor}
 \approx \sigma_\alpha(\anion p) \ n_\anion \ t \  \ ,
\eeq
so the ratio of hydrogen and NLSP optical depths is
\beq
\frac{\tau_\alpha(Xp)}{\tau_\alpha(ep)}
 = \frac{\sigma_\alpha(Xp)}{\sigma_\alpha(ep)}
   \frac{n_X}{n_e}
   \frac{t_{\rm bbn}}{t_{\rm cmb}} .
\eeq
The \xlya\ resonance cross section is, in the notation of ref~\cite{Peebles},
\beq
\sigma_\alpha = \frac{3}{8\pi} \lambda_\alpha^2 
  \frac{\galph^2}{(\omega-\omega_\alpha)^2 + \galph^2/4} \, ,
\eeq
where
$\lambda_\alpha \propto 1/E_\alpha \sim (\alpha^2 \mu)^{-1}$,
and $\Gamma_{2p \rightarrow 1s} \propto \mu$ is the decay rate.
Thermal broadening dominates over this width,
with $\delta \omega/\omega \sim v_T/c \sim \sqrt{T/m}$
where $m$ is the atomic mass.
Thus we have an effective mean cross section
\beq
\bar{\sigma}_\alpha \sim \lambda_\alpha^2 \frac{\galph^2}{\delta \omega^2}
  \sim \lambda_\alpha^2 \frac{\galph^2}{\omega_\alpha^2} \frac{m}{T}
  \propto \mu^{-2} \frac{m}{T}
  \propto \mu^{-2} \frac{m}{T_0} (1+z)^{-1} .
\eeq
where $z$ is the redshift.
The appropriate  number densities are the physical, not comoving,
values, and are set by $n_e \approx n_p$,
and by $n_X = Y_X n_p$, with
$n_p \sim n_{\rm B} \propto (1+z)^3$.
Finally, in the ordinary recombination case we have $z_{\rm cmb} \sim 1000$ and 
$t_{\rm cmb} \sim 400,000 \ \rm yr \sim 10^{13} \ \rm sec$, whereas in the NLSP
case we have $z_{\rm bbn} \sim 4 \times 10^8$ and $t_{\rm bbn} \sim 100 \ \rm sec$.

Putting the above information together, we have
\beqar
\frac{\tau_\alpha(Xp)}{\tau_\alpha(ep)}
  & = & \pfrac{m_e}{m_p}^2 \ 
        \pfrac{m_X}{m_p} \
        Y_X \pfrac{1+z_{\rm bbn}}{1+z_{\rm cmb}}^2 \
        \frac{t_{\rm bbn}}{t_{\rm cmb}} \\
  & \approx & 
  5 \times 10^{-7} \ \pfrac{Y_X}{0.01} \ \pfrac{m_X}{100 \ \rm GeV} .
\eeqar
Thus we see that the \xlya\ optical depth at NLSP 
recombination is much smaller than that of ordinary recombination.
However, the optical depth for ordinary recombination
is enormous, $\tau(ep) \sim 10^9$,
so
\beq
\tau_\alpha(Xp) \sim 500
  \ \pfrac{Y_X}{0.01} \ \pfrac{m_X}{100 \ \rm GeV} \ \ .
\eeq
We find an  optical depth against $(pX)$ absorption that is
much larger than unity.
Hence it would seem that this effect could also be important for the NLSP recombination case.

However,
in the case of NLSP recombination there is an additional
process to be considered that has no analogue in ordinary recombination,
namely the Compton scattering of \xlya\ photons on free electrons and positrons.  
In ordinary recombination, electrons act as both the dominant
photon scattering agent when they are unbound, and as the 
negatively-charged partners in the bound states. However, in our
$(AX^-)$ case, these roles are now separated to
$e^\pm$ and $X^-$ respectively.
The optical depth against electron scattering
can be estimated using the ordinary Thomson cross section $\sigma_{\rm T}$:
\beqar
\tau_{e\gamma} & = & n_{e,\rm net}^{\rm bbn} \sigma_{\rm T} t_{\rm bbn}
  \ge n_{\rm B}^{\rm bbn} \sigma_{\rm T} t_{\rm bbn} \\
  & \sim & 5 \times 10^{7} \
   \pfrac{T_{\rm bbn}}{100 \ \rm keV} .
\eeqar
This is a lower bound, because we have used
the {\em net} electron number $n_{e,\rm net} = n_{e^-}-n_{e^+} \simeq n_{\rm B}$,
whereas pairs dominate the {\em total} $e^\pm$ budget:
\beq
\frac{n_{e^-}+n_{e^+}}{n_{\rm B}} \sim  \eta^{-1} \pfrac{m_e}{T}^{3/2} e^{-m_e/T} \gg 1
\eeq
down to $T \sim m_e/\ln \eta^{-1} \simeq m_e/25 \sim 20 \ \rm keV$.
Note that during ordinary recombination, the optical
depth against Thompson scattering  drops below $\tau_{e\gamma} \sim 1$,
and cannot compete successfully with
resonant \lya\ scattering.

We see that an \xlya\ photon will typically suffer at least
$\sim \tau_{e\gamma}/\tau_\alpha(Xp) \sim 500$
Compton scatterings before encountering
a bound state that it could reionize.
Each of these scatterings degrades the photon energy,
pulling it out of resonance.
If the scattering were off a nonrelativistic electron,
the photon would lose energy according to the Compton 
formula
\beq
E_\gamma^\prime = \frac{E_\gamma}{1+\frac{E_\gamma}{m_e} (1-\cos \theta)} ,
\eeq
and we would expect an approximate mean energy loss per
scattering of 
\beq
\frac{\Delta E_\gamma}{E_\gamma} \sim \frac{E_\gamma}{m_e} .
\eeq
The Lyman photons of interest have $E(\xlya) = 3/4 \ B(AX)$,
and we have $B(\he4X) = 348$ keV
and $B(pX) = 25$ keV; each species recombines
at roughly $T \sim B/\ln \eta^{-1} \sim B/25$.

Thus \he4 recombination occurs when pairs are abundant,
whereas protons recombine when pairs have completely annihilated.
In either case, the Compton opacity dominates the
resonance opacity, so that Lyman photons scatter many times
before encountering a bound state.
Moreover, in the first scattering the Lyman photons suffer energy losses
$\Delta E/E \sim {\cal O}(1)$ for $(\he4X)$
and $\Delta E/E \sim {\cal O}(10^{-1})$ for $(pX)$.
The Lyman photons are thus thermalized rapidly,
long before they interact with any ground state
atoms. We conclude that NLSP recombination to the
ground state can occur unimpeded, unlike the case of ordinary recombination.

\section{Specification of the Supersymmetric Model Framework}

In this paper we analyze the possible implications of bound states of
massive metastable particles on Big-Bang Nucleosynthesis in the
context of the minimal supersymmetric extension of the Standard Model (MSSM).
In this model, there is a supersymmetric partner for each Standard Model particle,
and there are two Higgs doublet supermultiplets linked via a mixing parameter $\mu$.
The interactions are restricted to the same gauge and Yukawa interactions as in
the Standard Model, so the quantity $R = (-1)^{3B + L + 2S}$ is conserved
multiplicatively, where $B$ and $L$ are the baryon and lepton numbers, respectively,
and $S$ is the spin. As a consequence of $R$ conservation, the lightest supersymmetric particle (LSP)
is stable, and a candidate for cosmological dark matter.

We further assume the presence of soft supersymmetry-breaking fermion masses $m_{1/2}$,
scalar masses $m_0$ and trilinear parameters $A_0$ which are each universal at the
grand unification scale, a framework known as the constrained MSSM (CMSSM) \cite{cmssm1,Ellis:2012aa}.
In addition to $m_{1/2}, m_0$ and $A_0$, we treat the ratio of Higgs vacuum expectation
values, $\tan \beta$, as a free parameter. Motivated by the apparent discrepancy
between the experimental measurement \cite{g-2} of the anomalous magnetic moment of the muon,
$g_\mu -2$, and theoretical calculations within the Standard Model \cite{g-2th} we assume that the
MSSM Higgs mixing parameter $\mu$ is positive.
As specified, the CMSSM includes no prediction for the mass of the gravitino, $m_{3/2}$,
which we treat as a free and independent parameter.

We consider here the case in which the gravitino is the LSP,
and the next-to-lightest supersymmetric particle (NLSP) is the spartner of one of the
Standard Model particles. Possible candidates include the lightest neutralino $\chi$,
the lighter stau slepton ${\tilde \tau_1}$, the spartner of the right-handed electron or muon,
${\tilde e_R}$ or ${\tilde \mu_R}$, or the lighter stop squark ${\tilde t_1}$.
In the CMSSM as described above with a gravitino LSP, the most generic 
of these candidates for the NLSP are
the lightest neutralino $\chi$ and the lighter stau slepton ${\tilde \tau_1}$, and the
latter is the candidate we consider as an example of a charged metastable NLSP.

\section{Three-Body Stau Decays}
\label{app:stau3B}

We give here the matrix elements for $\stau^- \to \grav
\, \tau^-   Z $ and $\stau^- \to \grav 
\, \nu_\tau  W^-$, which are the three-body stau decay processes 
that are most relevant for our study. In the Feynman diagrams for these two processes, 
the vertices involving the outgoing gravitino are given in Appendix A of~\cite{Luo:2010he}, 
and the vertices involving only the MSSM fields are given in~\cite{Haber:1984rc} and~\cite{Gunion:1984yn}.
In the context of the CMSSM, left-right mixing needs to be taken into account
only for the third generation of sfermion fields. Neutrinos are treated the same way as in the Standard Model, 
i.e., as massless, purely left-handed neutrinos (and right-handed anti-neutrinos). 
Following Appendix B of~\cite{CEFLOS}, we write the relation between the mass eigenstates, 
$\stau_{1,2}$, and the interaction eigenstates, $\stau_{L,R}$, as 
\beqar
\label{eq:stau_rotation_matrix}
    \left( \begin{array}{c}
            \stau_L \\ \stau_R
        \end{array} \right) =
    \left( \begin{array}{cc}
            U_{\stau \, 1L} & U_{\stau \, 2L} \\
             U_{\stau \, 1R} & U_{\stau \,
             2R}
        \end{array} \right)
    \left( \begin{array}{c}
            \stau_1 \\ \stau_2
        \end{array} \right). 
\eeqar
In order to be quite general, we give expressions for $\stau_j^- \, (j=1, 2)$, where  
the stau NLSP is the lighter of the two mass eigenstates. 

At tree-level, the Feynman diagrams contributing to $\stau_j^- (p_1) \to \grav (p_2) \, \tau^- (p_3) \, Z (p_4)$ 
are the contact diagram and the $\tau^-$, $\stau_k^- \, (k=1, 2)$ and neutralino $\tilde{\chi}^0_k \, (k=1 - 4)$ 
exchange diagrams. The partial matrix elements for each of these diagrams (suppressing
spin and polarization indices) are
\beqar
\label{eq:stau3b_contact}
i \mathcal{M}_{\mathrm {contact}} 
&=& 
\bar \psi_\mu (p_2) \left( - \frac{i \sqrt 2}{\mpl} \frac{g}{\cosw} \right) 
\left[\left(T^3_f - Q_f \sin^2 \theta_W \right) U_{\stau \, jL} P_R + Q_f \sin^2 \theta_W U_{\stau \, jR} P_L \right] \eta^{\mu \nu} \, \nonumber \\
&& \cdot  \,  v(p_3)  \, \epsilon^*_\nu (p_4) , 
\eeqar

\beqar
\label{eq:stau3b_tau_exchange}
 i \mathcal{M}_{\stau_j^- \to \grav  \tau^{- *} \to \grav  Z  \tau^-} 
&=& 
\bar \psi_\mu (p_2) \left( - \frac{i \sqrt 2}{\mpl} \right) \left(U_{\stau \, jR} P_L -  U_{\stau \, jL} P_R \right) p^\mu_1 \, \frac{i \left(-\slashed{p}_1 + \slashed{p}_2 + m_\tau \right)}{\left(p_1 - p_2 \right)^2 - m_\tau^2}  \nonumber \\
&&  \cdot  \left(\frac{i g}{\cosw}\right) \gamma^\nu  \left[\left(T^3_f - Q_f \sin^2 \theta_W \right) P_R - Q_f \sin^2 \theta_W  P_L \right] \, v(p_3) \, \epsilon^*_\nu (p_4) , \nonumber \\ 
\eeqar

\beqar
\label{eq:stau3b_stau_exchange}
 i \mathcal{M}_{\stau_j^- \to Z  \stau_k^{- *} \to Z  \grav  \tau^-} 
&=& 
\bar \psi_\mu (p_2) \left( - \frac{i \sqrt 2}{\mpl} \right) \left(U_{\stau \, kR} P_L -  U_{\stau \, kL} P_R \right)  \left(p_1 - p_4\right)^\mu \frac{i}{\left(p_1 - p_4 \right)^2 - m_{\stau_k}^2} \nonumber \\
&& \cdot  \left(- \frac{i g}{\cosw}\right) \left[\left(T^3_f - Q_f \sin^2 \theta_W \right) U_{\stau \, jL} U_{\stau \, kL}^* - Q_f \sin^2 \theta_W  U_{\stau \, jR} U_{\stau \, kR}^*  \right] \nonumber \\
&& \cdot  \left(2 p_1 - p_4 \right)^\nu  v(p_3) \, \epsilon^*_\nu (p_4)  , 
\eeqar

\beqar
\label{eq:stau3b_chi_exchange}
 i \mathcal{M}_{\stau_j^- \to \tau^-  \tilde{\chi}^{0  *}_k \to \tau^-  \grav  Z} 
&=& 
\bar \psi_\mu (p_2) \left( - \frac{i}{\mpl} \right) \bigg[ \Big( H_L \, \eta^{\mu \nu} - \frac{1}{4} \, G_L \, [ \, \slashed{p}_4 \, , \gamma^\nu \, ] \, \gamma^\mu \Big) P_L \nonumber \\
&& + \Big( H_R \, \eta^{\mu \nu} - \frac{1}{4} \, G_R \, [ \, \slashed{p}_4 \, , \gamma^\nu \, ] \, \gamma^\mu \Big) P_R  \bigg] \frac{i \left(\slashed{p}_1 - \slashed{p}_3 + m_{\tilde{\chi}^{0}_k} \right)}{\left(p_1 - p_3 \right)^2 - m_{\tilde{\chi}^{0}_k}^2} \nonumber \\
&& \cdot \, i \big[ \left( C_L U_{\stau \, jL} + D_L U_{\stau \, jR} \right) P_L + \left( C_R U_{\stau \, jR} + D_R U_{\stau \, jL} \right) P_R \big] \, v(p_3) \, \epsilon^*_\nu (p_4) , \nonumber \\ 
\eeqar
where $\bar \psi_\mu (p_2)$ represents the outgoing gravitino with momentum
$p_2$, ${v}(p_3)$ represents the outgoing chiral fermion with
momentum $p_3$, $\epsilon^*_\nu (p_4)$ is the polarization four-vector for the outgoing gauge boson
with momentum $p_4$, $\eta^{\mu \nu}$ is
the flat-space Lorentz metric tensor, 
$H_L = m_Z \left(\cos \beta N^*_{k3} - \sin \beta N^*_{k4} \right)$, $G_L = \cosw N^*_{k2} - \sinw N^*_{k1}$, 
$H_R = H_L^{\,*}$, $G_R = G_L^{\,*}$, $C_L = - (g m_\tau N^*_{k3})/(\sqrt 2 m_W \cos \beta)$, 
$D_L = \sqrt 2 g N^*_{k1} \tan \theta_W Q_f$, $C_R = C_L^{\, *}$, 
$D_R = - \sqrt 2 g \big[ N_{k2} \, T^3_f + \tan \theta_W  N_{k1} \left(Q_f - T^3_f \right) \big]$, $T^3_f = -1/2$, 
$Q_f = -1$ and $N$ is the unitary matrix used to diagonalize the neutralino mass matrix (details can be found in~\cite{Gunion:1984yn}). 

For $\stau_j (p_1) \to \grav (p_2) \, \nu_\tau (p_3) \, W^- (p_4)$, the Feynman diagrams contributing to this process, 
corresponding to eq.~(\ref{eq:stau3b_contact}), (\ref{eq:stau3b_tau_exchange}), (\ref{eq:stau3b_stau_exchange}) 
and (\ref{eq:stau3b_chi_exchange}), are the contact diagram, and the $\tau^-$, $\tilde{\nu}_\tau$ and chargino 
$\tilde{\chi}^-_k \, (k=1, 2)$ exchange diagrams, respectively. The partial matrix elements are obtained by
making the substitutions $g \to g \sqrt 2 \cosw$, $T_f \to 1/2$ and $Q_f \to 0$ in eq.~(\ref{eq:stau3b_contact}) - 
(\ref{eq:stau3b_stau_exchange}), $m_{\stau_k} \to m_{\tilde \nu_\tau}$, $U_{\stau \, kR} \to 0$, $U_{\stau \, kL} \to 1$, 
$U_{\stau \, kR}^* \to 0$ and $U_{\stau \, kL}^* \to 1$ in eq.~(\ref{eq:stau3b_stau_exchange}) 
since there is no right-handed sneutrino in the MSSM. Also, in eq.~(\ref{eq:stau3b_chi_exchange}) 
$m_{\tilde{\chi}^{0}_k} \to m_{\tilde{\chi}^\pm_k}$, and now the coefficients are $H_L = \sqrt 2 m_W \cos \beta U^*_{k2}$, 
$G_L = U^*_{k1}$, $H_R = \sqrt 2 m_W \sin \beta V_{k2}$, $G_R = V_{k1}$, $C_L = D_L = 0$, 
$C_R = \left(g m_\tau U_{k2} \right) / \left( \sqrt 2 m_W \cos \beta \right) $ and $D_R = - g U_{k1}$, 
where $U$ and $V$ are the unitary matrices used to diagonalize the chargino mass matrix.

\end{document}